\documentclass[twocolumn]{aastex} 

\usepackage{xspace}
\usepackage{graphicx}
\usepackage{natbib}
\usepackage{amsmath}
\usepackage{booktabs}
\usepackage{multirow}
\usepackage{afterpage}

\newcommand{\hide}[1]{}

\newcommand{\lsim}{\ensuremath{\,\lesssim\,}\xspace}
\newcommand{\gb}{\ensuremath{{\it b}}\xspace}
\newcommand{\absb}{\ensuremath{\vert\,\gb\,\vert}\xspace}

\newcommand{\microns}{\ensuremath{\,\mu{\rm m}}\xspace}

\newcommand{\kpc}{\ensuremath{\,{\rm kpc}}\xspace}

\newcommand{\degree}{\ensuremath{^\circ}\xspace}

\newcommand{\arcsper}{\ensuremath{{^{\prime\prime}}}}

\newcommand{\mjy}{\ensuremath{\,{\rm mJy}}\xspace}
\newcommand{\mjyb}{\ensuremath{{\rm \,mJy\,beam^{-1}}}\xspace}

\newcommand{\rgal}{\ensuremath{\,R_G}\xspace}   

\newcommand{\hi}{{\rm H\,{\footnotesize I}}\xspace}
\newcommand{\hii}{{\rm H\,{\footnotesize II}}\xspace}

\def\fsec{{}^{\prime\prime}\mskip-9mu.\,}  
\def\fmin{{}^{\prime}\mskip-6mu.\,}       
\def\fd{{}^{\circ}\mskip-9mu.\,}          

\slugcomment{Accepted for publication in ApJ on 08/05/2017}
\shorttitle{IR and Radio Fluxes of \hii Regions}
\shortauthors{Makai et al.}

\begin{document}

\title{The Infrared and Radio Flux Densities of Galactic HII regions}

\author{Z.~Makai\altaffilmark{1}, L.~D.~Anderson\altaffilmark{1,2,3}, J.~L.~Mascoop\altaffilmark{1}, B.~Johnstone\altaffilmark{4}}

\altaffiltext{1}{Department of Physics and Astronomy, West Virginia University, Morgantown WV 26506}
\altaffiltext{2}{Adjunct Astronomer at the Green Bank Observatory, P.O. Box 2, Green Bank WV 24944}
\altaffiltext{3}{Center for Gravitational Waves and Cosmology, West Virginia University, Chestnut Ridge Research Building, Morgantown, WV 26505}
\altaffiltext{4}{Benjamin M. Statler College of Engineering and Mineral Resources, West Virginia University, Morgantown, WV 26506}

\begin{abstract}
We derive infrared and radio flux densities of all $\sim 1000$
known Galactic \hii regions in the Galactic longitude range $17\fd5 <
\ell < 65\degree$. Our sample comes from the Wide-Field Infrared
Survey Explorer (WISE) catalog of Galactic \hii regions
\citep{anderson2014}. We compute flux densities at six wavelengths in
the infrared ({\it Spitzer} GLIMPSE $8$\microns, WISE $12$\microns and
$22$\microns, {\it Spitzer} MIPSGAL $24$\microns, and {\it Herschel}
Hi-GAL $70$\microns and $160$\microns) and two in the radio (MAGPIS
$20\,$cm and VGPS $21\,$cm). All \hii region infrared flux densities
are strongly correlated with their $\sim 20\,$cm flux densities. All
\hii regions used here, regardless of physical size or
Galactocentric radius, have similar infrared to radio flux density
ratios and similar infrared colors, although the smallest regions 
($r<1\,$pc), have slightly elevated IR to radio ratios.
The colors $\log_{10}(F_{24\microns}/F_{12\microns}) \ge 0$
and  $\log_{10}(F_{70\microns}/F_{12\microns}) \ge 1.2$, and
$\log_{10}(F_{24\microns}/F_{12\microns}) \ge 0$ and
$\log_{10}(F_{160\microns}/F_{70\microns}) \le 0.67$ reliably select
\hii regions, independent of size. The infrared colors of $\sim
22\%$ of \hii regions, spanning a large range of physical sizes,
satisfy the IRAS color criteria of \citet{wood1989} for \hii regions,
after adjusting the criteria to the wavelengths used here. Since these
color criteria are commonly thought to select only ultra-compact 
\hii regions, this result indicates that the true ultra-compact \hii region
population is uncertain. Comparing with a sample of IR color
indices from star-forming galaxies, \hii regions show higher
$\log_{10}(F_{70\microns}/F_{12\microns})$ ratios.
We find a weak trend of decreasing infrared to $\sim 20\,$cm flux
density ratios with increasing $R_{gal}$, in agreement with previous
extragalactic results, possibly indicating a decreased dust
abundance in the outer Galaxy.
\end{abstract}


\section{Introduction}
\label{sec:intro}

High-mass stars form in dense molecular clouds located primarily in
spiral arms. Gas in the vicinity of these stars can be ionized by
ultra-violet photons from the high-mass stars, creating \hii
regions. Because these high mass stars are short-lived, \hii regions 
trace star formation in the current epoch. The radio and infrared (IR) 
emission from \hii regions is bright, allowing the study of star formation
across the entire Galaxy. \hii regions can therefore be used to study
global galactic properties.

The IR emission associated with \hii regions comes from dust, and much
of this dust is located in their photodissociation regions (PDRs)
\citep{harper1974}. The PDR is the boundary between the ionized gas of
the \hii region itself and the ambient interstellar medium. At the
wavelengths used here, $8.0$, $12$, $22$, $24$, $70$ and
$160$\microns, the IR emission is due to a variety of different dust
populations, as explained by \citep{robitaille2012}. At $8.0$ and
$12$\microns, dust emission is dominated by fluorescently excited
polycyclic aromatic hydrocarbons (PAHs), which also contribute to the
emission in the $22$ and $24$\microns bands. Small grains emit in the
$22$ and $24$\microns bands, and also contribute substantially to the
emission at $70$\microns. Large grains contribute to the $70$\microns
emission, but dominate the emission at $160$\microns. Although
emission in all these photometric bands is generically due to dust,
each band is sensitive to different dust populations excited by
different mechanisms. Also, the size and temperature of grains 
influence the observed IR spectral energy distribution (SED), i.e. 
more small grains can produce more $24$\microns emission that can 
shift the SED peak to the shorter wavelengths.

IR photometry of \hii regions began $\sim 40$ years ago, with
far-infrared (FIR) observations of the Orion and Omega nebulae
\citep{low1970}. More recently, \citet{phillips2008} performed IR
photometry of $58$ \hii regions from $3.6$\microns to $8.0$\microns,
using {\it Spitzer} Galactic Legacy Infrared Mid-Plane Survey
Extraordinaire \citep[GLIMPSE;][]{churchwell2001,benjamin2003}
data. They found that the PDRs associated with \hii regions are bright
at mid-IR wavelengths. Previous studies also found that IR flux density 
ratios, or ``colors,'' can be used to separate \hii regions from other
objects. \citet[][hereafter WC89]{wood1989} reported that ultra compact 
(UC) \hii regions have IR colors of $\log_{10}(F_{60\microns}/F_{12\microns})>1.30$ 
and $\log_{10}(F_{25\microns}/F_{12\microns})>0.57$, where $F_{\lambda}$ 
denotes the flux density value at wavelength $\lambda$. Infrared flux density 
ratios can also be used to separate \hii regions from planetary nebulae 
\citep[PNe;][]{anderson2012a}. IR colors can also be used to characterize 
the star formation rate of entire galaxies \citep[e.g.][]{temi2009}.

\hii regions in our Galaxy help us to better understand the IR radiation 
from galaxies in the local Universe. For example, IR flux density ratios 
can be used to investigate the star-formation and dust properties of local 
galaxies \citep[e.g.][]{helou1986,soifer1991,wang1991,sanders2003}. These 
and similar studies have found that the observed mid- and far-IR emission 
from galaxies contain contributions from warm and cooler dust components, 
and that this emission can be used to determine the star formation rate of 
galaxies.

Radio continuum emission from \hii regions is due to (thermal)
Bremsstrahlung radiation. Early radio continuum observations found
that most bright radio continuum sources are Galactic \hii regions
\citep[see e.g.][]{piddington1951,haddock1954,westerhout1958}.
\citet[][and references therein]{harper1974} noticed a linear
correlation between \hii region far-IR and radio ($2\,$cm) flux
densities. A strong correlation between IR emission (from ``warm''
dust emitting at $60$\microns) and radio continuum emission (at
$11\,$cm) for \hii regions was also reported in \citet{haslam1987} and
\citet{broadbent1989}.

Infrared and radio flux densities are also strongly correlated on galactic
scales. \citet{deJong1985} showed a strong linear correlation between
$6.3\,$cm radio continuum and $60$\microns far-infrared flux densities for a
sample of spiral, irregular, and dwarf galaxies. The radio continuum
emission at $6.3\,$cm is due to thermal emission associated with
high-mass star formation and non-thermal synchrotron emission. The
$60$\microns emission is from warm dust associated with star formation.
The correlation indicates that both radio continuum and far-infrared
emission trace star formation activity. This has also been shown by e.g. 
\citet[][]{tabatabaei2013} and by \citet[][and references therein]{mingo2016}.

Here, we examine the infrared and radio flux densities of a large
sample of inner-Galaxy \hii regions, determine relationships between
these flux densities, and search for variations with \hii region radius
and Galactocentric radius. In Section~\ref{sec:data}, we describe the
\hii region sample and the data sets used. We explain our aperture
photometry methodology in Section~\ref{sec:phot}, and examine
correlations between the IR and radio flux densities in
Section~\ref{sec:res}. The summary follows in Section~\ref{sec:sum}.
\section{Data}
\label{sec:data}

\subsection{Source Selection}
The WISE Catalog of Galactic \hii Regions contains all known and
candidate \hii regions in the Galaxy \citep{anderson2014}.  We use
catalog Version~$1.5$, available at
\url{http://www.astro.phys.wvu.edu/wise}. The catalog lists $\sim
1900$ known \hii regions that have measured ionized gas spectroscopic
lines (H$\alpha$ or radio recombination lines). There are an
additional $\sim 700$ ``grouped'' \hii regions that are part of large
star-forming complexes like W49 or W51, but which have not been
individually targeted for ionized gas spectroscopic observations. We
here use both the known and group \hii regions. The remaining $\sim
5800$ catalog entries are \hii region candidates that we do not
consider further. We restrict the \hii region sample to the Galactic
longitude range of $17\fd5 < \ell < 65$\degree and, because our
photometric data are limited in Galactic latitude, to $\absb <
2\fd5$. Our final sample of known and group \hii regions contains
$1011$ sources.

In the longitude and latitude zone of the present work, the WISE
catalog lists Heliocentric distances for $525$ known \hii regions and
$85$ group regions. It lists Galactocentric distances for $717$ known
regions and $126$ group regions. The group \hii region distances come
exclusively from molecular line experiments, and not from their
association with the known regions \cite[see][]{anderson2014}. Most
of the catalog distances are kinematic. Over the longitude range used
here, kinematic distances are relatively accurate, assuming the kinematic 
distance ambiguity is correctly resolved (T.V.~Wenger et al., $2017$, in 
prep.). Also, due to recent \hii region surveys \citep[e.g.][]
{anderson2011, anderson2014}, the sample is by a large margin more complete 
here than in the rest of the Galaxy.

\subsection{GLIMPSE}
\label{subsec:glimpse}
GLIMPSE is a \textit{Spitzer} legacy survey of the inner Galactic
plane, covering $-65\degree \lesssim \ell \lesssim 65\degree$, $|b|
\lsim 1\degree$. The data were taken with the Infrared Array Camera
\citep[IRAC,][]{fazio1998} in four different infrared bands
($3.6$\microns, $4.5$\microns, $5.8$\microns and $8.0$\microns) at
resolutions of $\sim 2$\arcsper. These emission bands contain strong
PAH features at $3.3$\microns, $6.2$\microns, $7.7$\microns and
$8.6$\microns, and many weaker PAH ``plateaus'' at slightly longer
wavelengths \citep{andrews2015}. Here we use only the $8.0$\microns
data, which for \hii regions is dominated by PAH emission.

Scattering within the focal plane causes higher measured flux
densities of extended sources with the IRAC instrument. To correct
this effect (which is wavelength dependent), we follow the {\it Spitzer}
recommendations\footnote{\url{http://irsa.ipac.caltech.edu/data/SPITZER/docs/irac/iracinstrumenthandbook/29/$\#$_Toc410728320}}
and apply an aperture correction to the $8.0$\microns flux densities,
based on the aperture size. This correction lowers the measured flux
densities values by a maximum of $35\%$ for an aperture of
$50$\arcsper.  Following the {\it Spitzer} instrument handbook
recommendation
\footnote{\url{http://irsa.ipac.caltech.edu/data/SPITZER/docs/irac/iracinstrumenthandbook/18/$\#$_Toc410728306}}, 
we did not apply a color correction factor to the GLIMPSE flux densities.

\subsection{WISE}
\label{subsec:wise}
The Wide-field Infrared Survey Explorer \citep[WISE;][]{wright2010}
mapped the entire sky at four wavelengths: $3.4$\microns,
$4.6$\microns, $12$\microns, and $22$\microns. The angular resolutions
are $6.1$\arcsper, $6.4$\arcsper, $6.5$\arcsper and $12$\arcsper with
the $5\sigma$ sensitivities of $0.08$\mjy, $0.11$\mjy, $1$\mjy and
$6$\mjy, respectively. We use also here the $12$\microns and $22$\microns
bands (which we also refer to as W3 and W4, respectively). The
$12$\microns emission mechanism is similar to that of the
$8.0$\microns GLIMPSE data, in that it is also sensitive to PAH
features, at $11.2$\microns, $12.7$\microns and $16.4$\microns
\citep[e.g.][]{roser2015,tielens2008}. As the WISE data have DN units, we
used the DN-to-Jy conversion factors of $2.9045 \times 10^{-6}$ and
$5.2269 \times 10^{-6}$ in cases of $12$\microns and $22$\microns,
respectively\footnote{For more information, see 
\url{http://wise2.ipac.caltech.edu/docs/release/prelim/expsup/wise_prelrel_toc.html}}.
We use the color-corrections of \citet{wright2010}, assuming a
spectral index of $\alpha = 0$. This correction raises the W3 flux
densities by $9.1\%$ and the W4 flux densities by $1.0\%$.

The W4 $22$\microns bandpass is similar to that of the {\it Spitzer} 
MIPS instrument used for the $24$\microns MIPSGAL survey described below. 
Both $\sim 20$\microns data are sensitive to stochastically-heated very 
small grains (VSGs) within the \hii region plasma, and also to dust grains 
within the PDRs (PAHs are prominant contributors of $24$\microns emission, 
\citet{robitaille2012}). \citet{deharveng2010} showed that roughly half of 
the dust emission traced by the MIPSGAL $24$\microns emission originates 
from the interior of \hii regions.

\subsection{MIPSGAL}
\label{subsec:mipsgal}
MIPSGAL is a {\it Spitzer} Galactic plane survey using the Multiband
Infrared Photometer for {\it Spitzer} \citep[MIPS;][]{rieke2004}
photometer \citep{carey2005}. Like GLIMPSE, it covers $-65\degree
\lesssim \ell \lesssim 65\degree$, $|b| \lsim 1\degree$.  We use the
$24$\microns MIPSGAL data here, which has a resolution of $6$\arcsper.
MIPSGAL saturates at $1700\,$MJy sr$^{\rm -1}$ in cases of
extended sources at $24$\microns and, contrary to the IRAC bands,
MIPSGAL flux densities have a negligible correction factor for scattering in
the focal plane \citep{cohen2009}. We use the (small) color
correction factor given in the MIPS Instrument Handbook\footnote{
\url{http://irsa.ipac.caltech.edu/data/SPITZER/docs/mips/mipsinstrumenthandbook/51/\#\_Toc288032329}}.
This correction is appropriate for $100$ to $1000\,$K dust and raises the
MIPSGAL flux densities by $3.5\%$.

\subsection{Hi-GAL}
The Herschel infrared Galactic Plane Survey
\citep[Hi-GAL;][]{molinari2010,molinari2016} used the PACS
\citep{poglitsch2010} and SPIRE \citep{griffin2010} instruments on
board the {\it Herschel Space Observatory} \citep{pilbratt2010} to map
the entire Galactic plane within $|b| \le 1\degree$. The photometric
bands are centered at $70$\microns, $160$\microns (PACS), and
$250$\microns, $350$\microns and $500$\microns (SPIRE). We use here
only the $70$\microns and $160$\microns PACS data. Although the dust
associated with \hii regions do emit in the longer-wavelength data
\citep{anderson2012a}, at these wavelengths the associated emission is
difficult to disentangle from the background. The IRAS $60$\microns
and the PACS $70$\microns bands trace emission from the same dust
components, and include contributions from both very small grains and
large grains \citep{paladini2012}. At $160$\microns the emission is
almost entirely due to the large grains. The Hi-GAL point source
sensitivities are $0.5\,$Jy\,beam$^{-1}$ and $4.1\,$Jy\,beam$^{-1}$ in complex fields
and the spatial resolutions are $6\fsec7$ and $11^{\prime\prime}$, in
the $70$\microns and $160$\microns bands, respectively
\citep{molinari2010}. We applied the color corrections given in the
\textit{PACS Photometer -- Colour Corrections}
document\footnote{Website: 
\url{http://herschel.esac.esa.int/twiki/bin/view/Public/PacsCalibrationWeb}}.
Our correction is appropriate for $30$ to $100\,$K dust and increases
the $70$\microns and $160$\microns flux densities by $1\%$.

\subsection{MAGPIS}
The Multi-Array Galactic Plane Imaging Survey (MAGPIS) $20\,$cm data
\citep{helfand2006} covers a portion of the first Galactic quadrant
($5\degree < \ell < 48\fd5$, $|b| < 0\fd8$). These data were created
using multiple VLA configurations in addition to Effelsburg
single-dish data, and so are sensitive to a range of spatial
scales. The MAGPIS point source detection threshold is $\sim 2$\mjy,
excluding bright extended emissions, and the angular resolution is
$\sim 6$\arcsper. As the MAGPIS data possess higher angular resolution
than the VGPS data ($\sim 6^{\prime\prime}$ vs. $1^{\prime}$; see
below), the MAGPIS data can be used to more accurately separate the
emission from compact \hii regions from that of the background.

\citet{helfand2006} reported a possible inaccurate flux density scale
for large sources. They compared the MAGPIS flux densities of $25$
known SNRs with values from the literature compiled by
\citet{green2004}. They found a reasonably good correlation between
the flux density values, but with overestimation of the true flux
density by a factor of $2$ due to the backgrounds, and contaminating
sources \citep[some possible reasons are given in Figure~$6$
  of][]{helfand2006}.

\subsection{VGPS}
\label{sybsec:vgps}
We also use radio continuum data from the Very Large Array (VLA)
Galactic Plane Survey (VGPS), which mapped the $21\,$cm emission from
neutral atomic hydrogen (\hi). The VGPS survey covers $18\degree <
\ell < 67\degree$, $|b| < 2\fd5$ with $1^{\prime}$ resolution and a
sensitivity of $11\,$\mjyb \citep{stil2006}. In addition to the
\hi data, the VGPS produced the radio continuum data used here using
line-free portions of the spectra. They filled in the continuum
zero-spacing using the Effelsberg data from \citet{reich1986} and \citet{reich1990}.

\subsection{HRDS}
\label{subsec:hrds}
We also use $3\,$cm radio continuum data from the Green Bank Telescope
(GBT) \hii Region Discovery Survey \citep[HRDS;][]{bania2010}. The
original HRDS covers $-17\degree < \ell < 67\degree$,
$|b| < 1\degree$ \citep{anderson2011}. The HRDS extension 
\citep{anderson2014} covered the entire sky north of a declination of 
$-45\degree$, which is equivalent to $-20\degree \leqslant \ell \leqslant 
270\degree$ at $b = 0\degree$. The HRDS continuum was created using 
total-power cross scans in RA and Dec for each source. For the current 
twork, we only use sources from HRDS whose continuum emission profile 
could be modeled by a single Gaussian, and whose peak emission as derived 
from the cross scans was within $10\arcsec$ of the targeted position.

\begin{figure*}
   \centering
   \includegraphics[width=0.24\textwidth]{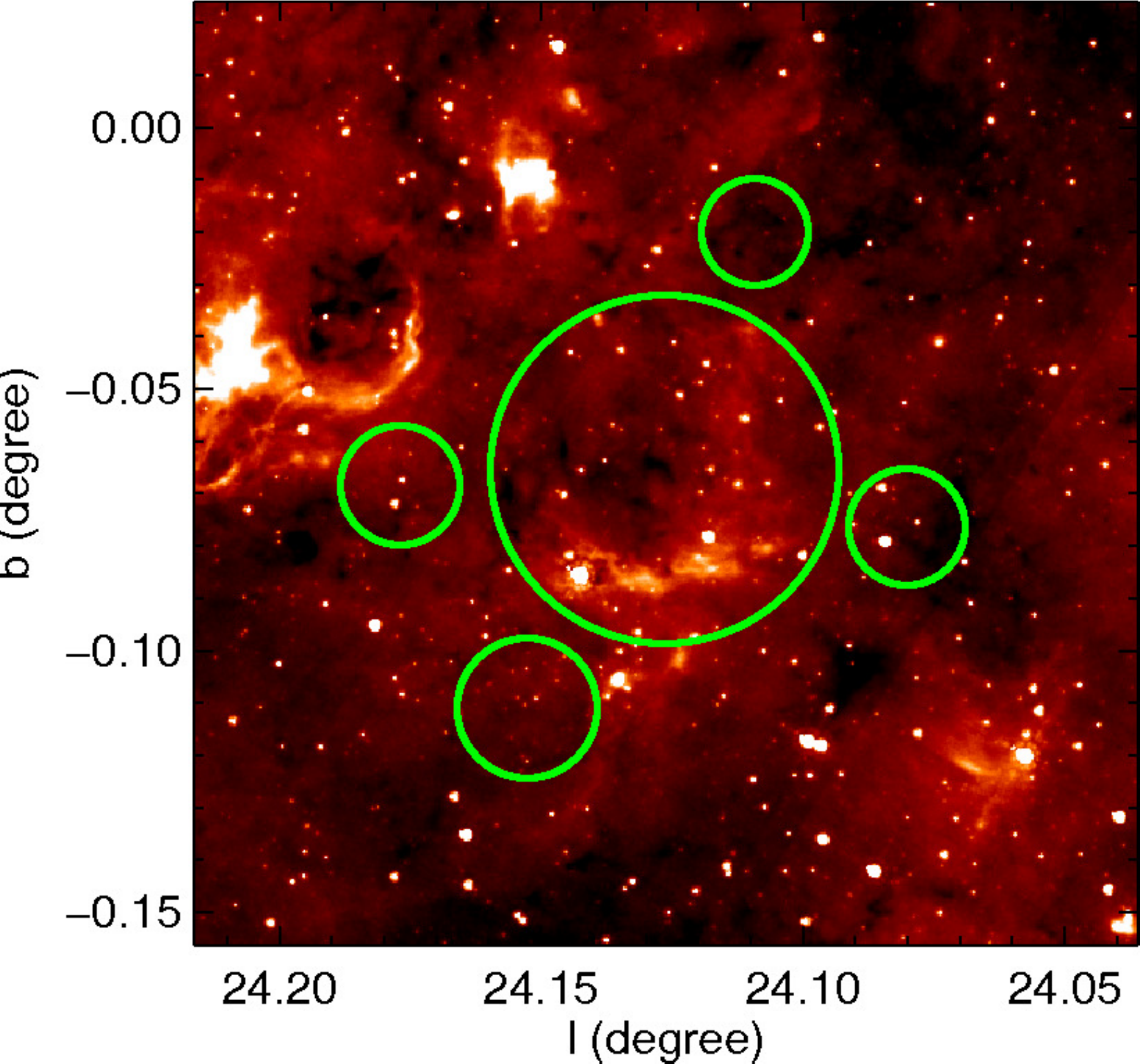}
   \includegraphics[width=0.24\textwidth]{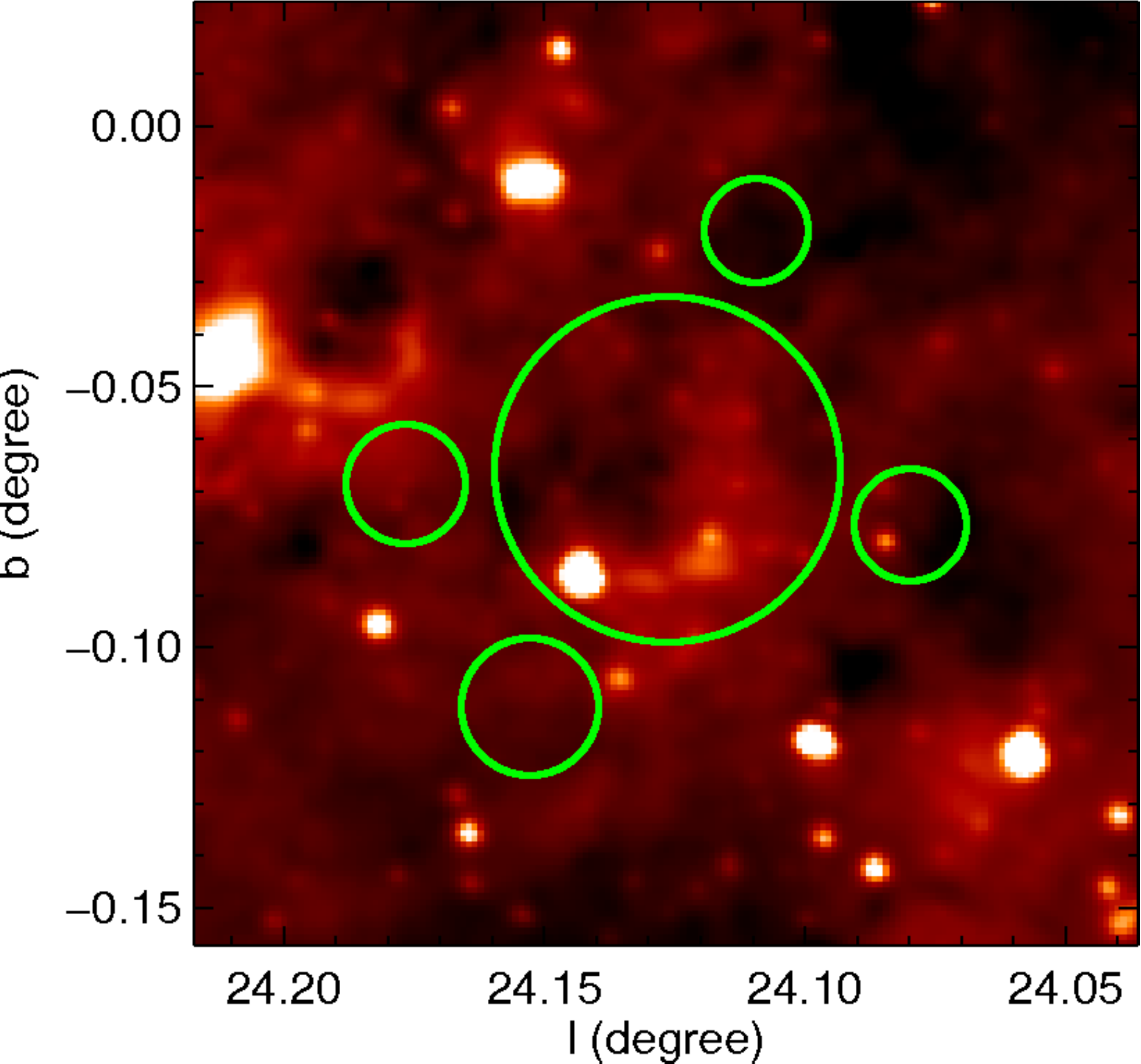}
   \includegraphics[width=0.24\textwidth]{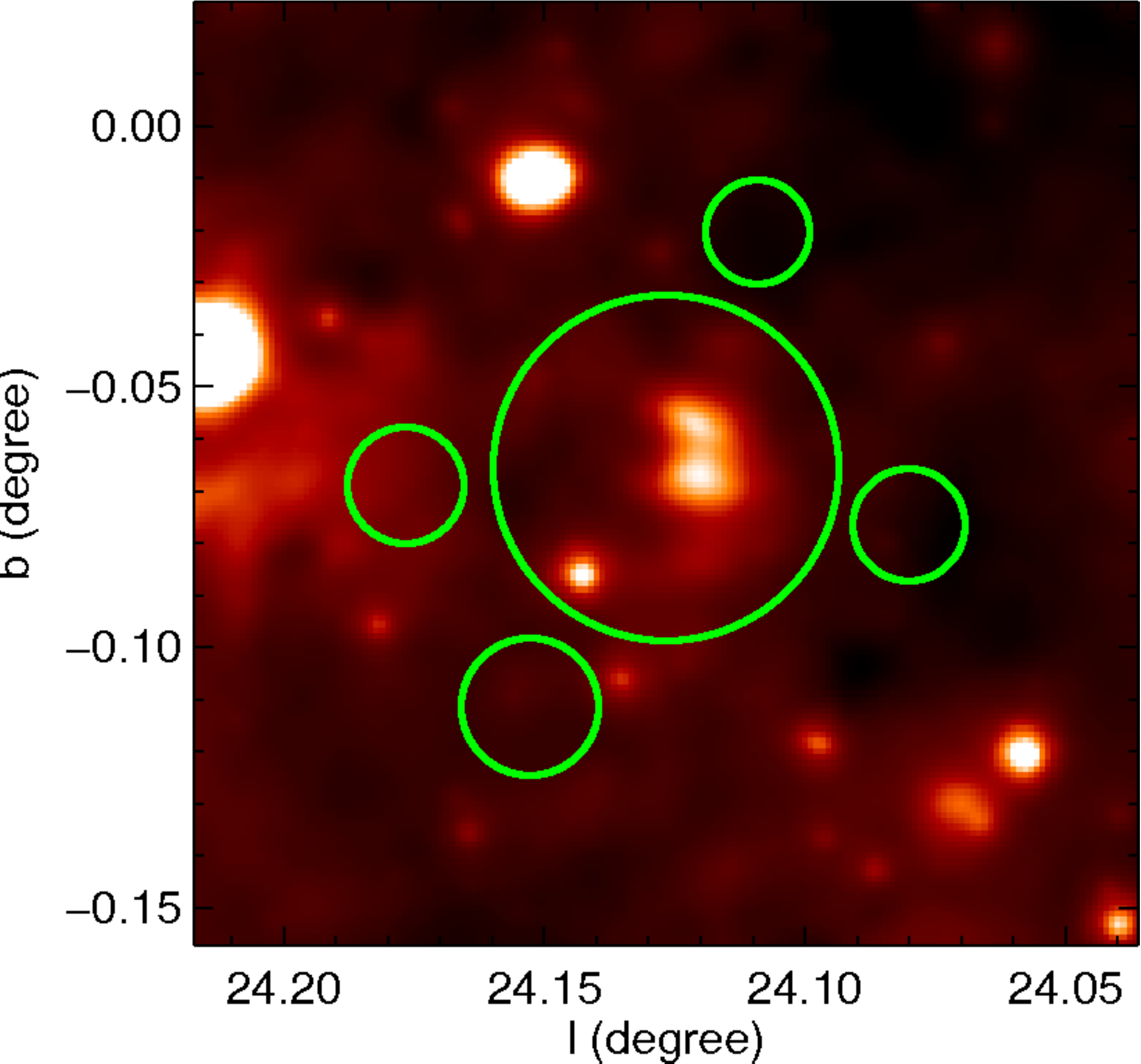}
   \includegraphics[width=0.24\textwidth]{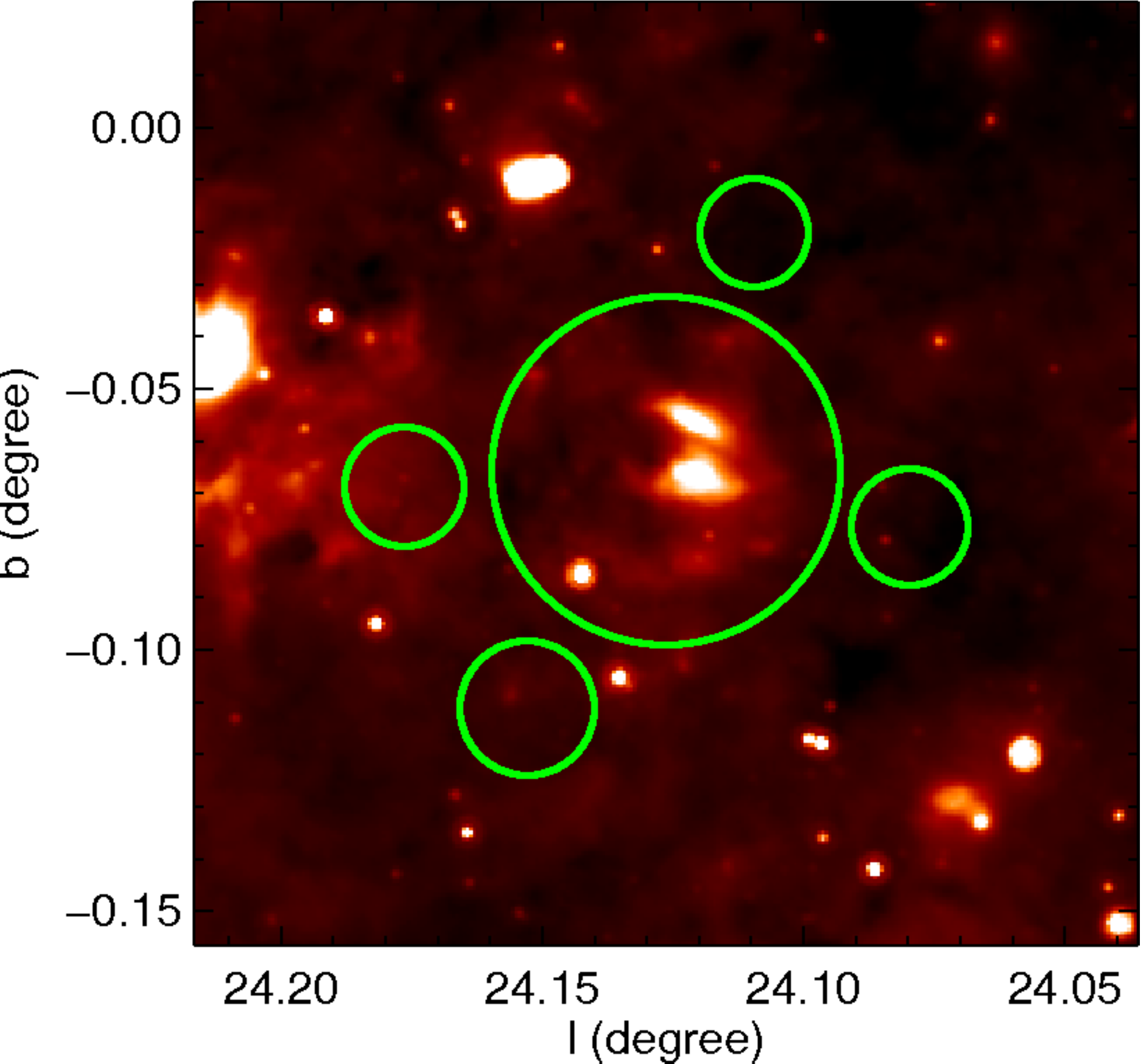}
   \includegraphics[width=0.24\textwidth]{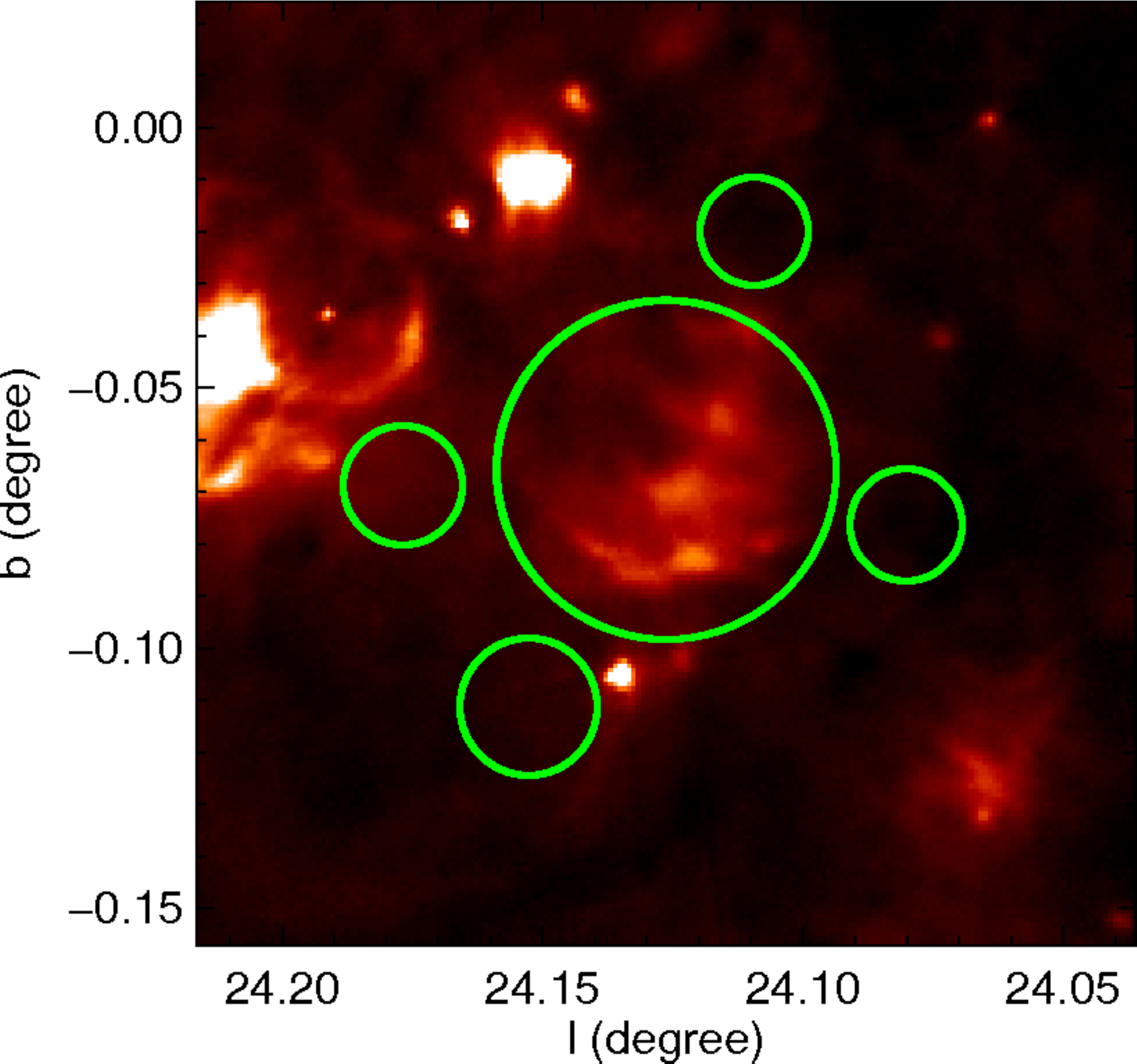}
   \includegraphics[width=0.24\textwidth]{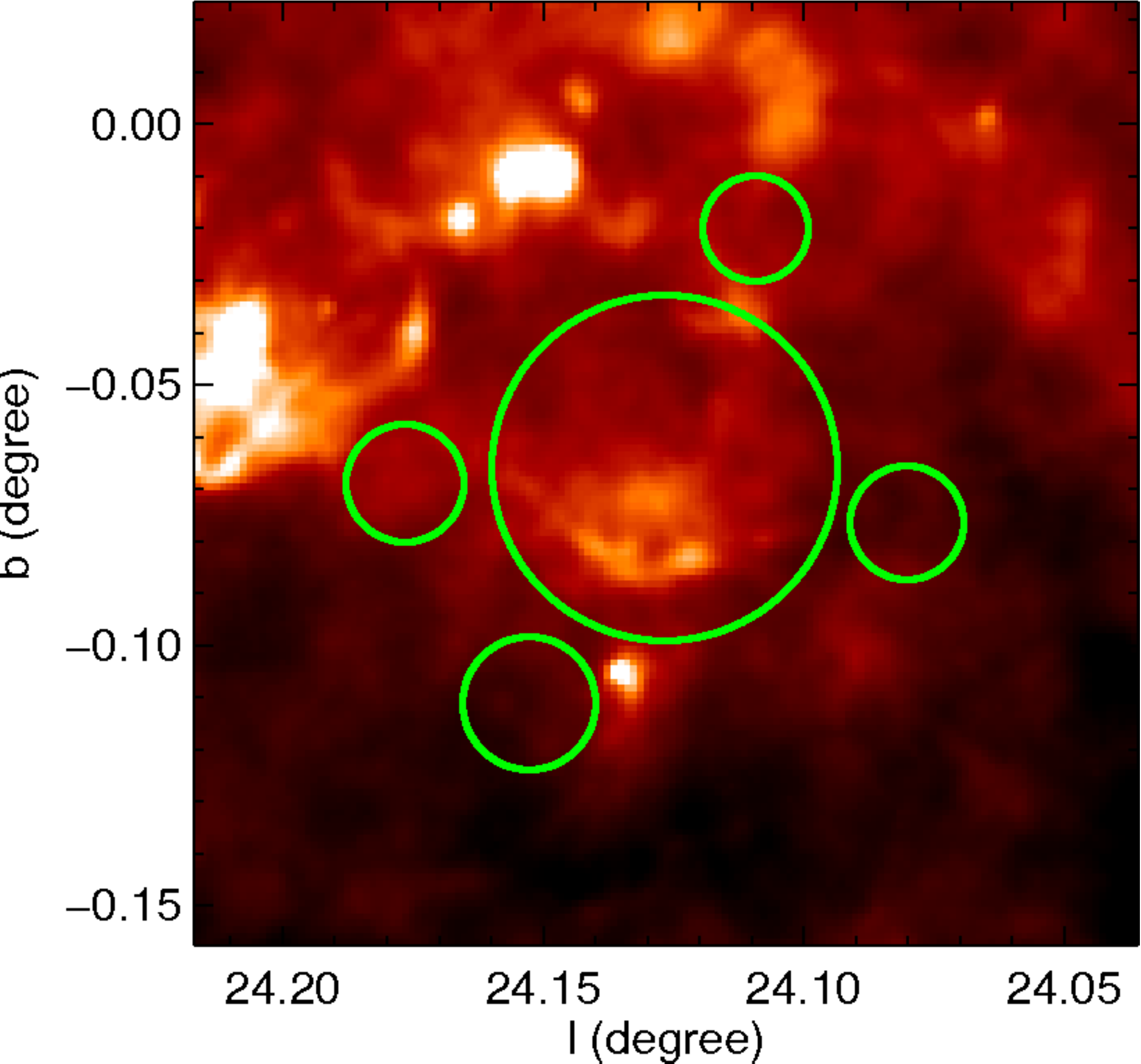}
   \includegraphics[width=0.24\textwidth]{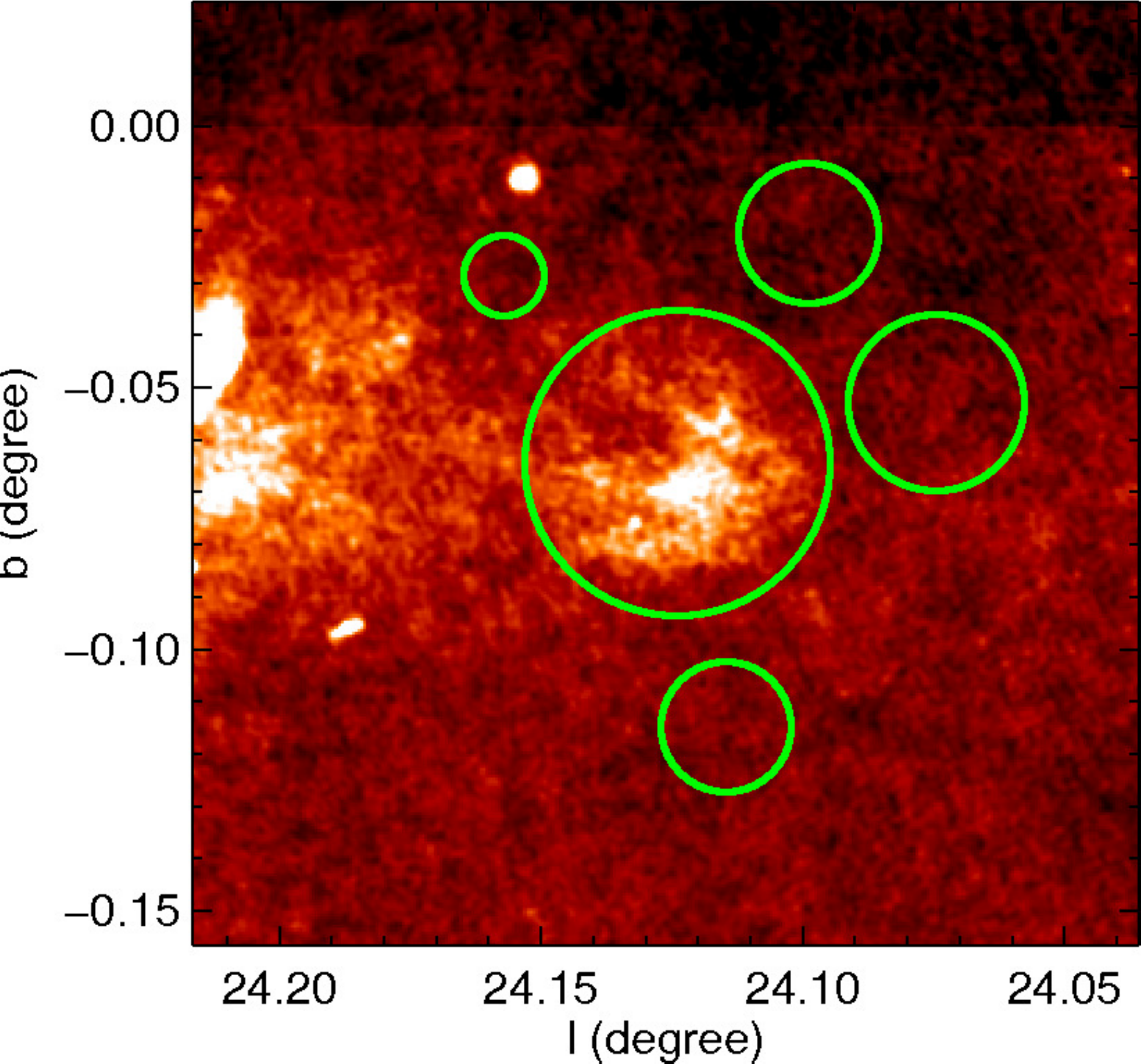}
   \includegraphics[width=0.24\textwidth]{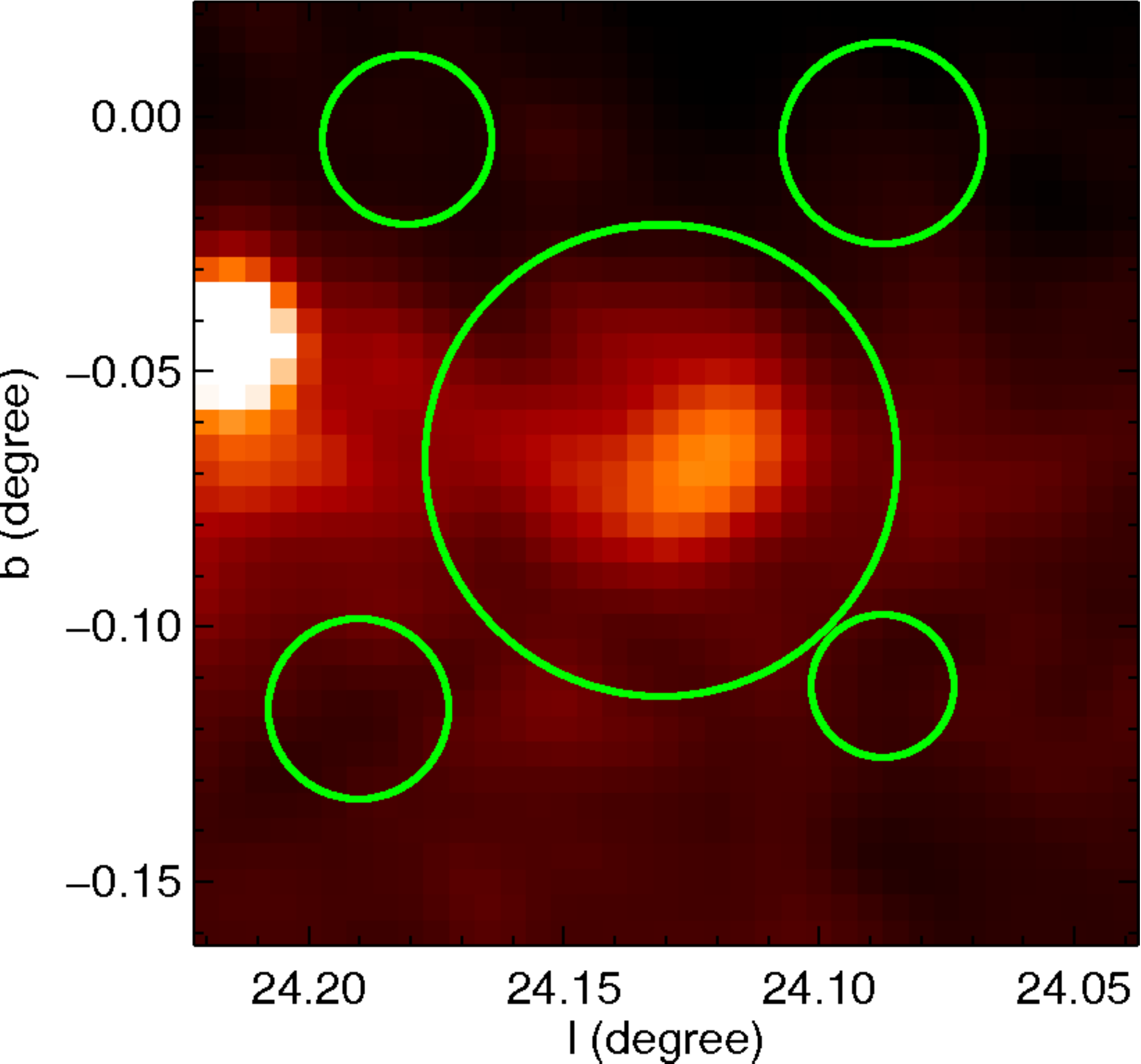}
   \caption{Example apertures for one \hii region,
     G024.126$-$00.066. \textit{Top row, left to right:} GLIMPSE
     $8.0$\microns, W3 $12$\microns, W4 $22$\microns, and MIPSGAL
     $24$\microns data. \textit{Bottom row (from left to right):}
     Hi-GAL $70$\microns, Hi-GAL $160$\microns, MAGPIS $20\,$cm and
     VGPS $21\,$cm data. The largest circle represents the source
     aperture, and the four smaller apertures are used for the
     background flux density derivation. The different background
     levels necessitate three different aperture sets, one for all IR
     data, one for MAGPIS, and one for the VGPS. \label{fig:ir_images}}
\end{figure*}

\section{Aperture photometry}
\label{sec:phot}
Radio continuum emission traces the ionized gas content of an \hii
region. Mid- to far-infrared emission traces dust that is co-spatial
with the ionized gas (notably the $\sim 20$\microns and
$70$\microns emission from very small grains), and also from the PDR
(the $\sim 10$\microns and $160$\microns emission from PAHs and
large grains, respectively). Hereafter, we use language suggesting
that the radio continuum and MIR emission are both from the \hii
region, although strictly speaking they come from different parts of
the star formation region.

We perform aperture photometry using the Kang software\footnote{\url{http://www.bu.edu/iar/kang/}}. 
Since many \hii regions in our sample are located close to objects that can
contaminate the emission of the source and/or have irregular
morphology, it is important to have a flexible aperture shape and
size. When computing aperture photometry, Kang apportions flux density from
partial pixels, which can be important for smaller regions.

We define one source aperture and four background apertures for each
source (see Figure~\ref{fig:ir_images}). We use the angular extent of
each region from the WISE catalog to guide our source aperture
definitions. In many cases, we must remove contaminating point
sources, which we do manually. We also remove any other contaminating
\hii regions that may fall within the source aperture. Our goal with
the background apertures is to accurately sample the background in
four cardinal directions, while avoiding extended emission from nearby
sources. Our methodology is identical to that of \citet{anderson2012a}.

Because \hii regions have different sizes at infrared and radio
wavelengths, and because the backgrounds differ between the infrared
and radio, we define multiple sets of apertures. We define one set of
source and background apertures for all the infrared data, one set for
MAGPIS data, and one set for the VGPS data. Although the radio
backgrounds should in principle be similar, the two radio aperture sets
are required due to the very different spatial resolutions.

We perform aperture photometry using the equation
\begin{equation}
   S_{\nu} = S_{\nu,0} - \frac{B_{\nu}}{N_{B}} \times N_{S},
\end{equation}
where $S_{\nu}$ is the source flux density (after background subtraction),
$S_{\nu,0}$ is flux density within the source aperture (without background
subtraction), $B_{\nu}$ is the flux density in the background aperture,
$N_{B}$ is the number of pixels within the background aperture, and
$N_{S}$ is the pixel number within the source aperture. With this
method, we subtract the mean flux density of the background aperture from
every pixel in the source aperture. Because we have four background
apertures, we have four values of $S_\nu$. We use in all subsequent
analysis the mean of these four values, and take as the uncertainty
their standard deviation.

\subsection{Handling uncertainties}
\label{subsec:phot_unc}

The choice of source aperture size is somewhat subjective, and results 
in additional photometric uncertainties. We attempt to quantify this 
uncertainty using the \hii region G$035.126-00.755$ and GLIMPSE 
$8.0$\microns data. For this region, we compare the flux density derived 
using source apertures of four different sizes. We find that as the 
source aperture size increases, the background-subtracted flux density also 
increases, with the largest apertures measuring flux densities nearly 
$20\%$ higher than the smallest apertures. The uncertainties on the 
largest aperture flux densities are $\sim 30\%$, whereas they are 
$\sim 10\%$ for the smallest apertures. As the source apertures increase in 
size, they sample more background emission, and are therefore more 
sensitive to the choice of background apertures. The GLIMPSE data have 
the strongest background variations, and other wavelengths may show a 
smaller effect. We conclude that the choice of source aperture size has 
an effect on the derived flux densities that is comparable to that
of the photometric uncertainties derived from our four background apertures.

For clarity, we generally do not show photometric uncertainties in
subsequent plots. We do show an analysis of the uncertainties in
Appendix~\ref{appdx:fr_err}. Based on this analysis, $50\%$ of the
data have $\lsim 35\%$ fractional uncertainties, while over $90\%$ of
the data have fractional uncertainties $\lsim 200\%$, at \textit{all}
wavelengths (see Figures~\ref{fig:appdx1}, ~\ref{fig:appdx2} and
Table~\ref{tab:frac_errors} in Appendix~\ref{appdx:fr_err}).

There are a number of reasons why we cannot compute a flux density for
a given souce at a given wavelength. Many sources are confused, and
cannot be separated from nearby \hii regions. In such cases, we do
not compute aperture photometry measurements. Some sources are simply
not detected at a given wavelength. The sky coverage for each survey
is different and so excludes some sources.

We also exclude flux densities for sources that have more than $0.1\%$
of all pixels that are so strongly saturated that they have a value of
``NaN''. This limit was found by \citet{anderson2012a} to be the best
value when discriminating between sources whose flux densities were
seriously impacted by saturation and those that were not. Due to
saturation, we remove $18$, $21$, $58$, $71$, $2$ and $3$ data points
that correspond to $\sim 2.0\%$, $\sim 2.3\%$, $\sim 6.1\%$, $\sim
7.5\%$, $\sim 0.2\%$ and $\sim 0.3\%$ data loss from GLIMPSE
($8$\microns), W3 ($12$\microns), W4 ($22$\microns), MIPSGAL
($24$\microns) and Hi-GAL ($70$\microns and $160$\microns) surveys,
respectively. The radio continuum surveys (MAGPIS and VGPS) do not
suffer from saturation.

\subsection{Source radius estimation}
\label{subsec:size_est}
The derivation of the size of an \hii region is not straightforward
because \hii regions are not necessarily spherical, and because the
\hii region boundary can be difficult to define. We are interested in
examining trends in \hii region flux densities and flux density ratios
as a function of \hii region radius, and therefore we need a reliable
method for estimating \hii region radii.

\begin{figure}
   \centering
   \includegraphics[width=0.47\textwidth]{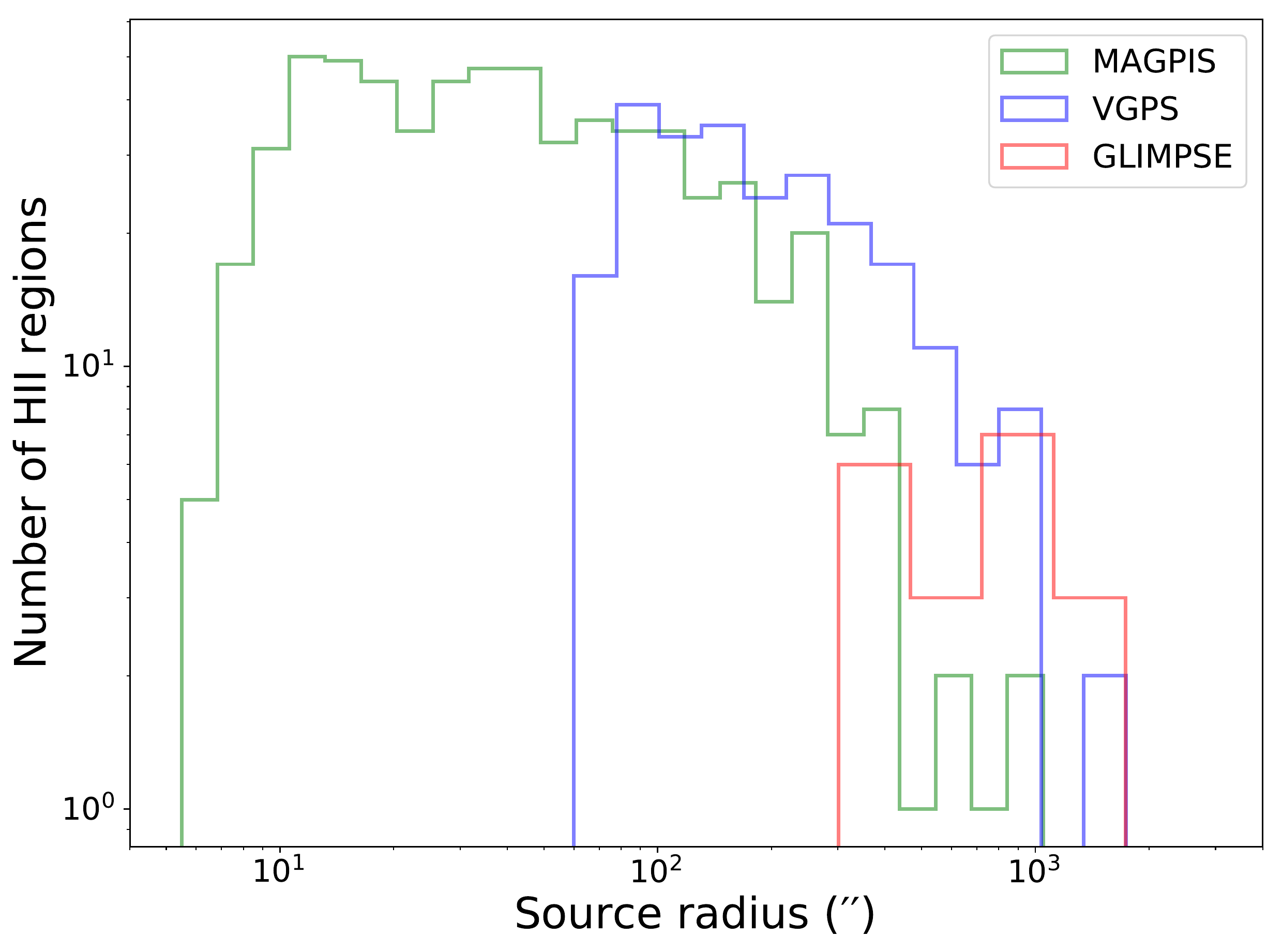}
   \caption{Distribution of the angular radii of \hii regions.
       We estimate the radius of each \hii region in the sample using
       MAGPIS (green), VGPS (blue), or IR (red) apertures (see
       Section~\ref{subsec:size_est}). The radii of the smallest
       regions are mostly measured using MAGPIS data, whereas many of
       the largest regions are measured using GLIMPSE
       data. \label{fig:size_distr}}
\end{figure}

We primarily use the MAGPIS radio apertures to estimate \hii region
radii. Since it has an angular resolution of $\sim6\arcsec$ and traces
the ionized gas, this is the most appropriate survey. Some larger
diffuse regions are not detected in MAGPIS and therefore do not have
MAGPIS apertures. For these, and for regions outside the range of the
MAGPIS survey, we instead use the VGPS $21\,$cm apertures as long as the
source has a diameter twice that of the VGPS resolution. For large
regions, the MIR and radio continuum radii are similar. For the small
number of regions with radii $>5^{\prime}$ where we are also missing
MAGPIS and VGPS data, we use the infrared aperture radius. In total,
we estimate $868$ \hii region radii; $609$, $242$ and $17$ using the
MAGPIS, VGPS and GLIMPSE apertures, respectively
(Figure~\ref{fig:size_distr}).

The radii we use in subsequent analyses are those corresponding to a
circle of the same size as the aperture area (the apertures are not
necessarily circular). In much of the following analysis, we use 
physical radii computed using the \hii region distances compiled in the 
WISE catalog.

It is important to remember that these radii are only approximate,
having been defined by-eye from radio continuum (and in some cases IR)
data. Furthermore, the physical radii used later rely on accurate
distances for the \hii regions. The angular and physical radii are
therefore not appropriate for detailed analyses of individual regions.
We use these radii only statistically here.

\section{Results}
\label{sec:res}
We give the aperture photometry results in Table~\ref{tab:phot_result}, 
which includes the names, flux density values with their $1\sigma$ 
uncertainties, angular radii, and physical radii.

We follow the same fitting method throughout the remainder of the paper. 
For each fit throughout the paper,  we fit a power law of the form 
$F_{Y} = AF^{\alpha}_{X}$, where the terms $F_{X}$ and $F_{Y}$ denote the 
flux density values plotted on the x- and y-axes. Each fit is determined 
using the least-squares method, and weights are computed using Orthogonal 
Distance Regression (ODR) method \citep{boggs1990} that takes into account 
errors for both independent and dependent variables. Importantly, we perform 
the fit by taking the base-ten logarithm of the data and fitting a linear 
regression. This is equivalent to fitting the power law form above, but 
removes the bias toward the highest flux density values found when fitting the 
data itself using the least squares method. We compute the $R^{2}$ values as 
the goodness of fit statistic, using the linear fit and the base-ten logarithm 
of the data.

\floattable
\begin{deluxetable}{lDDDDDDDDDc}
\tabletypesize{\scriptsize}
\rotate
\tablecaption{HII Region Flux Densities. \label{tab:phot_result}}
\tablehead{
\colhead{Source} & \multicolumn2c{$\mathrm{\quad F_{8\microns}}$} & \multicolumn2c{$\mathrm{\quad F_{12\microns}}$} & \multicolumn2c{$\mathrm{\quad F_{22\microns}}$} & \multicolumn2c{$\mathrm{\quad F_{24\microns}}$} & \multicolumn2c{$\mathrm{\quad F_{70\microns}}$} & \multicolumn2c{$\mathrm{\quad F_{160\microns}}$} & \multicolumn2c{$\mathrm{\quad F_{20\,}}$cm} & \multicolumn2c{$\mathrm{\quad F_{21\,}}$cm} & \multicolumn2c{Radius} & \colhead{Radius} \\
 & \multicolumn{2}{c}{\quad [Jy]} & \multicolumn{2}{c}{\quad [Jy]} & \multicolumn{2}{c}{\quad [Jy]} & \multicolumn{2}{c}{\quad [Jy]} & \multicolumn{2}{c}{\quad [Jy]} & \multicolumn{2}{c}{\quad [Jy]} & \multicolumn{2}{c}{\quad [Jy]} & \multicolumn{2}{c}{\quad [Jy]} & \multicolumn{2}{c}{\quad [$^{\prime\prime}$]} & [pc]}
\decimals
\startdata
G018.076$+$00.068 &  8.3$\pm$0.5  &  9.0$\pm$0.3  & 22.5$\pm$0.8  & 22.6$\pm$1.1  & 343.0$\pm$33.2  &  932.3$\pm$160.8 & 0.3$\pm$0.1 & 0.3$\pm$0.1 &  48.1 &  2.74 \\
G018.096$-$00.322 & 24.1$\pm$12.5 & 25.0$\pm$17.4 & 64.3$\pm$24.8 & 72.1$\pm$25.5 & 714.4$\pm$391.1 & 1849.7$\pm$950.5 & 0.2$\pm$0.1 & 0.3$\pm$0.1 &  41.7 &  \nodata \\
G018.152$+$00.090 &  9.4$\pm$0.9  &  8.4$\pm$0.4  &  9.5$\pm$1.8  &  9.8$\pm$2.0  & 253.3$\pm$26.7  &  736.2$\pm$65.1  & 0.2$\pm$0.1 & 0.3$\pm$0.1 &  91.5 &  1.83 \\
G018.195$-$00.171 & 10.1$\pm$5.3  & 13.7$\pm$6.1  & 32.5$\pm$11.5 & 33.1$\pm$10.1 & 355.5$\pm$116.4 &  692.2$\pm$203.3 & 0.8$\pm$0.2 & 0.6$\pm$0.2 &  62.2 &  3.73 \\
G018.218$+$00.397 &  7.9$\pm$4.7  &  9.7$\pm$7.2  & 14.4$\pm$4.4  & 13.8$\pm$4.2  & 208.6$\pm$79.7  &  590.4$\pm$812.9 & 0.6$\pm$0.2 & 0.5$\pm$0.2 & 167.8 & 13.28 \\
G018.329$+$00.024 &  0.7$\pm$2.5  &  1.8$\pm$1.6  &  4.7$\pm$1.3  &  5.0$\pm$1.1  &  78.0$\pm$26.0  &  294.6$\pm$214.2 & 0.3$\pm$0.1 & 0.1$\pm$0.1 &  96.9 &  5.71 \\
G018.451$-$00.016 &  4.8$\pm$0.7  &  5.2$\pm$0.7  & 15.2$\pm$12.9 & 18.3$\pm$4.3  & 350.1$\pm$175.3 &  746.9$\pm$972.7 & 0.3$\pm$0.1 &  . \nodata  &  60.8 &  3.50 \\
G018.461$-$00.003 &  1.3$\pm$0.1  &  2.3$\pm$0.3  &  8.8$\pm$1.2  & 13.2$\pm$2.5  & 853.3$\pm$11.8  & 1382.9$\pm$13.6  & 0.1$\pm$0.1 &  . \nodata  &  11.9 &  0.68 \\
G018.584$+$00.344 &  2.9$\pm$0.3  &  2.8$\pm$1.2  &  4.6$\pm$1.0  &  5.2$\pm$0.9  &  71.5$\pm$10.4  &  162.5$\pm$47.5  & 0.1$\pm$0.1 &  . \nodata  &  30.8 &  2.24 \\
G018.630$+$00.309 &  3.9$\pm$0.4  &  3.7$\pm$0.7  &  5.1$\pm$0.4  &  5.9$\pm$0.3  & 115.4$\pm$3.5   &  276.4$\pm$51.1  & 0.1$\pm$0.1 & 0.1$\pm$0.1 &  32.6 &  \nodata \\
G018.657$-$00.057 & 28.1$\pm$4.0  & 28.0$\pm$4.9  & 65.8$\pm$4.0  & 54.2$\pm$4.8  & 949.3$\pm$60.1  & 2274.7$\pm$236.2 & 0.2$\pm$0.1 & 0.2$\pm$0.1 &  40.3 &  2.44 \\
G018.677$-$00.236 &  8.7$\pm$1.0  & 10.0$\pm$2.0  & 21.0$\pm$3.9  & 22.8$\pm$3.6  & 350.7$\pm$37.2  &  856.3$\pm$71.6  & 0.4$\pm$0.1 & 0.3$\pm$0.2 &  36.9 &  \nodata \\
G018.710$+$00.000 &  0.6$\pm$0.1  &  1.7$\pm$0.2  &  8.2$\pm$0.5  &  7.8$\pm$0.2  & 156.0$\pm$2.9   &  282.2$\pm$13.0  & 0.1$\pm$0.1 & 0.1$\pm$0.1 &  12.2 &  \nodata \\
G018.741$+$00.250 &  8.7$\pm$1.2  &  9.3$\pm$1.0  & 27.1$\pm$0.4  & 26.4$\pm$0.7  & 339.7$\pm$5.5   &  769.7$\pm$47.8  & 0.2$\pm$0.1 & 0.3$\pm$0.1 &  33.4 &  2.30 \\
G018.750$-$00.535 &  3.8$\pm$3.0  &  2.9$\pm$2.7  &  8.9$\pm$2.3  &  9.5$\pm$2.2  & 101.1$\pm$51.3  &  145.3$\pm$162.2 & 0.2$\pm$0.1 & 0.1$\pm$0.1 &  75.4 &  \nodata \\
\enddata
\end{deluxetable}

\subsection{IR flux density accuracy}
\label{subsec:ir_fluxes}
We compare our \hii region flux densities with those from the second version
of the IRAS Point Source Catalog (PSC) to inspect the flux density
calibrations and the aperture photometry methodology. We select
isolated \hii regions that can be found in the IRAS catalogs, and
have measured IRAS flux densities at $12$\microns, $25$\microns, and
$60$\microns.

\begin{figure}
   \centering
   \includegraphics[width=0.47\textwidth]{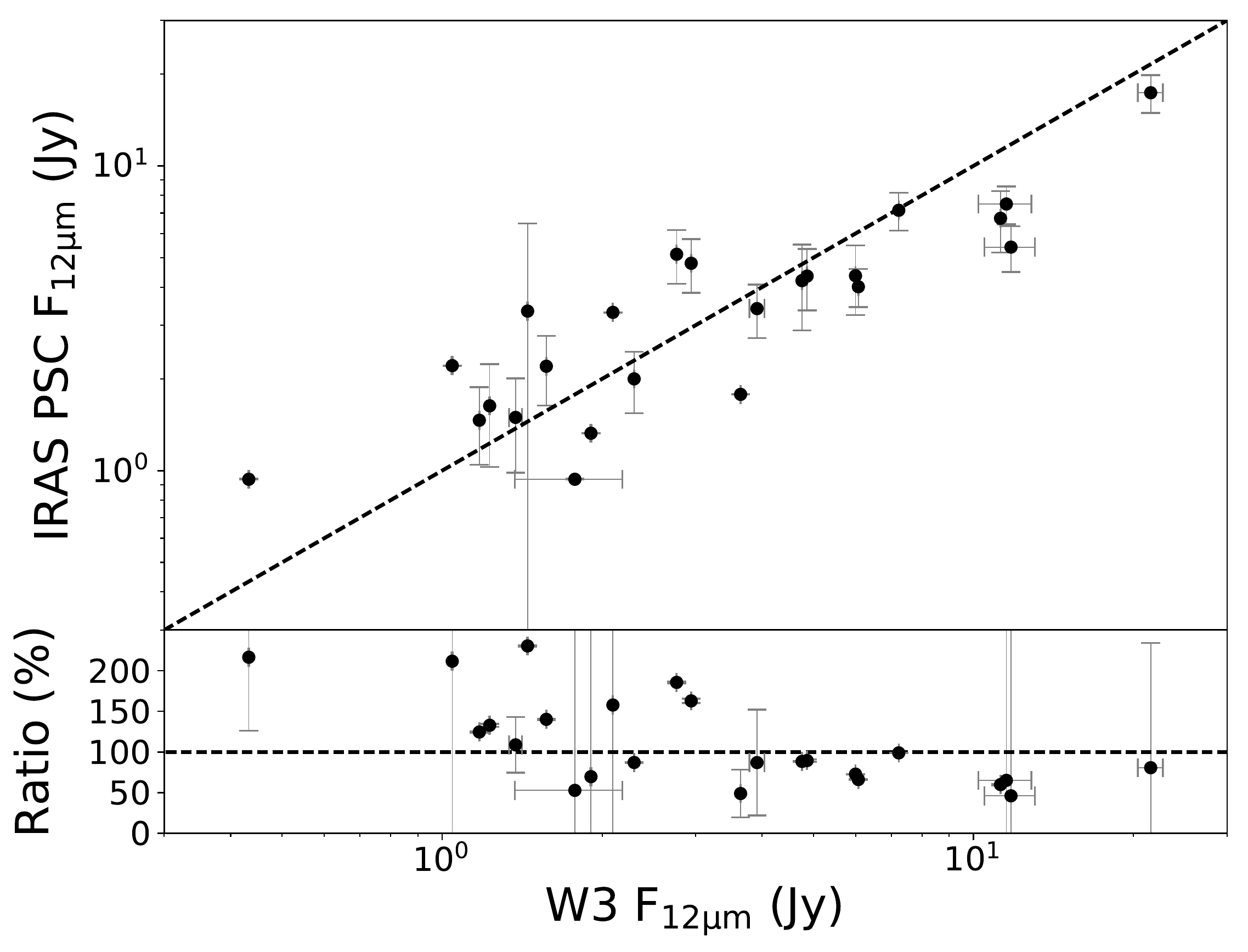}
   \includegraphics[width=0.47\textwidth]{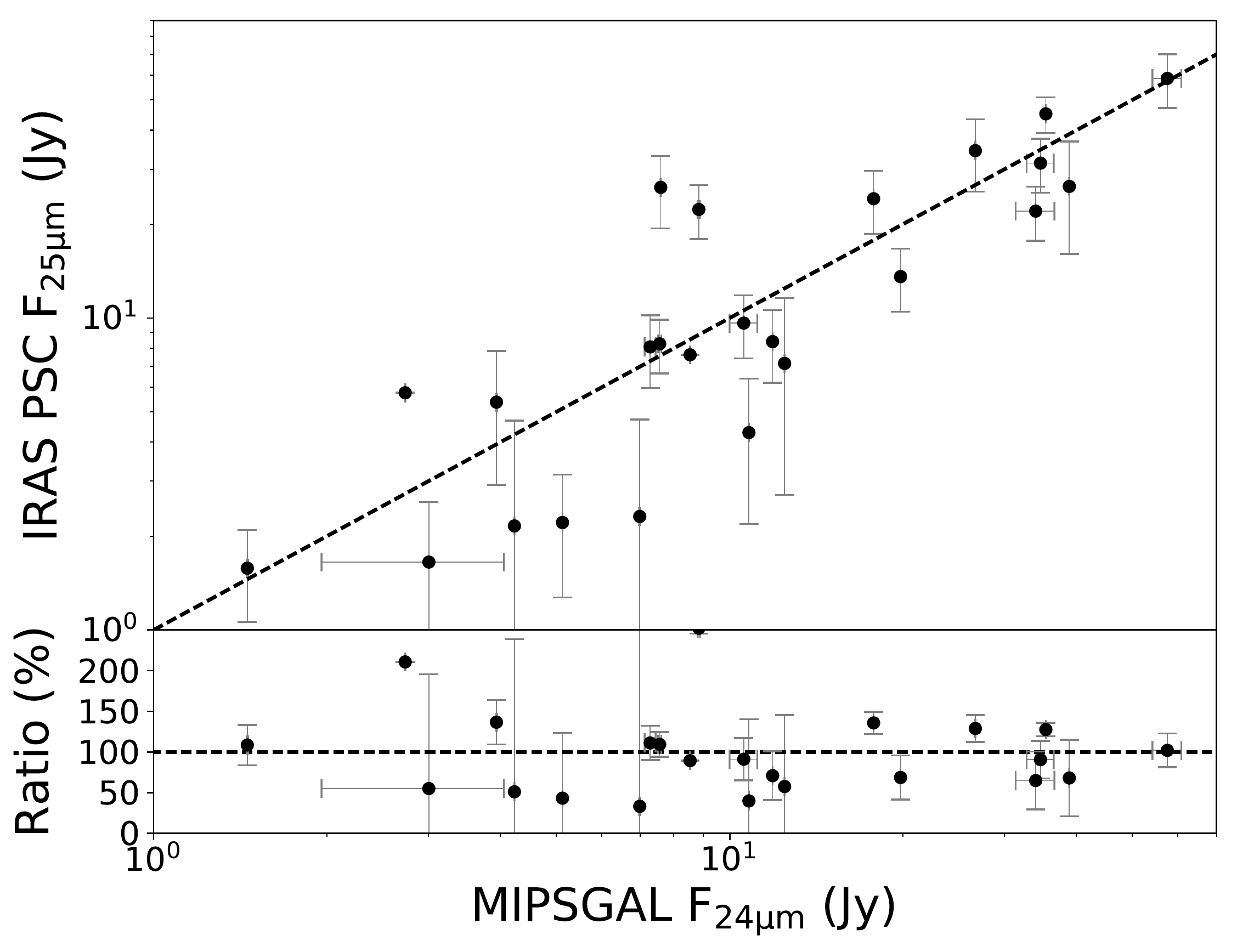}
   \includegraphics[width=0.47\textwidth]{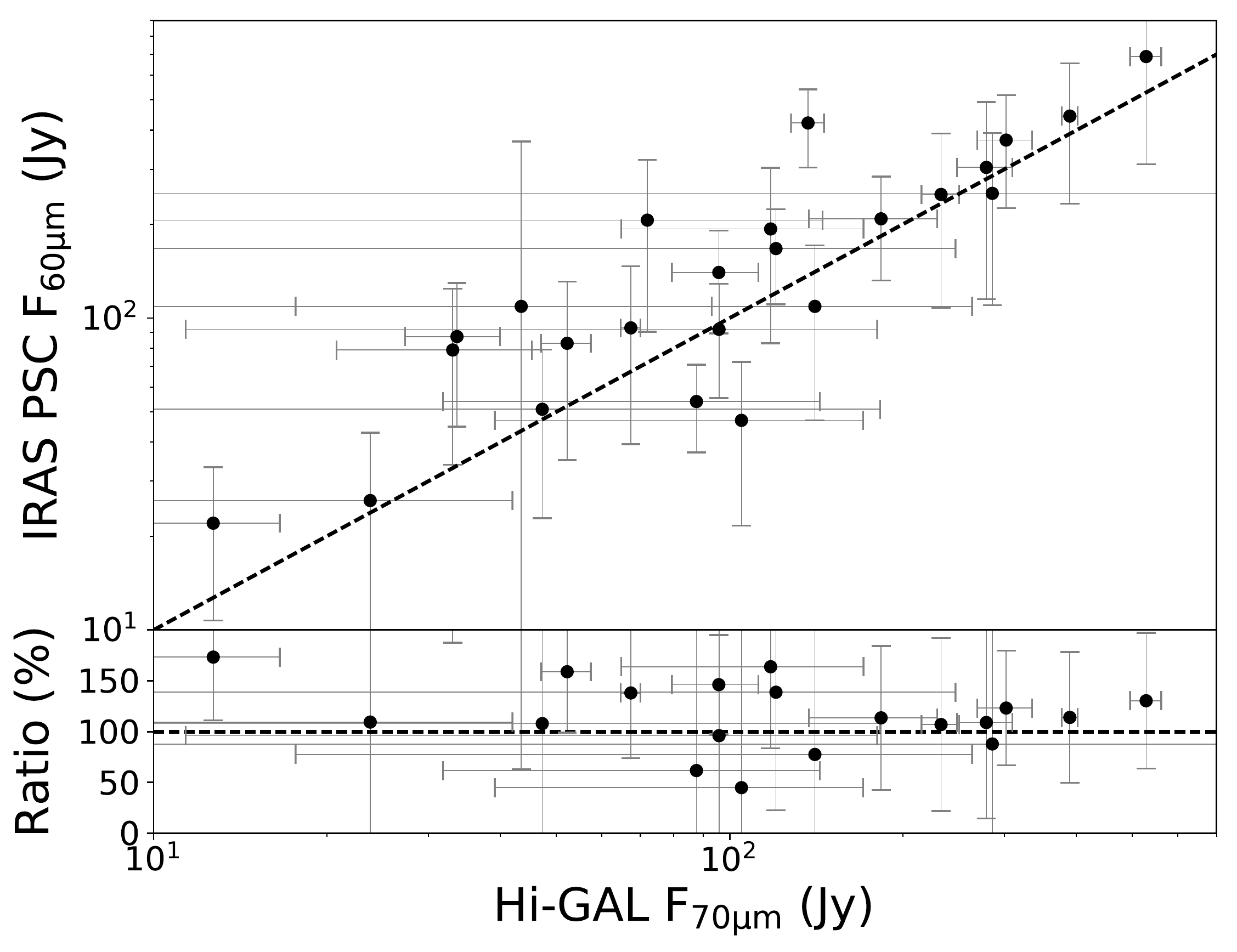}
   \caption{Comparison of IRAS PSC ($12$\microns) and W3 fluxes
     densities (top subplot), IRAS PSC $25$\microns and MIPSGAL
     ($24$\microns) flux densities (middle subplot), and IRAS PSC
     $60$\microns and Hi-GAL ($70$\microns) flux densities (bottom
     subplot). The black dashed lines mark the 1:1 ratio. The lower
     panels show the deviation from the 1:1 ratio. The $3\sigma$
     photometric uncertainties are represented by the errorbars. The
     various flux densities are strongly correlated, with no obvious
     systematic offset. The significant scatter is likely due to the
     photometric uncertainties. \label{fig:iras_comp1}}
\end{figure}

\begin{figure}
   \centering
   \includegraphics[width=0.47\textwidth]{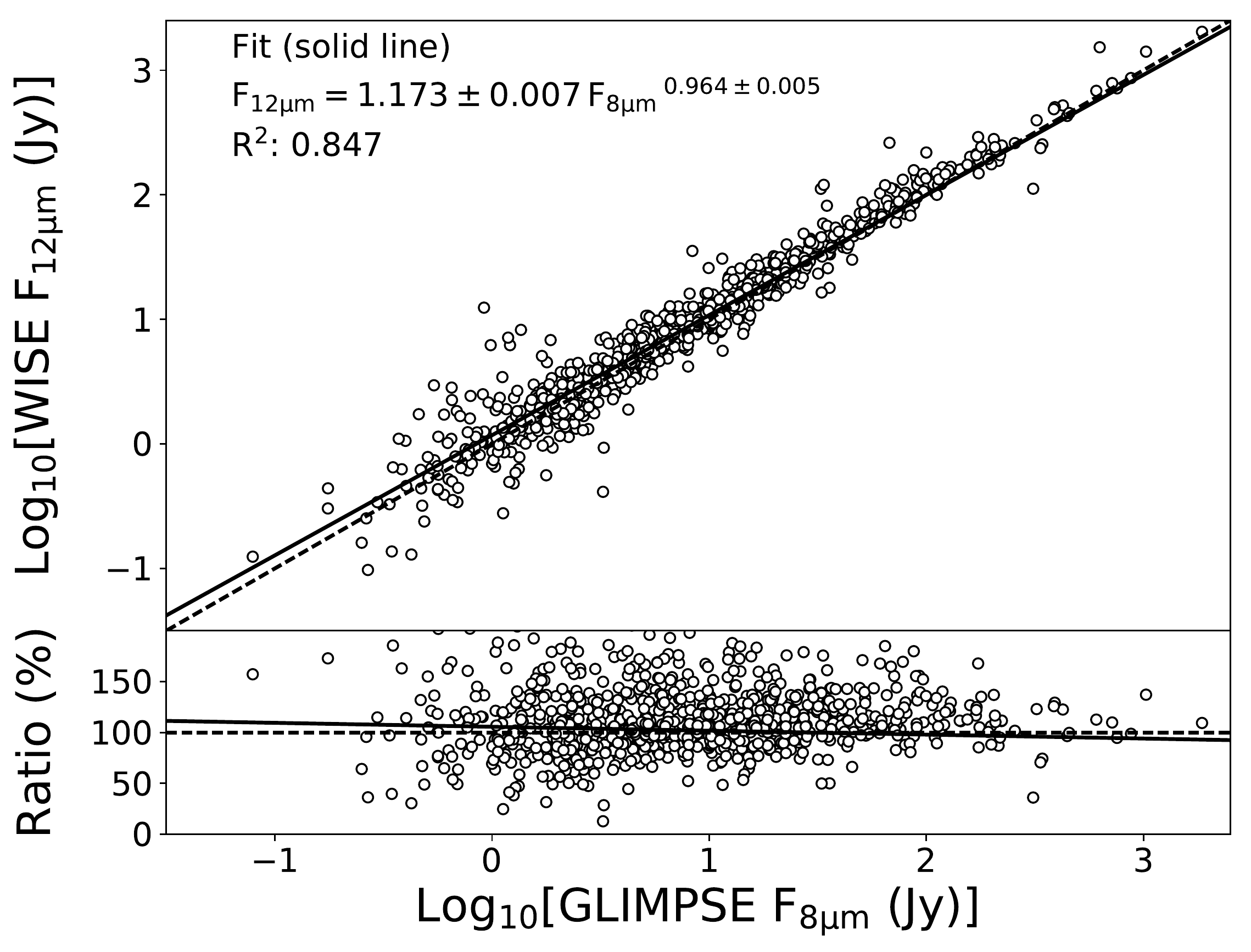}
   \includegraphics[width=0.47\textwidth]{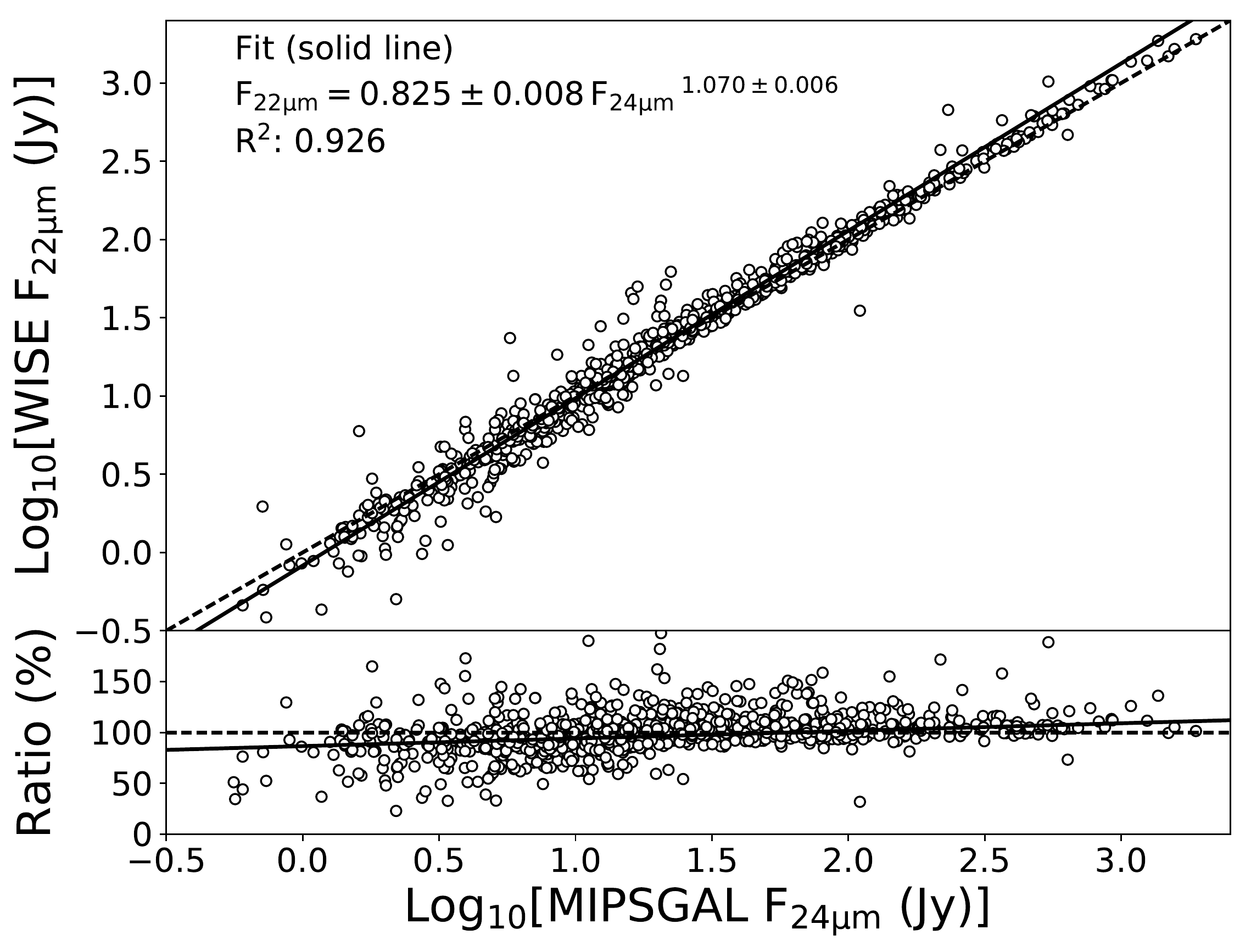}
   \caption{Correlations between similar IR photometric bands. The
     upper subplot shows the GLIMPSE and W3 flux densities, while the
     lower subplot shows the correlation between MIPSGAL and W4 flux
     densities. The black solid lines represent the best fit.
     As in Figure~\ref{fig:iras_comp1}, the lower panels represent the
     variations of the data points (in $\%$) from the unity line ratio
     (black dashed lines). \label{fig:glmi_w3w4}}
\end{figure}

We compare our derived values with those from the IRAS PSC using the
W3 $12$\microns versus IRAS $12$\microns data, the MIPSGAL
$24$\microns versus IRAS $25$\microns, and the Hi-GAL $70$\microns versus 
IRAS $60$\microns data (Figure~\ref{fig:iras_comp1}). There is no systematic 
offset in any of these relationships, although there is large scatter. 
The standard deviation of the ratio of IRAS PSC $F_{12\microns}$ and W3
$F_{12\microns}$ is $\sim58\%$ from unity. In case of the IRAS PSC
$F_{25\microns}$ and MIPSGAL $F_{24\microns}$ ratio, the spread of
the data points is larger, with a standard deviation of $\sim
81\%$. The ratio of IRAS PSC $F_{60\microns}$ and Hi-GAL
$F_{70\microns}$ has a standard deviation of $\sim 51\%$.

GLIMPSE $8.0$\microns and W3 $12$\microns flux densities, as well as
MIPSGAL $8.0$\microns and W4 $22$\microns flux densities, are strongly
correlated (Figure~\ref{fig:glmi_w3w4}): the slopes of the fits are 
$0.964\pm0.005$ and $1.070\pm0.006$ for GLIMPSE versus W3 and MIPSGAL 
versus W4, respectively. Since both the GLIMPSE
and W3 data trace PAH emission, the strong linear correlation is
unsurprising. The W4 and MIPSGAL bandpasses are similar, and therefore
the strong correlation between these quantities is also
unsurprising. The standard deviations from the unity line ratio are
$\sim 60\%$ and $\sim 34\%$ for the W3 ($12$\microns)/GLIMPSE
($8$\microns) ratio and W4 ($22$\microns)/MIPSGAL ($24$\microns)
ratio, respectively.

\subsection{Radio flux density accuracy}
\label{subsubsec:radio_fluxes}

Since the VGPS and MAGPIS surveys are at approximately the same
wavelength, their flux densities should also be highly correlated.
\citet{helfand2006}, however, reported a possible large-scale
calibration problem for the MAGPIS data. They computed the MAGPIS flux
densities for $25$ known SNRs and compared those values with those
given in \citet{green2004}. They found that MAGPIS overestimates flux
densities, but that the discrepancy is reduced by subtracting
$0.07\,$Jy arcmin$^{-2}$.

\begin{figure}
   \centering
   \includegraphics[width=0.47\textwidth]{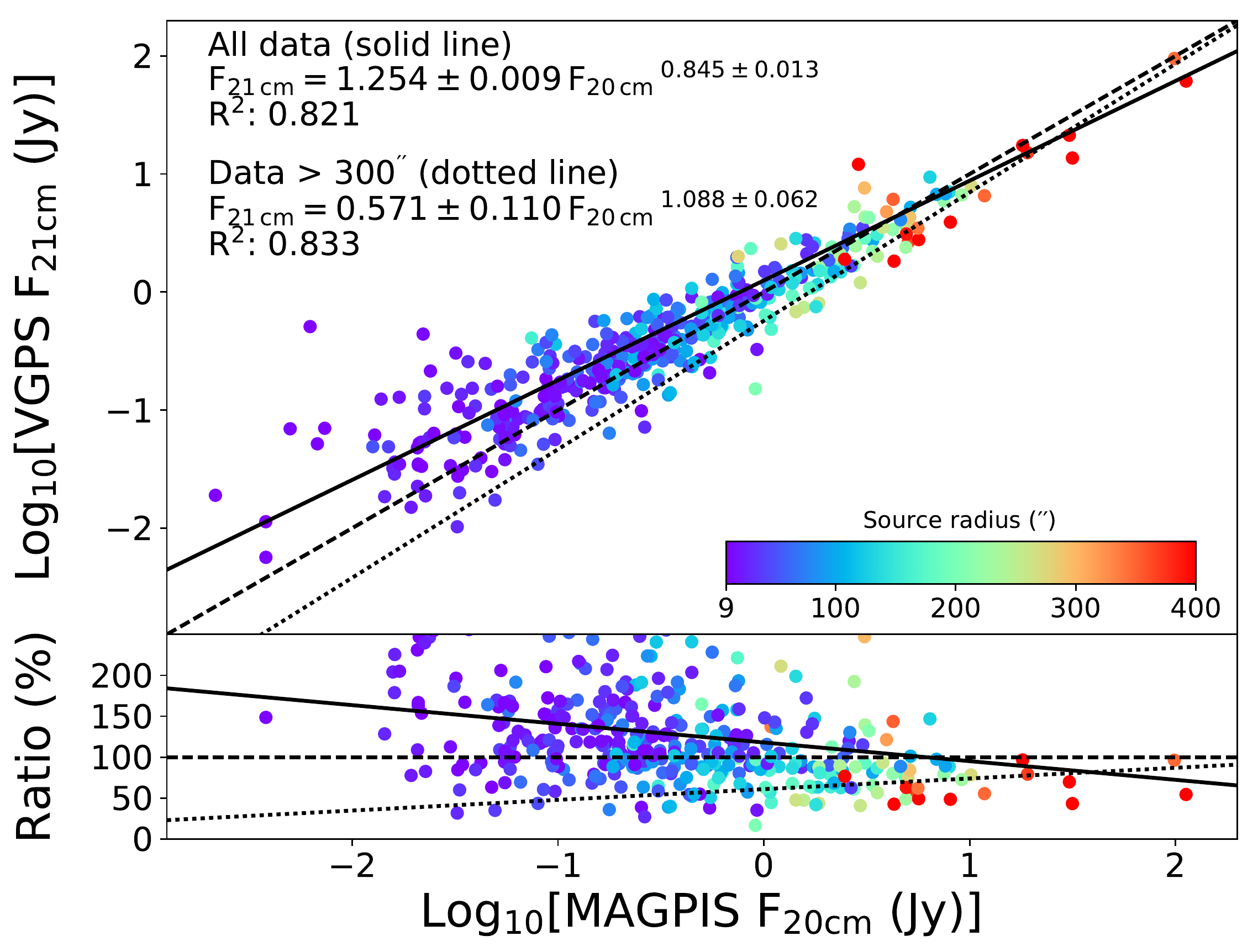}
   \includegraphics[width=0.47\textwidth]{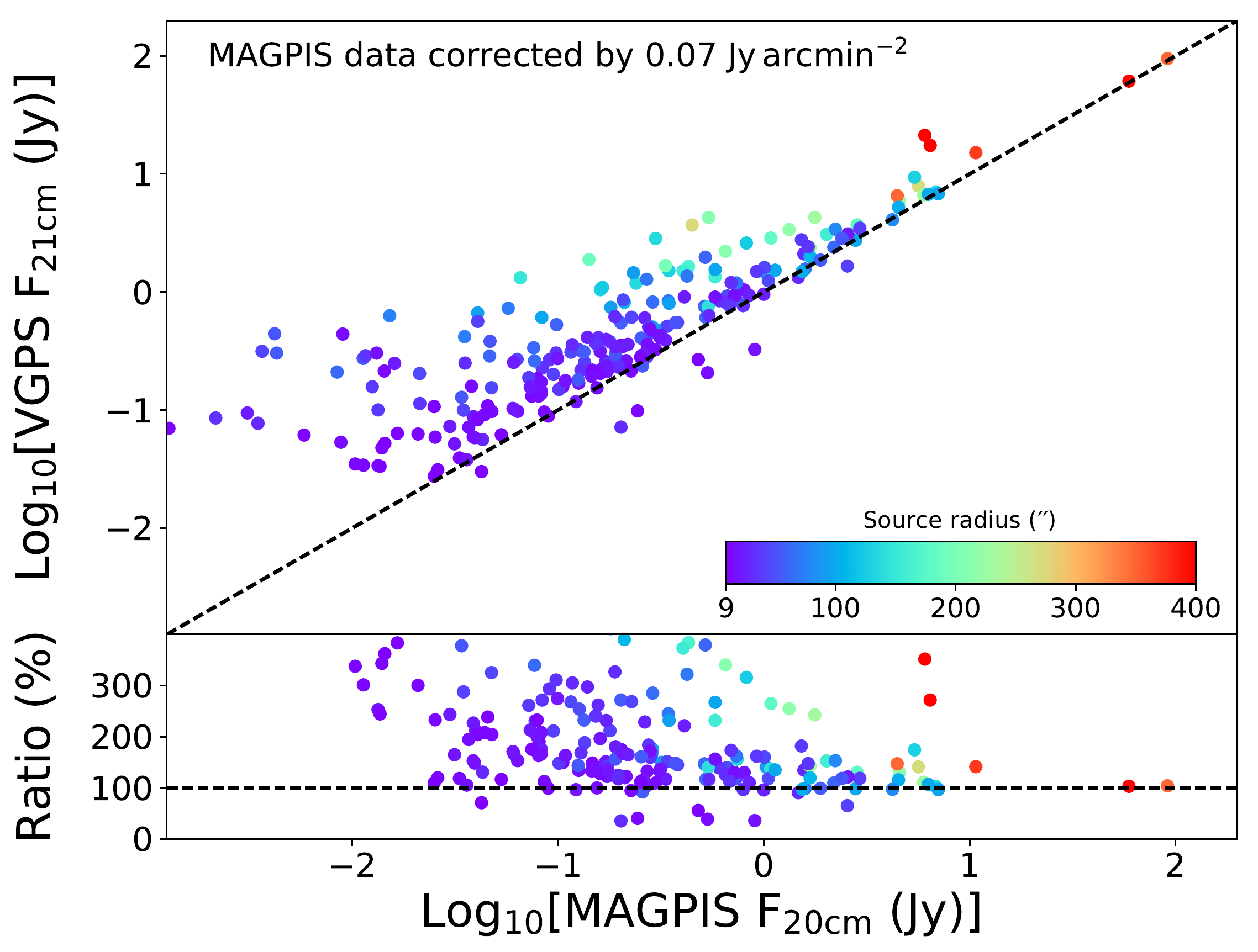}
   \includegraphics[width=0.47\textwidth]{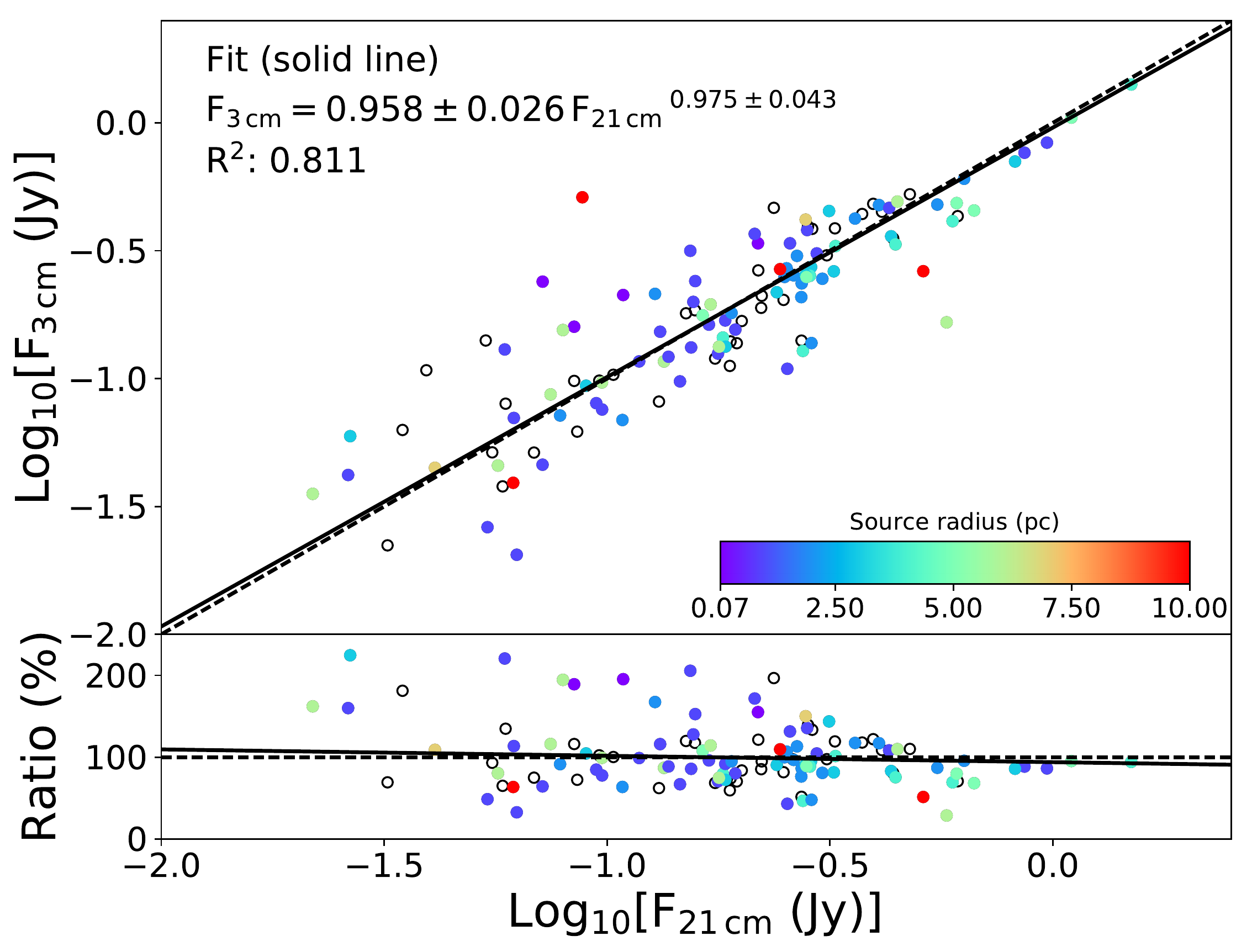}
   \caption{The correlations between VGPS and MAGPIS, and the HRDS and VGPS 
     flux densities. The black dashed lines represent the 1:1 ratio. 
     \textbf{Top:} shows a fit to all data points (solid line) 
     and a fit to data from regions with radii $> 6^{\prime}$ (dotted lines). 
     \textbf{Middle:} comparison with the MAGPIS flux densities corrected by a 
     factor of $0.07\,\text{Jy}\,\text{arcmin}^{-2}$ (see text). \textbf{Bottom:}
     The open circles represent sources without distance measurements (no physical 
     radius). The solid black line shows the best fit to the highest quality 
     data (colored and empty circles). The correlation suggests negligible radio 
     optical depth effects.
     \label{fig:vgps_magpis}}
\end{figure}

Contrary to \citet{helfand2006}, we find that the MAGPIS and VGPS flux
density values do not have a significant offset
(Figure~\ref{fig:vgps_magpis}). This result also holds independently
for large \hii regions $>6^{\prime}$ in radius, orange--red dots on
Figure~\ref{fig:vgps_magpis}. In the sample of SNRs used in
\citet{helfand2006}, there are $16$ SNRs that have radii larger than
$6^{\prime}$. There are also $16$ \hii regions in our sample with
radii $>6^{\prime}$. The slightly different fit to large \hii regions 
(dotted line) does not influence our results: the lack of an offset 
remains for both smaller and larger \hii regions.

Decreasing the flux densities by the factor of $0.07\,\text{Jy arcmin}^{-2}$
(middle subplot on Figure~\ref{fig:vgps_magpis}) results in a much
poorer correlation. The standard deviation from the $1:1$ line is
$\sim 7$ times higher ($\sim 27.4\,$Jy) after applying the correction
factor of $0.07\,\text{Jy arcmin}^{-2}$ to our MAGPIS data points
(bottom panel of the middle subplot on Figure~\ref{fig:vgps_magpis}),
than before ($\sim 4.3\,$Jy; bottom panel of the top subplot on
Figure~\ref{fig:vgps_magpis}). Worse, the flux densities for $42\%$ of the
regions become negative if the factor is applied, and only $5$ of the $16$
large regions has a positive flux density value.

Some scatter in the relationships may come from
radio optical depth effects. At low radio frequencies, the radio
continuum emission from \hii regions can be optically thick. The
wavelength of this optically thick transition is determined by the
electron density, and therefore young compact \hii regions are more
likely to be optically thick at a given frequency. Comparing $3\,$cm
(HRDS) and $21\,$cm (VGPS) flux densities, it seems that the optical
depth of regions does not strongly affect their flux densities
(bottom subplot on Figure~\ref{fig:vgps_magpis}). We only compare flux densities for 
\hii regions with HRDS-derived radii that are within $50\%$ of the values 
derived here, which ensures that we are sampling the same radio continuum 
sources. From Figure~\ref{fig:vgps_magpis}, it is clear that the
strength of the correlation is the same, regardless of \hii region
physical radius. Because the exponent of the fit to the data is close
to unity ($0.975\pm0.043$), we conclude that statistically the optical 
depth of \hii regions at $21\,$cm does not have a strong effect on the 
measured flux densities.

\subsection{Combined radio data set}
\label{subsec:combradio}
For much of the following analysis, we use a combined MAGPIS and VGPS
data set. The two flux densities are strongly correlated (see
Figure~\ref{fig:vgps_magpis}), and are at approximately the same
wavelength. Some large diffuse sources are detected in the VGPS that
are not in MAGPIS and many compact sources are unconfused in MAGPIS,
but are confused in the VGPS. The combination allows us to determine
radio flux densities for the largest possible sample. The combined
data set has VGPS flux densities if those of MAGPIS are not available,
and MAGPIS flux densities if those of the VGPS are not available. For
sources that have both VGPS and MAGPIS flux densities, we use the
average. We denote this combined data set using ``radio'', 
i.e. the combined flux density is given by the term $F_{radio}$.

\begin{figure*}
   \centering
   \includegraphics[width=0.47\textwidth]{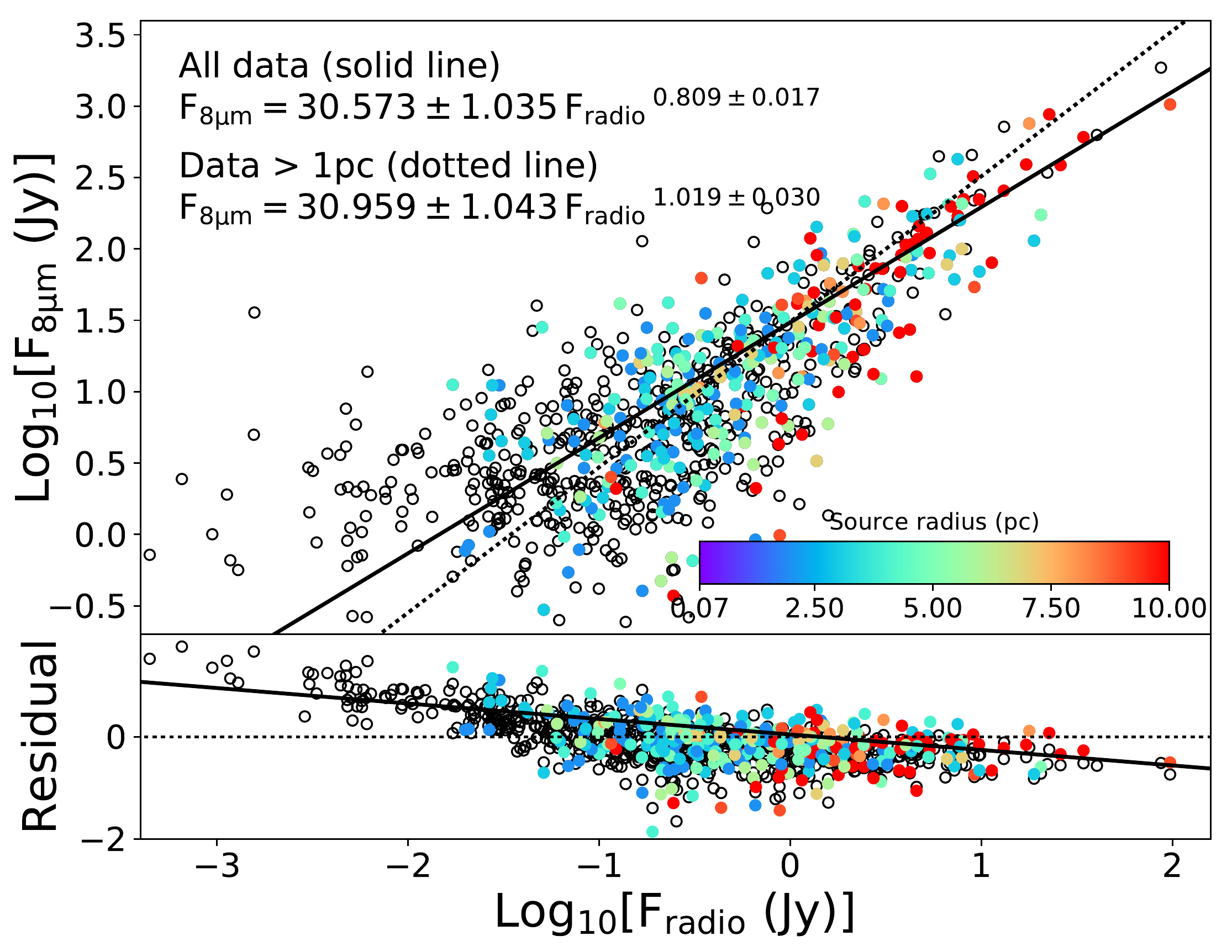}
   \includegraphics[width=0.47\textwidth]{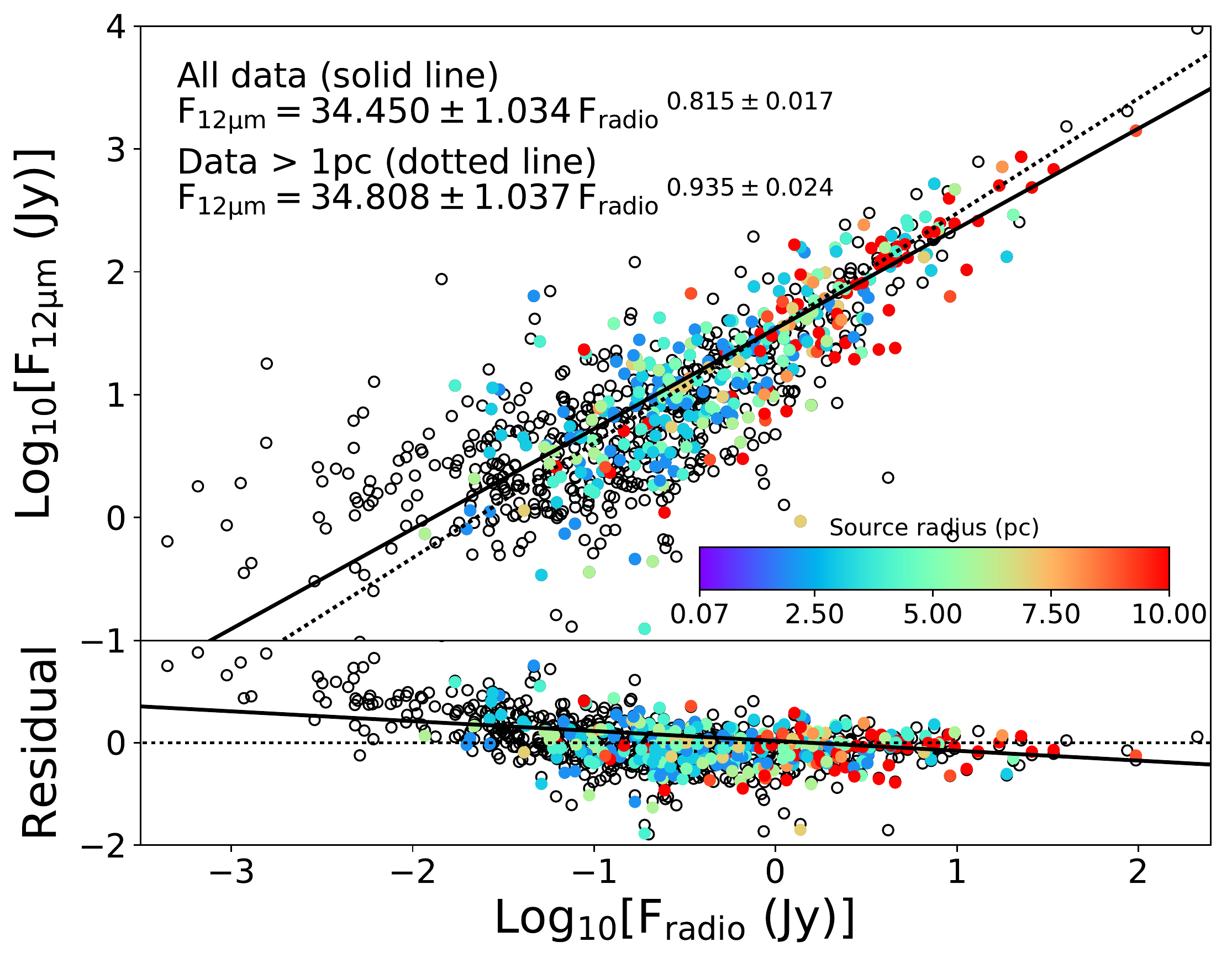}
   \medskip
   \includegraphics[width=0.47\textwidth]{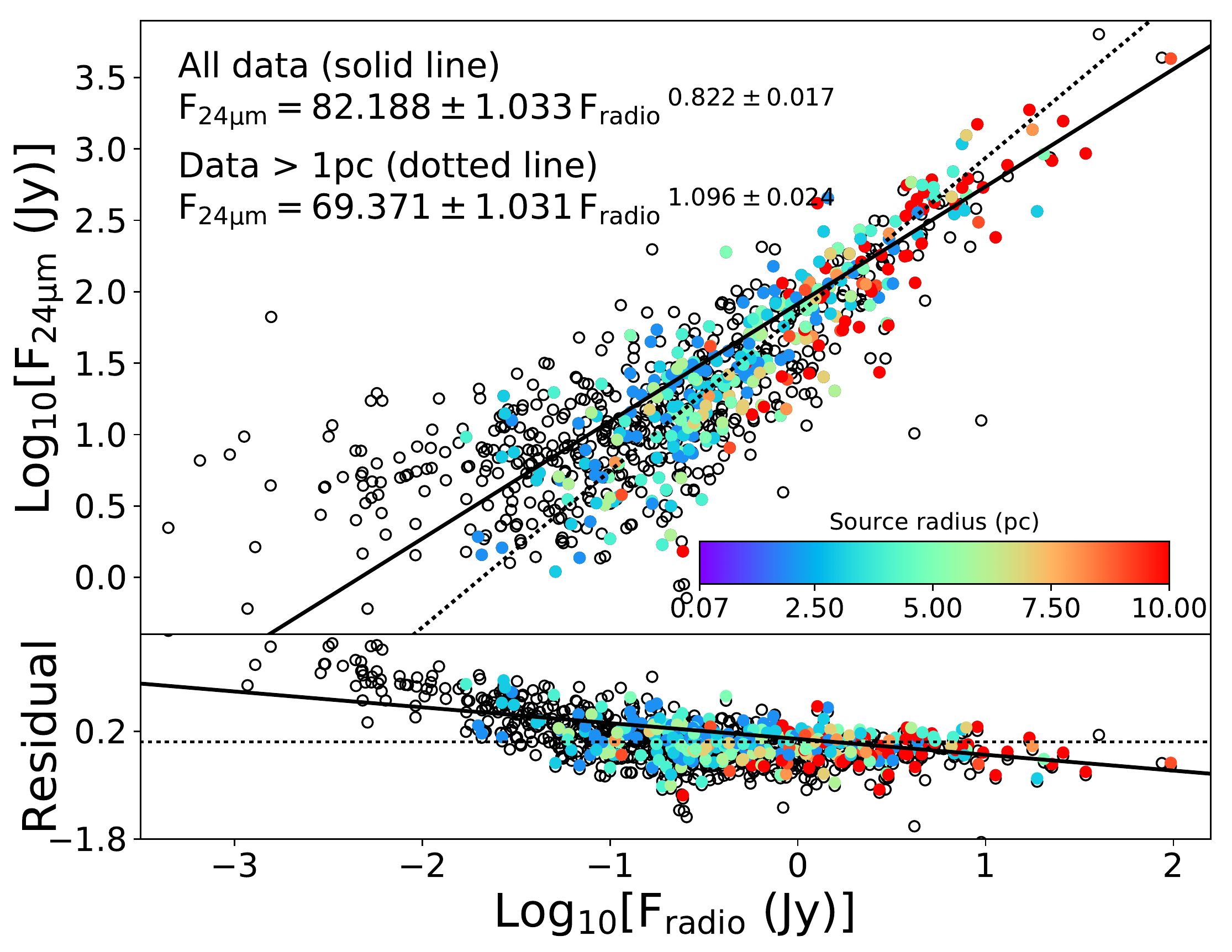}
   \includegraphics[width=0.47\textwidth]{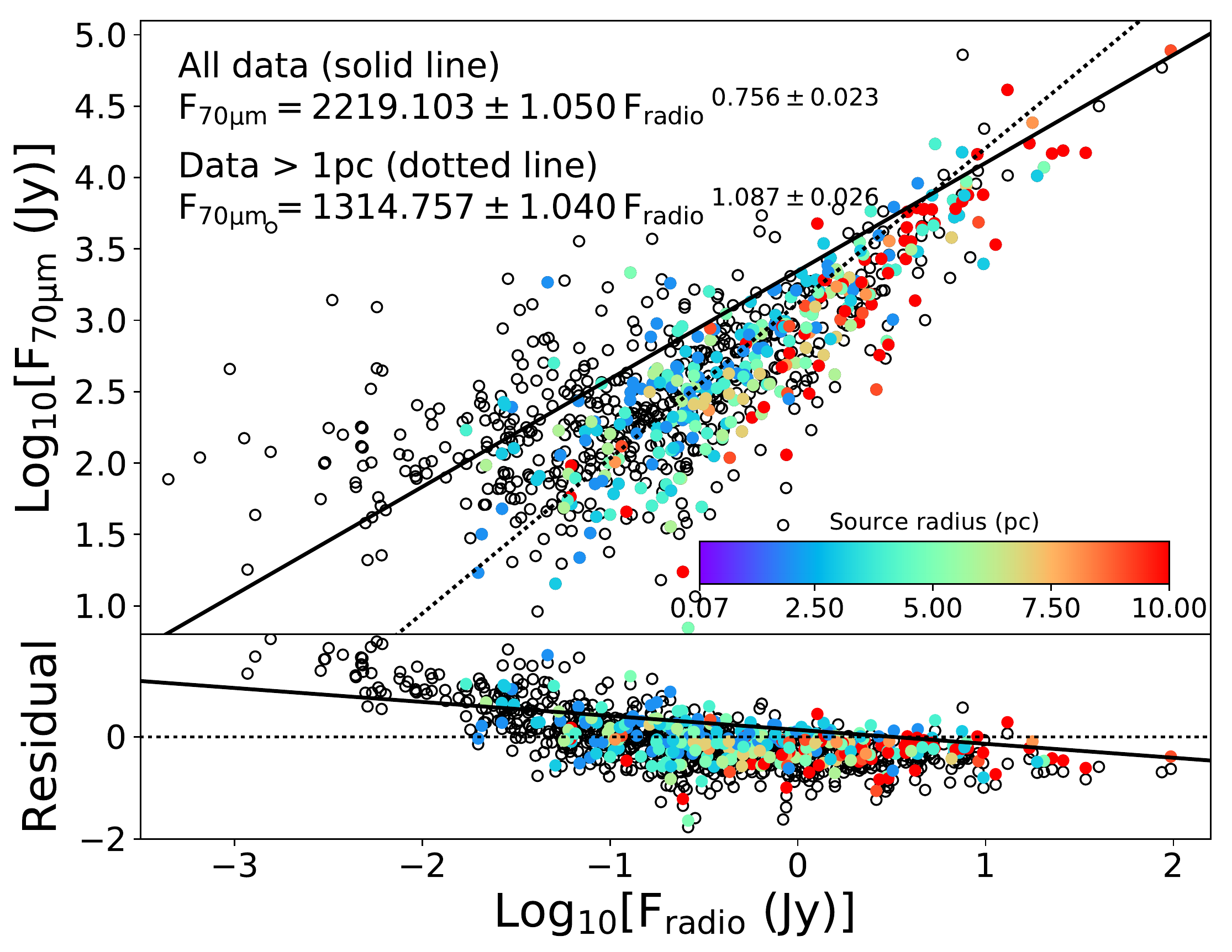}
   \medskip
   \includegraphics[width=0.47\textwidth]{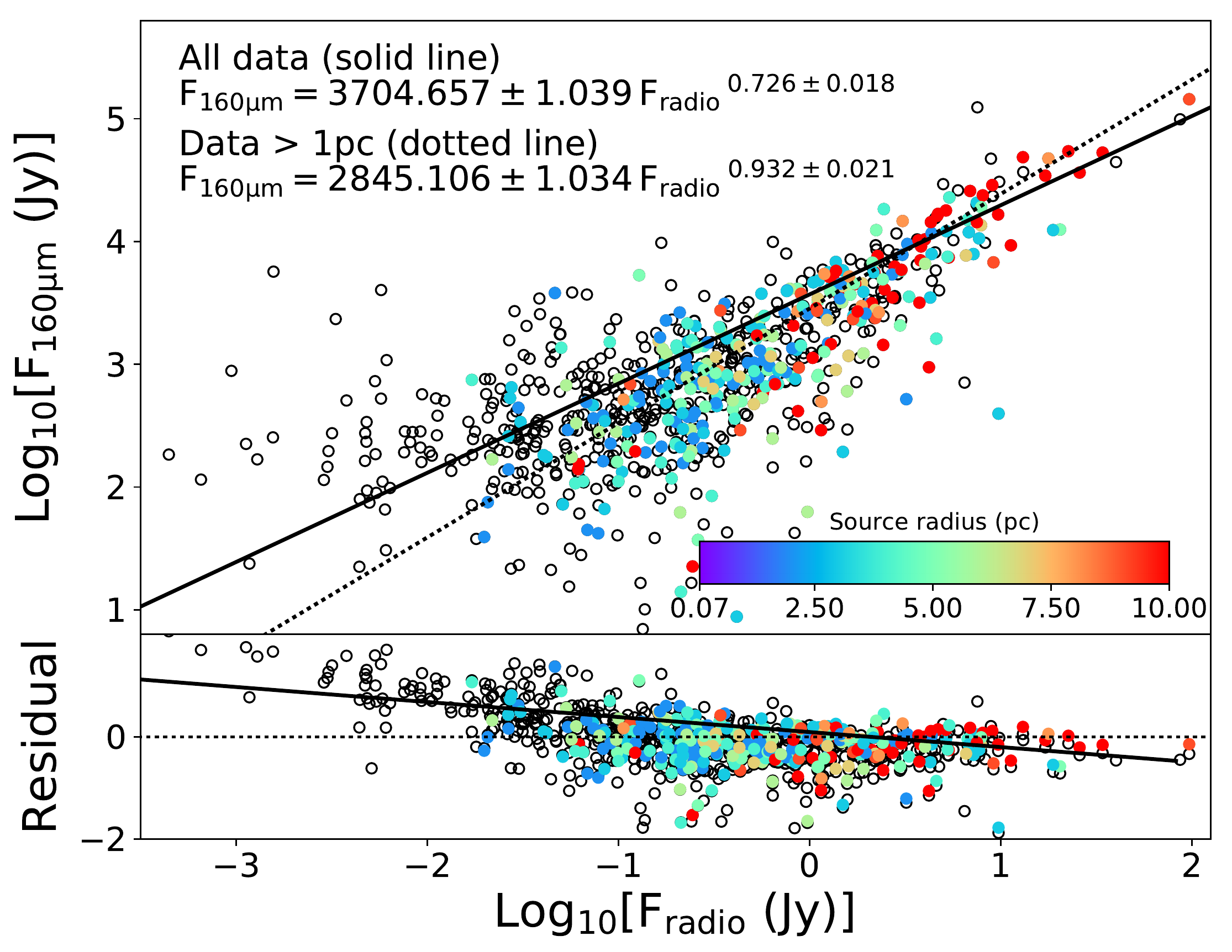}
   \caption{Comparisons of IR and radio flux densities for \hii
     regions. The symbol colors reflect the physical radius in
     parsecs. The open circles are for sources that have no distance
     measurements, and hence unknown physical radii. The solid
     lines are the best fit to all data points. The dotted 
     line marks the fit to the colored data points larger than $1\,$pc. 
     The larger scatter at the lower flux densities probably caused 
     by the photometric uncertainties. The label 
     $F_{radio}$ denotes the combined $20\,$cm and $21\,$cm 
     radio continumm data (see text). \label{fig:ir_radio}}
\end{figure*}

\subsection{IR and radio flux density correlations}
\label{subsec:ir_radio}
We investigate correlations between the IR and radio flux densities in
Figure~\ref{fig:ir_radio}, and summarize the results in
Table~\ref{tab:fit_results}, Figures~\ref{fig:boxplot1b} and \ref{fig:boxplot2}.

The infrared and radio flux densities are strongly correlated, albeit 
with higher scatter at the lower flux density values. We perform each 
fit twice, once to all data (solid line in Figure~\ref{fig:ir_radio}), 
and once to data from regions with radii $r > 1\,$pc (dotted line in 
Figure~\ref{fig:ir_radio}). The latter fit by default excludes all regions 
lacking known distances. The fits to all data all have similar power law 
exponents in the range $\sim 0.7-0.8$. The power law exponents for regions 
with radii $r > 1\,$pc are all near unity. These results show that the 
smallest regions have on average higher IR to radio flux density ratios 
compared with larger regions. Additionally, there are many regions with 
low radio flux densities and high IR to radio

\begin{deluxetable}{ccDDc}
\tablecaption{Correlation between \hii\ region radio and IR flux densities\tablenotemark{a} \label{tab:fit_results}}
\tablehead{
\colhead{$\lambda\,(IR)$} & \colhead{Data} &\multicolumn2c{$A$} & \multicolumn2c{$\alpha$} & \colhead{$R^{2}$}}
\decimals
\startdata
\multirow{2}{*}{8\microns}   & All      & 30.573$\pm$1.035   & 0.809$\pm$0.017 & 0.654 \\
                             & $>1\,$pc & 30.959$\pm$1.043   & 1.019$\pm$0.030 & 0.385 \\
\multirow{2}{*}{12\microns}  & All      & 34.450$\pm$1.034   & 0.815$\pm$0.017 & 0.670 \\
                             & $>1\,$pc & 34.808$\pm$1.037   & 0.935$\pm$0.024 & 0.721 \\                             
\multirow{2}{*}{24\microns}  & All      & 82.188$\pm$1.033   & 0.822$\pm$0.017 & 0.716 \\
                             & $>1\,$pc & 69.371$\pm$1.031   & 1.096$\pm$0.024 & 0.526 \\
\multirow{2}{*}{70\microns}  & All      & 2219.103$\pm$1.050 & 0.756$\pm$0.023 & 0.421 \\
                             & $>1\,$pc & 1314.757$\pm$1.040 & 1.087$\pm$0.026 & 0.369 \\
\multirow{2}{*}{160\microns} & All      & 3704.657$\pm$1.039 & 0.726$\pm$0.018 & 0.718 \\
                             & $>1\,$pc & 2845.106$\pm$1.034 & 0.932$\pm$0.021 & 0.598 \\
\enddata
\tablenotetext{a}{Fits made using equation $F_{radio} = A F^{\alpha}_{\lambda\,(IR)}$, where $F_{radio}$ denotes the combined $20\,$cm and $21\,$cm radio continuum data.}
\end{deluxetable}

\noindent{flux density ratios that lack physical size measurements; 
we speculate that these too have small physical sizes.}

We observed similarly strong correlations between the IR 
and radio data with an exception of the $70$\microns data that has 
slightly weaker correlation. These strong correlations suggest that 
IR emissions can be used as a star-formation tracer, observed even 
at longer wavelengths \citep[e.g.][]{calzetti2010}. When considering 
only the regions with radii $r > 1\,$pc, the GLIMPSE $12.0$\microns 
and Hi-Gal $160$\microns data show the strongest relationships. Since 
the radio emission is the most direct tracer of high-mass star-formation, 
these results hint that infrared emission can be used as a suitable 
proxy, but that the accuracy of the IR to radio ratio can be improved 
by excluding the smallest regions.

\subsection{Variation of IR and radio properties with HII region radius}
\label{subsec:fluxratios_size}
The size of an \hii region is related to the ionizing photon flux
\citep{stromgren1939}. \hii regions expand as they age, but
\hii regions hosting more massive ionizing stars expand faster
\citep[cf.][]{spitzer1978}. On average, therefore, a population of
larger \hii regions would host stars with higher ionizing photon
fluxes densities compared with a population of smaller \hii regions.

\begin{figure*}
   \centering
   \includegraphics[width=\textwidth]{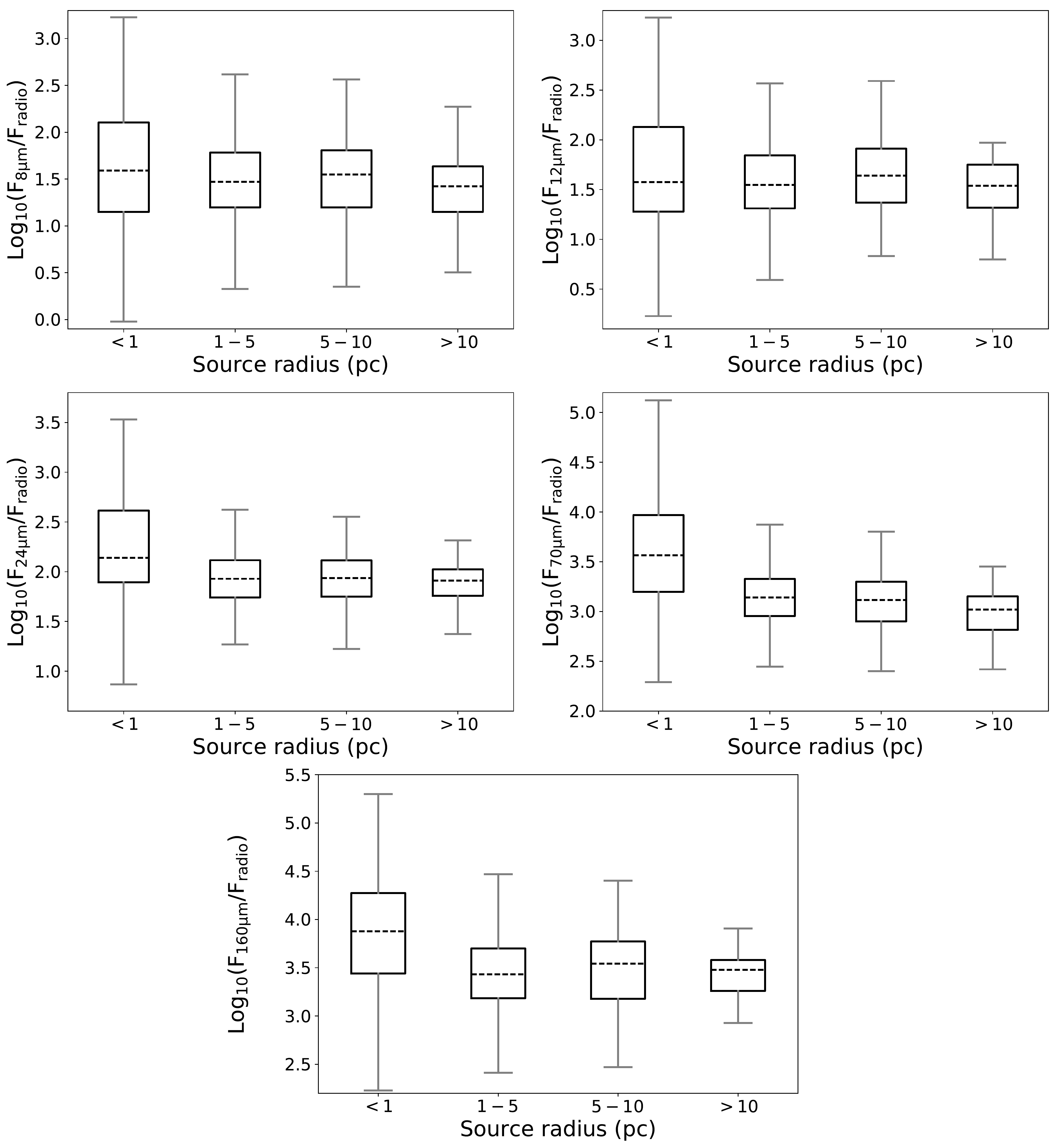}
   \caption{``Box and whiskers'' plots for the IR/radio ratios in four
     size bins. The dashed lines mark the median values (second
     quartile). The lower and upper limits of the boxes mark the
     $25^{th}$ percentile (first quartile; Q1) and $75^{th}$
     percentile (third quartile; Q3) of the data, respectively. We use
     for the ``whiskers'': $Q1-1.5 \times IQR$ (for the lower
     limit) and $Q3+1.5 \times IQR$ (for the upper limit),
     where IQR is the interquartile ($IQR=Q3-Q1$). 
     The plots show relatively flat IR to radio ratios as a 
     function of \hii region size, although the ratio is elevated for 
     the smallest \hii regions $< 1\,$pc at the longer IR wavelengths. 
     The label $F_{radio}$ denotes the combination of the $20\,$cm 
     and the $21\,$cm radio continuum data (see Section \ref{subsec:combradio}). \label{fig:boxplot1b}}
\end{figure*}

\begin{figure*}
   \centering
   \includegraphics[width=0.9\textwidth]{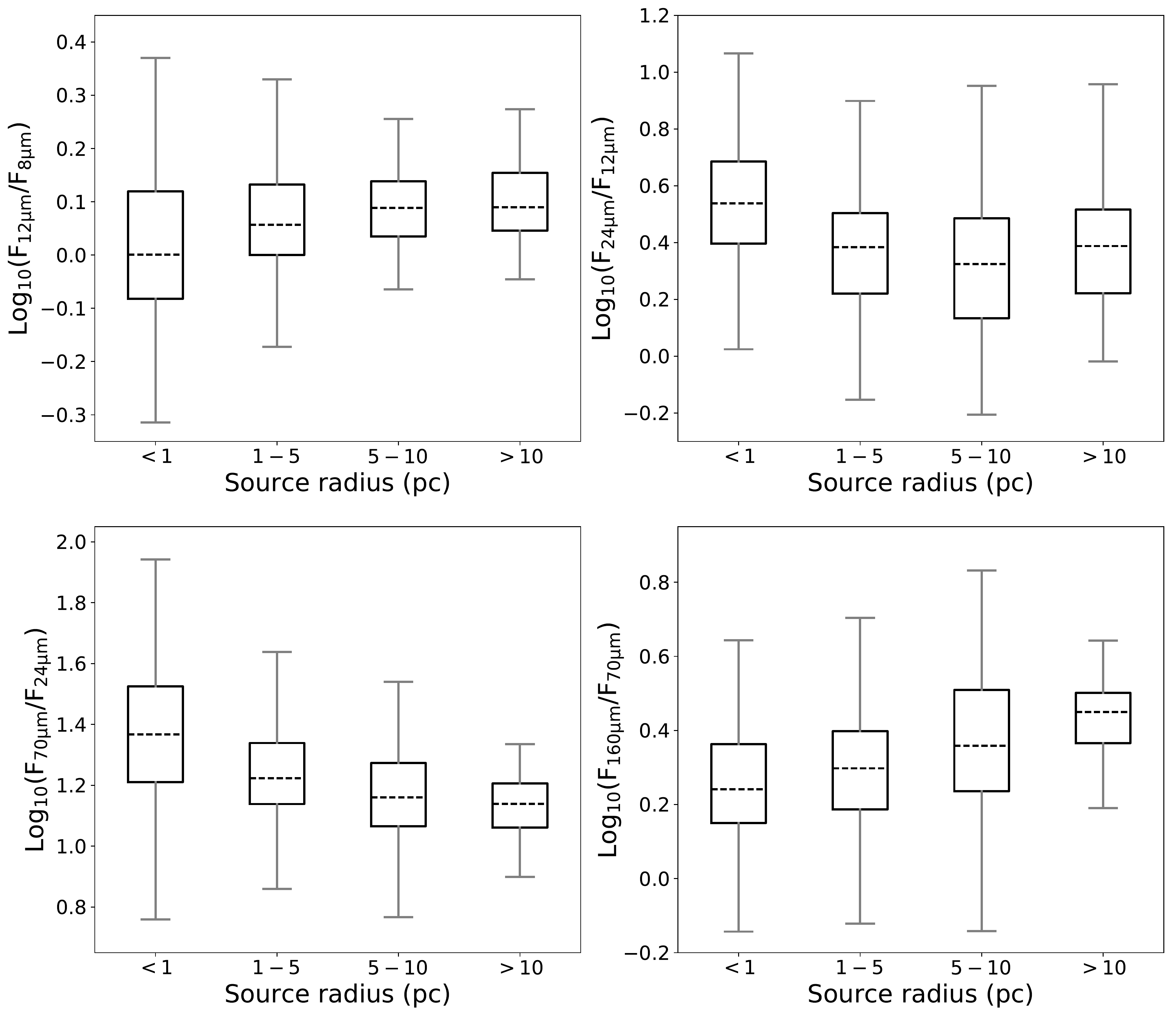}
   \caption{Box and whiskers plots as in Figure~\ref{fig:boxplot1b}
     for the IR colors in four size bins. The $12$\microns to
     $8$\microns and $24$\microns to $12$\microns ratios do not show a
     clear trend. The $70$\microns to $24$\microns ratio decreasing
     with increasing \hii region radius, while the $160$\microns to
     $70$\microns ratio increases with \hii region
     radius. \label{fig:boxplot2}}
\end{figure*}

\begin{figure*}
   \centering
   \includegraphics[width=0.9\textwidth]{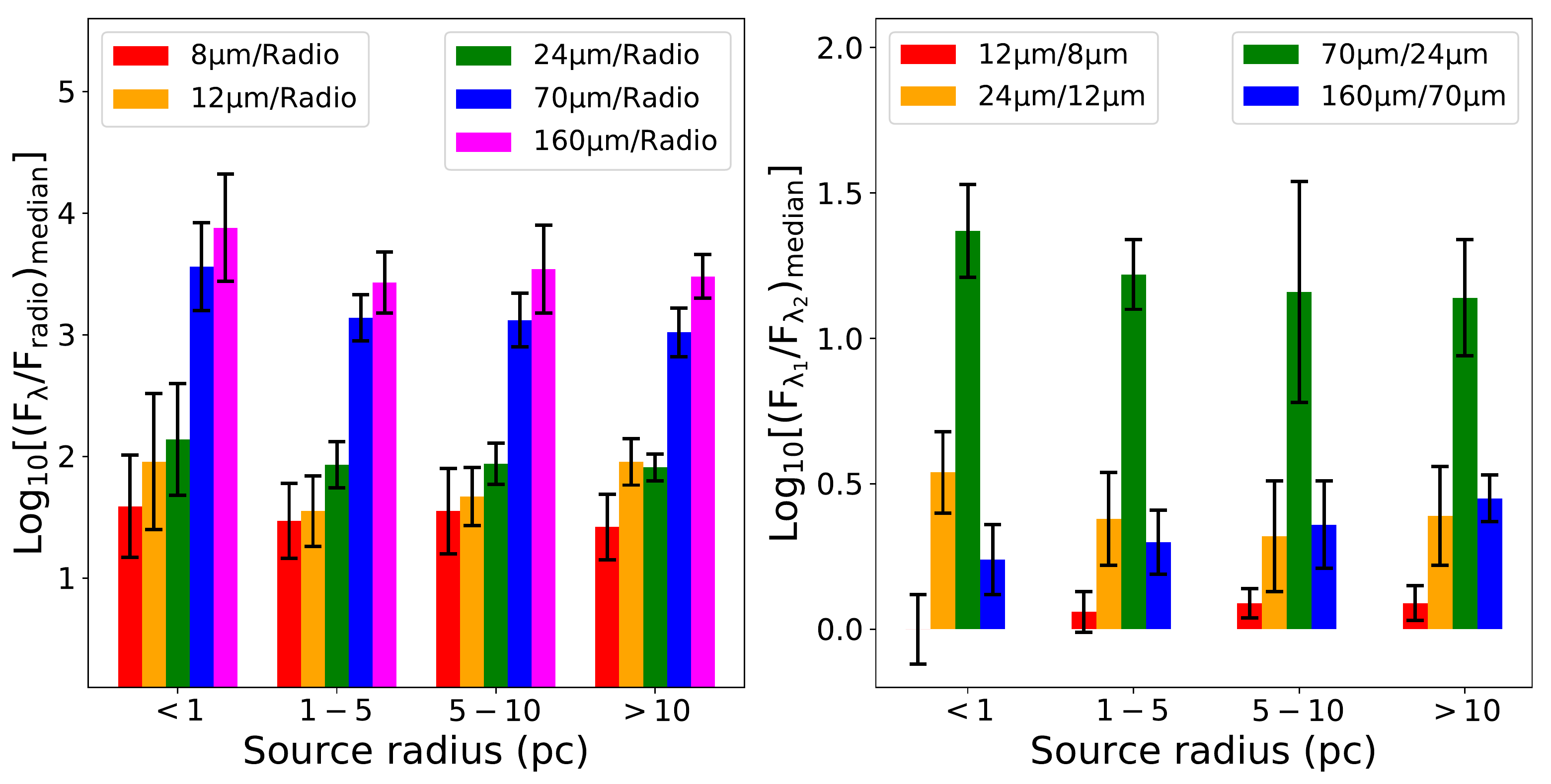}
   \caption{Graphical summary of the median values from
     Table~\ref{tab:bxplt_res2}. The standard deviation of the data
     points is marked with the error bars. The ``radio'' term
     means the combination of the $20\,$cm and the $21\,$cm radio
     continuum data, see Section \ref{subsec:combradio}. On the
     left plot, we show the ratios of the IR and radio flux densities.
     For the smallest regions, the IR to radio ratios are higher at
     the longest wavelengths, and the dispersion is greater.  The
     right plot shows the distribution of median values of color
     indices. This plot suggests that all the investigated IR flux
     density ratios are unchanged within the errors, independent of
     \hii region size.\label{fig:barplot1}}
\end{figure*}

We study the distribution of the IR to radio flux density ratios in
Figure~\ref{fig:boxplot1b}, and the IR flux density ratios (color
indices) in Figure~\ref{fig:boxplot2}, all as a fuction of \hii region
radius. We summarize the results of these two Figures in
Table~\ref{tab:bxplt_res2}, and graphically in
Figure~\ref{fig:barplot1}. In this analysis, for clarity we do not
analyze the $22$\microns data since the flux densities are essentially
the same as those at $24$\microns (see Figure~\ref{fig:glmi_w3w4}).
Table~\ref{tab:bxplt_res2} gives the median values of the flux
densities (dashed lines on Figures~\ref{fig:boxplot1b}), the standard
deviations, the color criteria, that is, the range of the ``whiskers''
(see subtitle of Figure~\ref{fig:boxplot1b}) in Figures~\ref{fig:boxplot1b} 
and \ref{fig:boxplot2}, and the number of data points in four different 
source size bins. The lines ``All'' show the same values but for 
\textit{all} data points, independent of the bins.

The IR to radio flux density ratios in Figure~\ref{fig:boxplot1b} and 
Table~\ref{tab:bxplt_res2} show that the median IR to radio ratio has 
no strong dependence on \hii region size. We note, however, that the 
median IR to radio ratios for $24$\microns, 70\micron, and $160$\microns 
is elevated for the smallest regions with $r < 1$\,pc. Additionally, 
the spread of IR to radio ratios is larger for all IR wavelengths in 
this smallest size bin.

We investigate the IR colors of \hii regions as a function of \hii
region radius in Figure~\ref{fig:boxplot2}, summarize the results in
Figure~\ref{fig:barplot1}, and list the results in
Table~\ref{tab:bxplt_res2}. In this table, we give the median values
of the flux densities (red lines on Figures~\ref{fig:boxplot1b} and
\ref{fig:boxplot2}), the standard deviations, the color criteria
(range between the whiskers) and the number of data points in four
different source size bins. The lines ``All'' show the same values but
for \textit{all} data points, independent of the bins. Despite the
strong correlation of \hii region flux densities and IR to radio
ratios with \hii region sizes, IR colors are not a strong function of
\hii region size. The two exceptions to this trend are the 
$\log_{10}(F_{70\microns}/F_{24\microns})$ ratio, which does decrease 
with increasing \hii region size, and the 
$\log_{10}(F_{160\microns}/F_{70\microns})$ ratio, which increases 
with increasing \hii region size. Because the $70$\microns data are 
sensitive to both small and large grains, the dependence of this ratio 
on \hii region size may imply that the larger regions have a larger 
small grain population relative to large grains. It is also worth 
noting that the $\log_{10}(F_{160\microns}/F_{70\microns})$ ratio is 
slightly above unity for all sizes, which implies that the peak of the 
spectral energy distribution is closer to $160$\microns than $70$\microns 
\citep{anderson2012a}.

\citet{dopita2006} suggested that infrared color-color diagrams of \hii regions 
can be used as a pressure diagnostic tool. They produced color-color diagrams 
(Figure~$11$ in their paper) using theoretical spectral energy distributions 
(SEDs) passed through the transmission function of the MIPS instrument on \textit{Spitzer}.
We were not able to compare the theoretical models introduced by \citet{dopita2006} 
with our observational data points because their SEDs peak at $\sim 30$\microns; a 
peak this low is not observed.

\subsection{Do the UC HII region color criteria only select UC HII regions?}
\label{subsub:uc}

It has been known for some time that IR colors can be used to identify
\hii regions \citep{wood1989b,hughes1989,zoo1990,white1991,helfand1992,becker1994}.
The IR criteria, however, have been derived almost exclusively using
IRAS point sources. At the IRAS angular resolutions of $0\fmin5$,
$0\fmin5$, $1\fmin0$ and $2\fmin0$ at $12$\microns,

\newpage

\begin{deluxetable*}{lcDDcc}
\tabletypesize{\small}
\tablecaption{Median IR and radio flux density ratios as a function of \hii region radius \label{tab:bxplt_res2}}
\tablehead{
\colhead{Flux density ratio} & \colhead{Radius} & \multicolumn2c{$\mathrm{Median}$} & \multicolumn2c{$\mathrm{\sigma}$} & \colhead{Color criteria} & \colhead{$\#$ of sources} \\
 & [pc] & \multicolumn{2}{c}{} & \multicolumn{2}{c}{} & $\mathrm{(Q1-1.5 \times IQR}$ --- $\mathrm{Q3 + 1.5 \times IQR)}$\tablenotemark{*} & }
\decimals
\startdata
\multirow{5}{*}{$\mathrm{\log_{10}(F_{8\microns}/F_{\rm radio})}$} & $r < 1$ & $1.59$ & $0.42$ & $-0.02 - 3.23$ & $141$ \\
  & $1 < r < 5$  & $1.47$ & $0.31$ & $0.33 - 2.62$  & $237$ \\
  & $5 < r < 10$ & $1.55$ & $0.35$ & $0.35 - 2.56$  & $87$ \\
  & $r > 10$     & $1.42$ & $0.27$ & $0.50 - 2.27$  & $43$ \\
  & All          & $1.52$ & $0.34$ & $-0.02 - 3.23$ & $508$ \\
  \multicolumn{7}{c}{} \\
\multirow{5}{*}{$\mathrm{\log_{10}(F_{12\microns}/F_{\rm radio})}$} & $r < 1$ & $1.57$ & $0.56$ & $0.23 - 3.23$ & $137$ \\
  & $1 < r < 5$  & $1.55$ & $0.29$ & $0.59 - 2.57$ & $238$ \\
  & $5 < r < 10$ & $1.67$ & $0.24$ & $0.83 - 2.59$ & $98$ \\
  & $r > 10$     & $1.54$ & $0.19$ & $0.80 - 1.97$ & $48$ \\
  & All          & $1.59$ & $0.32$ & $0.23 - 3.23$ & $521$ \\
  \multicolumn{7}{c}{} \\
\multirow{5}{*}{$\mathrm{\log_{10}(F_{24\microns}/F_{\rm radio})}$} & $r < 1$ & $2.14$ & $0.46$ & $0.87 - 3.53$ & $136$ \\
  & $1 < r < 5$  & $1.93$ & $0.19$ & $1.27 - 2.62$ & $232$ \\
  & $5 < r < 10$ & $1.94$ & $0.17$ & $1.22 - 2.55$ & $90$ \\
  & $r > 10$     & $1.91$ & $0.11$ & $1.37 - 2.31$ & $42$ \\
  & All          & $1.99$ & $0.25$ & $0.87 - 3.53$ & $500$ \\
    \multicolumn{7}{c}{} \\
\multirow{5}{*}{$\mathrm{\log_{10}(F_{70\microns}/F_{\rm radio})}$} & $r < 1$ & $3.56$ & $0.36$ & $2.29 - 5.12$ & $142$ \\
  & $1 < r < 5$  & $3.14$ & $0.19$ & $2.45 - 3.87$ & $240$ \\
  & $5 < r < 10$ & $3.12$ & $0.22$ & $2.40 - 3.80$ & $95$ \\
  & $r > 10$     & $3.02$ & $0.20$ & $2.42 - 3.45$ & $44$ \\
  & All          & $3.20$ & $0.24$ & $2.29 - 5.12$ & $521$ \\
  \multicolumn{7}{c}{} \\
\multirow{5}{*}{$\mathrm{\log_{10}(F_{160\microns}/F_{\rm radio})}$} & $r < 1$ & $3.88$ & $0.44$ & $2.23 - 5.30$ & $140$ \\
  & $1 < r < 5$  & $3.43$ & $0.25$ & $2.41 - 4.47$ & $239$ \\
  & $5 < r < 10$ & $3.54$ & $0.36$ & $2.47 - 4.40$ & $91$ \\
  & $r > 10$     & $3.48$ & $0.18$ & $2.93 - 3.91$ & $42$ \\
  & All          & $3.59$ & $0.31$ & $2.23 - 5.30$ & $512$ \\
\hline
\multirow{5}{*}{$\mathrm{\log_{10}(F_{12\microns}/F_{8\microns})}$} & $r < 1$ & $0.00$ & $0.12$ & $-0.31 - 0.37$ & $327$ \\
  & $1 < r < 5$  & $0.06$ & $0.07$ & $0.17 - 0.33$  & $267$ \\
  & $5 < r < 10$ & $0.09$ & $0.05$ & $-0.06 - 0.26$ & $91$ \\
  & $r > 10$     & $0.09$ & $0.06$ & $-0.05 - 0.27$ & $46$ \\
  & All          & $0.06$ & $0.07$ & $-0.31 - 0.37$ & $550$ \\
  \multicolumn{7}{c}{} \\
\multirow{5}{*}{$\mathrm{\log_{10}(F_{24\microns}/F_{12\microns})}$} & $r < 1$ & $0.54$ & $0.14$ & $0.02 - 1.07$ & $144$ \\
  & $1 < r < 5$  & $0.38$ & $0.16$ & $-0.15 - 0.90$ & $262$ \\
  & $5 < r < 10$ & $0.32$ & $0.19$ & $-0.21 - 0.95$ & $92$ \\
  & $r > 10$     & $0.39$ & $0.17$ & $-0.02 - 0.96$ & $44$ \\
  & All          & $0.41$ & $0.16$ & $-0.21 - 1.07$ & $542$ \\
  \multicolumn{7}{c}{} \\
\multirow{5}{*}{$\mathrm{\log_{10}(F_{70\microns}/F_{24\microns})}$} & $r < 1$ & $1.37$ & $0.16$ & $0.76 - 1.94$ & $145$ \\
  & $1 < r < 5$  & $1.22$ & $0.12$ & $0.86 - 1.64$ & $262$ \\
  & $5 < r < 10$ & $1.16$ & $0.38$ & $0.77 - 1.54$ & $93$ \\
  & $r > 10$     & $1.14$ & $0.20$ & $0.90 - 1.34$ & $45$ \\
  & All          & $1.23$ & $0.21$ & $0.76 - 1.94$ & $545$ \\
  \multicolumn{7}{c}{} \\
\multirow{5}{*}{$\mathrm{\log_{10}(F_{160\microns}/F_{70\microns})}$} & $r < 1$ & $0.24$ & $0.12$ & $-0.14 - 0.64$ & $150$ \\
  & $1 < r < 5$  & $0.30$ & $0.11$ & $-0.12 - 0.70$ & $273$ \\
  & $5 < r < 10$ & $0.36$ & $0.15$ & $-0.14 - 0.83$ & $96$ \\
  & $r > 10$     & $0.45$ & $0.08$ & $0.19 - 0.64$  & $44$ \\
  & All          & $0.35$ & $0.11$ & $-0.14 - 0.83$ & $563$ \\
\enddata
\tablenotetext{*}{$Q1$ and $Q3$ are the $25^th$ and $75^th$ percentile of the data, respectively, and $IQR=(Q3-Q1)$ is the interquartile.}
\end{deluxetable*}

\noindent{$25$\microns, $60$\microns and $100$\microns respectively, the 
use of these IRAS criteria only identify relatively compact sources. WC89 
suggested that the IR colors of ultra-compact (UC) \hii regions are distinct
from other sources found in the IRAS PSC, which was supported by high
resolution VLA observations \citep{wood1989b}. They found that UC \hii
regions have color indices of
$\log_{10}(F_{25\microns}/F_{12\microns}) \ge 0.57$ and
$\log_{10}(F_{60\microns}/F_{12\microns}) \ge 1.22$. Based on the
number of IRAS PSC sources that satisfy these criteria, they suggested
that there are $1650$ UC \hii regions in the Galaxy. By examining the
IRAS colors of optical \hii regions, \citet{hughes1989} found that
similar colors select UC \hii regions:
$\log_{10}(F_{25\microns}/F_{12\microns}) \ge 0.4$ and
$\log_{10}(F_{60\microns}/F_{25\microns}) \ge 0.25$ and estimated that
there are $\sim 2300$ UC \hii regions in the Galaxy.}

It is often assumed that the WC89 criteria identify UC \hii regions,
but perhaps not more evolved nebulae \citep[e.g.][]{cohen2007}. Although there is 
evidence that all \hii regions have similar IRAS colors \citep{hughes1989,anderson2012a}, 
this has never been proven for a sample as large as ours. \citet{chini1987} found 
in a study of $66$ compact \hii regions that all had a similar energy distributions 
in the range $1-1300$\microns. \citet{leto2009} found that UC \hii
regions cannot be identified solely by their infrared colors due to
the contamination of other classes of Galactic sources. The number of
UC \hii regions in the Galaxy therefore remains somewhat uncertain.

\begin{figure*}
   \centering
   \includegraphics[width=0.49\textwidth]{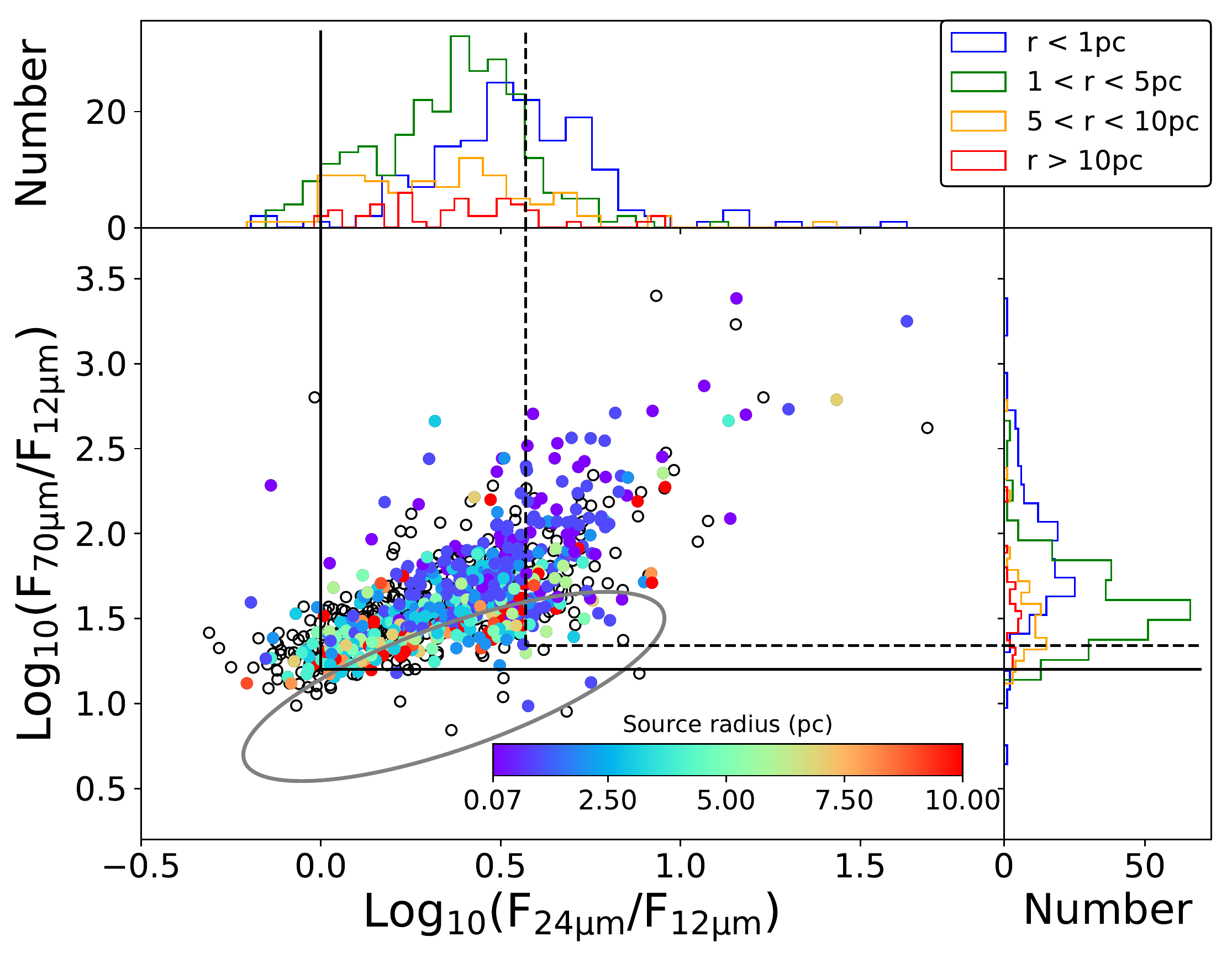}
   \includegraphics[width=0.49\textwidth]{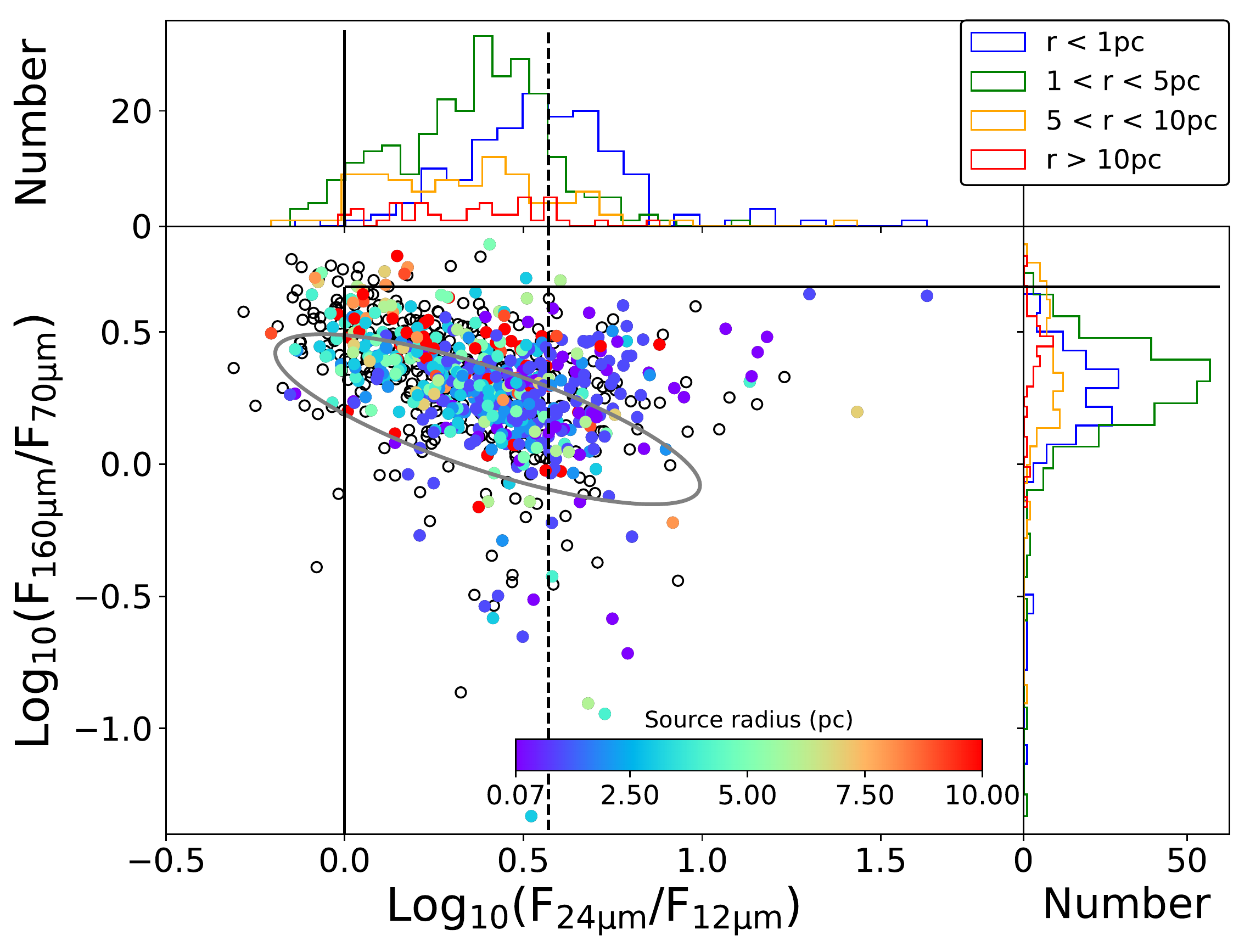}
   \caption{\hii region color-color diagrams similar to those in
     WC89. The color scale shows the physical size of the \hii regions
     in parsecs while the empty black circles represent the sources
     lacking size measurements. On the left subplot, the dashed lines
     represent the color-corrected infrared color indices
     $\log_{10}(F_{25\microns}/F_{12\microns}) \ge 0.57$ and
     $\log_{10}(F_{70\microns}/F_{12\microns}) \ge 1.34$ used to
     separate the UC \hii regions from other point sources (WC89).
     On the right subplot, the dashed line shows the 
     $\log_{10}(F_{25\microns}/F_{12\microns}) \ge 0.57$ limit. The 
     gray ellipses show areas where $\sim 95\%$ of the galaxies from 
     \citet{sanders2003} are located (see text).
     The solid lines represent the criteria suggested by our data: 
     $\log_{10}(F_{24\microns}/F_{12\microns}) \ge 0$ and
     $\log_{10}(F_{70\microns}/F_{12\microns}) \ge 1.2$ (left subplot), 
     $\log_{10}(F_{24\microns}/F_{12\microns}) \ge 0$ and 
     $\log_{10}(F_{160\microns}/F_{70\microns}) \le 0.67$ (right subplot).
     These color indices can therefore be used as an approximate IR limit 
     for \hii regions, \textit{independent} of size. \label{fig:cc1}}
\end{figure*}

In Figure~\ref{fig:cc1} we show similar color-color diagrams to those
of WC89 (their Figure 1b): in place of their
  $\log_{10}(F_{25\microns}/F_{12\microns})$ and
  $\log_{10}(F_{60\microns}/F_{12\microns})$ color indices we show
  $\log_{10}(F_{24\microns}/F_{12\microns})$ and
  $\log_{10}(F_{70\microns}/F_{12\microns})$ color indices. Since our
$70$\microns data sample a slightly different portion of the spectral
energy distribution compared with IRAS $60$\microns data, we must
scale the WC89 criteria used to identify UC \hii regions. The
PACS
guide\footnote{\url{http://herschel.esac.esa.int/twiki/pub/Public/PacsCalibrationWeb/cc_report_v1.pdf}}
shows that the PACS $70$\microns data should be multipled by $0.76$ to
convert to the IRAS $60$\microns bandpass, assuming $30$ to $100\,$K
dust. The WC89 criterion of $\log_{10}(F_{60\microns}/F_{12\microns})
\ge 1.22$ therefore is equivalent to
$\log_{10}(F_{70\microns}/F_{12\microns}) \ge 1.34$, which we show on
the plot. Because of the similarity of the bandpasses, we do not scale
the $12$\microns or $25$\microns flux densities. As we do not have
flux densities at $100$\microns, we are not able to reproduce the
$\log_{10}(F_{25\microns}/F_{12\microns})$ versus
$\log_{10}(F_{100\microns}/F_{60\microns})$ color-color plot from
WC89. Instead, we show $\log_{10}(F_{24\microns}/F_{12\microns})$
versus $\log_{10}(F_{160\microns}/F_{70\microns})$.

Figure~\ref{fig:cc1} shows that the color citeria established by
WC89 are most sensitive to the smallest regions, but do not uniquely 
identify UC \hii regions. In the size bins defined earlier, 
$\sim 43\%$, $\sim 12\%$, $\sim 16\%$ and $\sim 11\%$ of colored data points
satisfy the $\log_{10}(F_{24\microns}/F_{12\microns})$ versus 
$\log_{10}(F_{70\microns}/F_{12\microns})$ color criteria of WC89 for 
source radius bins of $r<1$pc, $1<r<5$pc, $5<r<10$pc and $r>10$pc, 
respectively. 
For the color indices $\log_{10}(F_{24\microns}/F_{12\microns})$ versus 
$\log_{10}(F_{160\microns}/F_{70\microns})$, due to the uncertainty of 
converting the IRAS color criteria to the wavelengths used here, we can not 
reliably give percentages. Although there are no definite \hii region 
criteria for the $\log_{10}(F_{24\microns}/F_{12\microns})$ versus
$\log_{10}(F_{160\microns}/F_{70\microns})$ color indices, the
distribution of data points in our plot looks similar to that of WC89,
and suggest no visible trend with \hii region size. 

The new IR color indices defined here (solid lines in Figure~\ref{fig:cc1}), 
$\log_{10}(F_{24\microns}/F_{12\microns}) \ge 0$ and 
$\log_{10}(F_{70\microns}/F_{12\microns}) \ge 1.2$, are satisfied by 
$\sim 96\%$, $\sim 94\%$, $\sim 95\%$ and $\sim 98\%$ of the colored data 
points for source radii bins of $r<1$pc, $1<r<5$pc, $5<r<10$pc and $r>10$pc, 
respectively. For the IR color limits of $\log_{10}(F_{24\microns}/F_{12\microns}) 
\ge 0$ and $\log_{10}(F_{160\microns}/F_{70\microns}) \le 0.67$, these numbers 
are $\sim 98\%$, $\sim 94\%$, $\sim 87\%$ and $\sim 91\%$ for the same source 
radii bins.

\citet{dreher1981} suggested that the number of UC \hii regions is
much higher in our Galaxy then we would expect if we assume that their
sizes are determined by a free expansion at the thermal sound speed,
leading to the so-called ``lifetime problem''. The estimates for the
number of UC \hii regions in the Galaxy by WC89 and
\citet{hughes1989} only exacerbated the lifetime problem. There have
been a number of solutions to the lifetime problem proposed. For
example, \citet[][and references therein]{peters2010a,peters2010b},
showed that the \hii regions do not expand monotonically and
isotropically (during the main accretion phase). This effect causes
flickering of the radio continuum emission, which has been observed in
Sgr~B2 by \citet{depree2014}. It remains unclear, however, if the
lifetime problem is as acute as suggested due to uncertainties in the
number of UC \hii regions in the Galaxy. The similarity of all \hii
region IR colors, regardless of radius, implies that perhaps the
number of UC \hii regions in the Galaxy has been overestimated, and
therefore the lifetime problem is much less severe than sometimes
assumed.

The radii we use here cannot be directly compared with those derived
previously for UC \hii regions. \citet{churchwell2002} state that UC
\hii regions have diameters $\lsim 0.1\,$pc. These sizes, however,
correspond to the densest parts of ionized gas, sampled with
high-resolution radio interferometric observations. Our sizes include
all the emission from the regions, including any possible extended
envelopes \citep{kim2002}. Nevertheless, the fact that nearly
$25\%$ of the \hii regions in our sample satisfy the color
indices, and that our sample includes all known Galactic \hii regions,
means that the criteria cannot uniquely identify UC \hii
regions. The low total number of \hii regions satisfying the WC89 
criteria, and the fact that these criteria do not select only UC \hii 
regions, implies a significant uncertainty in the number of UC \hii 
regions in the Galaxy derived using the WC89 criteria.

Previous studies of IR colors of galaxies showed characteristic 
values for the IRAS bands. For example, \citet{helou1986} examined galaxies 
with no quasar-like nucleus and found a clear decreasing trend in 
$\log_{10}(F_{60\microns}/F_{100\microns})$ as the ratio 
$\log_{10}(F_{12\microns}/F_{25\microns})$ increases. This indicates a higher 
star-formation rate at higher values of $\log_{10}(F_{60\microns}/F_{100\microns})$ 
and lower values of $\log_{10}(F_{12\microns}/F_{25\microns})$. Similar results 
have been reported by \citet{soifer1991,wang1991} and \citet{sanders2003}.

In Figure~\ref{fig:cc1}, we investigated the IR flux density ratio pairs of 
$\log_{10}(F_{24\microns}/F_{12\microns})$ versus 
$\log_{10}(F_{160\microns}/F_{70\microns})$, and 
$\log_{10}(F_{24\microns}/F_{12\microns})$ versus 
$\log_{10}(F_{70\microns}/F_{12\microns})$, using $\sim 600$ galaxies from 
\citet{sanders2003}. Their sample has galaxies of morphological types irregular 
and spiral, including starbursting. They found that the main contributor of 
the observed thermal emission from galaxies is the cool dust ($T_{dust} \sim 15-70\,$K). 
However, as \citet{sanders2003} pointed out, a significant amount of emission 
from warm dust (peaking around $25$\microns) has also been observed from galaxies. 
\citet{sanders2003} also reported an inverse correlation between 
$\log_{10}(F_{12\microns}/F_{25\microns})$ and 
$\log_{10}(F_{60\microns}/F_{100\microns})$, which was interpreted as the result 
of efficient destruction of small grains due to the increasing total galaxy IR 
luminosity.

As we do not have the same IR wavelengths as IRAS, we require a correction factor 
to be able to compare with previous results. Using the PACS 
guide\footnote{\url{http://herschel.esac.esa.int/twiki/pub/Public/PacsCalibrationWeb/cc_report_v1.pdf}}, 
and assuming dust temperature range of  $30 - 100\,$K, we divided the flux density 
in the $60$\microns bandpass by a correction factor ($0.76$) to convert it to the 
$70$\microns bandpass. To convert the IRAS $100$\microns flux densities to the 
$160$\microns bandpass, we used SED templates of different type of galaxies from 
\citep{sieben2007}: $F_{160\microns} \approx F_{100\microns}/cf$ where 
$cf \approx 1.55$ is the averaged correction factor derived from the SEDs. Even 
if this method may not be sophisticated for qualitative studies, we believe, it is 
good enough for statistical purposes. About $\sim 95\%$ of the galaxies from 
\citet{sanders2003} are located within the gray ellipses in Figure~\ref{fig:cc1}.

We conclude that the color indices of star forming galaxies are similar to those of 
\hii regions. The $\log_{10}(F_{24\micron})/\log_{10}(F_{12\micron})$ color index of 
\hii regions is essentially the same as that of galaxies. There are offsets compare to 
the IR colors of Galactic \hii regions ($\sim 0.6\,$dex and $\sim 0.2\,$dex in ratios of 
$\log_{10}(F_{70\microns}/F_{12\microns})$ and $\log_{10}(F_{24\microns}/F_{12\microns})$, 
respectively). Since the flux density of \hii regions peaks near $70$\microns, that this 
ratio differs between galaxies and \hii regions is not unexpected.

\subsection{Variation of IR and radio properties with $R_{gal}$}
\label{subsec:fluxratios_rgal}

\begin{figure*}
   \centering
   \includegraphics[width=\textwidth]{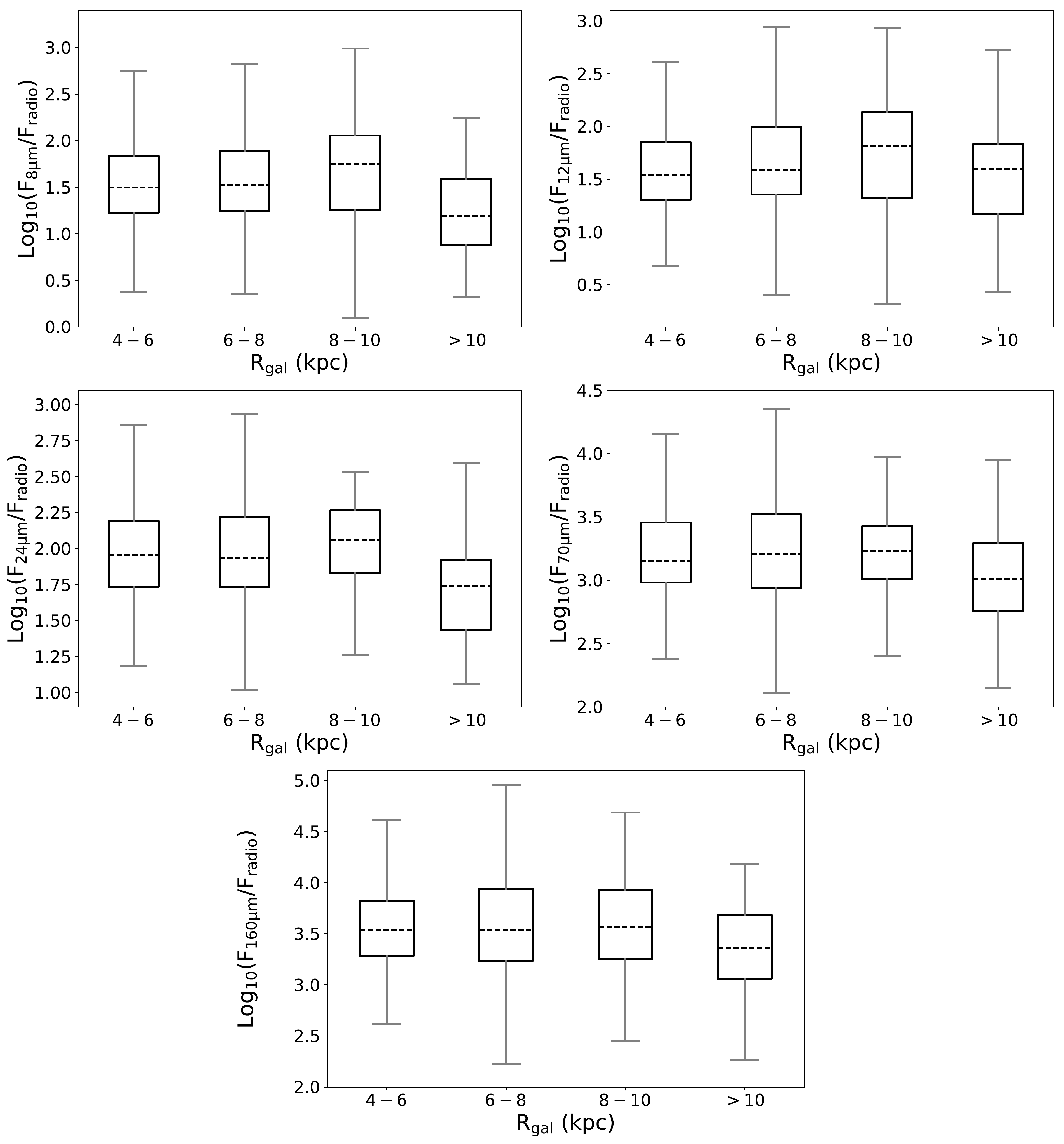}
   \caption{Box and whiskers plots as in Figure~\ref{fig:boxplot1b}
     for the IR/radio flux density ratios in four Galactocentric radii
     bins. The plots, show no trend in IR/radio ratios for \hii
     regions with $R_{gal}$, although the regions with
     $R_{gal}>10\,$kpc do show consistently lower ratios. The
     label $F_{radio}$ denotes the combination of the $20\,$cm
     and the $21\,$cm radio continuum data (see Section
     \ref{subsec:combradio}). \label{fig:boxplot3b}}
\end{figure*}

\begin{figure*}
   \centering
   \includegraphics[width=0.9\textwidth]{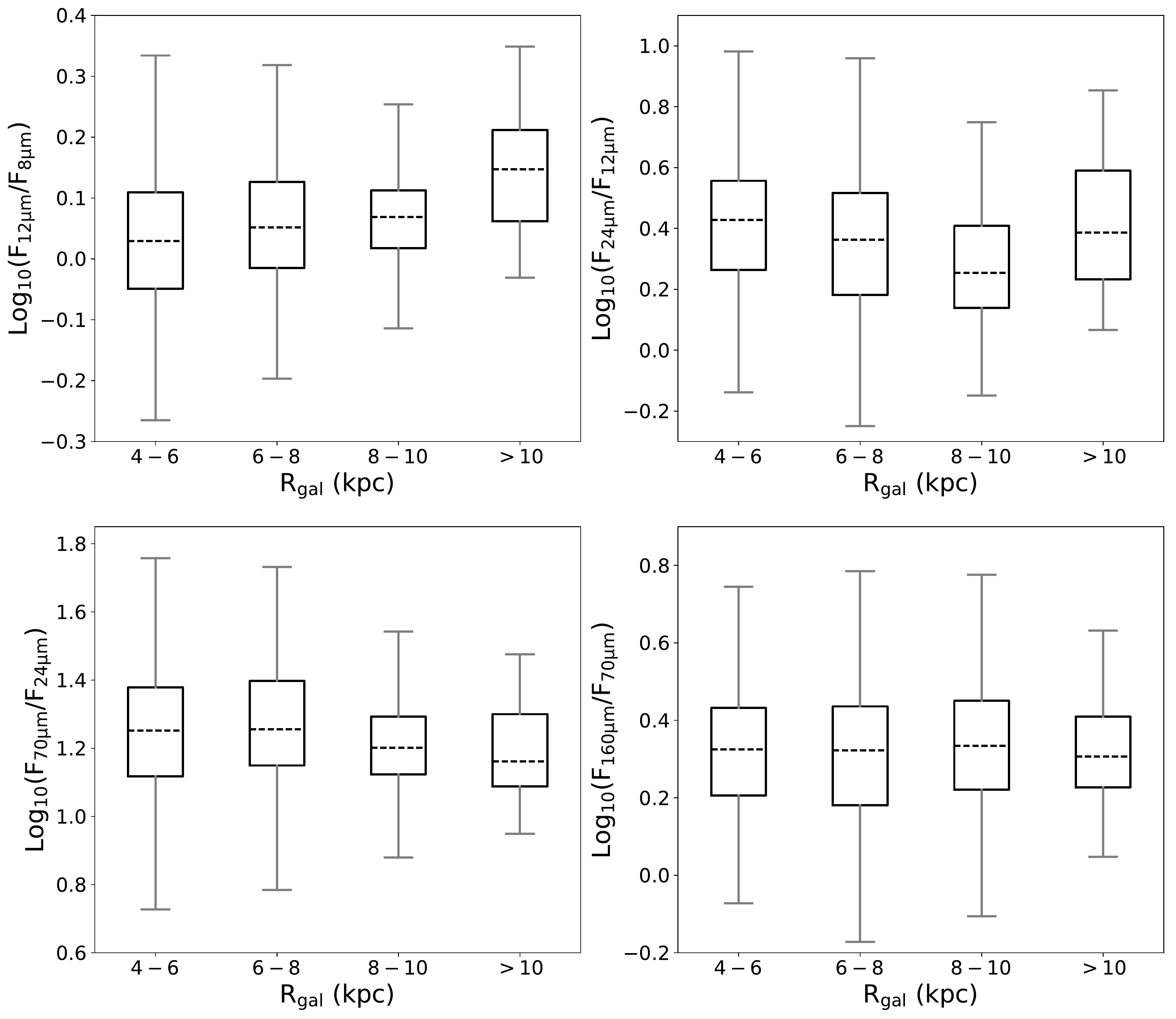}
   \caption{Box and whiskers plots as in Figure~\ref{fig:boxplot1b}
     for IR colors in four Galactocentric radii bins. The plots, show
     similar infrared colors for all \hii regions, regardless of
     $R_{gal}$, except the $\log_{10}(F_{12\microns}/F_{8\microns})$ 
     ratio which shows slightly increasing trend, and the 
     $\log_{10}(F_{24\microns}/F_{12\microns})$ ratio that shows some 
     fluctuations. \label{fig:boxplot4}}
\end{figure*}

We analyze the IR to radio flux density ratios
(Figure~\ref{fig:boxplot3b}), and the IR flux density ratios
(Figure~\ref{fig:boxplot4}), as

\newpage

\begin{deluxetable*}{lcDDcc}
\tabletypesize{\small}
\tablecaption{The median IR to radio, and IR flux density ratios as a function of $R_{gal}$ \label{tab:bxplt_res4}}
\tablehead{
\colhead{Flux density ratios} & \colhead{$\mathrm{R_{gal}}$} & \multicolumn2c{$\mathrm{Median}$} & \multicolumn2c{$\mathrm{\sigma}$} & \colhead{Color criteria} & \colhead{$\#$ of sources} \\
 & [kpc] & \multicolumn{2}{c}{} & \multicolumn{2}{c}{} & $\mathrm{(Q1-1.5 \times IQR}$ --- $\mathrm{Q3 + 1.5 \times IQR)}$\tablenotemark{*} & }
\decimals
\startdata
\multirow{5}{*}{$\mathrm{\log_{10}(F_{8\microns}/F_{\rm radio})}$} & $4 < R_{gal} < 6$ & $1.50$ & $0.34$ & $0.38 - 2.75$ & $322$ \\
  & $6 < R_{gal} < 8$  & $1.52$ & $0.37$ & $0.35 - 2.83$ & $260$ \\
  & $8 < R_{gal} < 10$ & $1.75$ & $0.50$ & $0.10 - 2.99$ & $48$ \\
  & $R_{gal} > 10$     & $1.20$ & $0.32$ & $0.33 - 2.25$ & $41$ \\
  & All                & $1.43$ & $0.38$ & $0.10 - 2.99$ & $658$ \\
  \multicolumn{7}{c}{} \\
\multirow{5}{*}{$\mathrm{\log_{10}(F_{12\microns}/F_{\rm radio})}$} & $4 < R_{gal} < 6$ & $1.54$ & $0.29$ & $0.68 - 2.61$ & $327$ \\
  & $6 < R_{gal} < 8$  & $1.59$ & $0.41$ & $0.41 - 2.94$ & $264$ \\
  & $8 < R_{gal} < 10$ & $1.82$ & $0.50$ & $0.32 - 2.93$ & $46$ \\
  & $R_{gal} > 10$     & $1.59$ & $0.25$ & $0.44 - 2.72$ & $41$ \\
  & All                & $1.64$ & $0.36$ & $0.32 - 2.94$ & $678$ \\
  \multicolumn{7}{c}{} \\
\multirow{5}{*}{$\mathrm{\log_{10}(F_{24\microns}/F_{\rm radio})}$} & $4 < R_{gal} < 6$ & $1.96$ & $0.23$ & $1.18 - 2.86$ & $324$ \\
  & $6 < R_{gal} < 8$  & $1.94$ & $0.28$ & $1.02 - 2.93$ & $259$ \\
  & $8 < R_{gal} < 10$ & $2.06$ & $0.23$ & $1.26 - 2.53$ & $45$ \\
  & $R_{gal} > 10$     & $1.74$ & $0.30$ & $1.06 - 2.60$ & $25$ \\
  & All                & $1.93$ & $0.26$ & $1.02 - 2.93$ & $653$ \\
    \multicolumn{7}{c}{} \\
\multirow{5}{*}{$\mathrm{\log_{10}(F_{70\microns}/F_{\rm radio})}$} & $4 < R_{gal} < 6$ & $3.15$ & $0.31$ & $2.38 - 4.16$ & $330$ \\
  & $6 < R_{gal} < 8$  & $3.21$ & $0.28$ & $2.11 - 4.35$ & $268$ \\
  & $8 < R_{gal} < 10$ & $3.23$ & $0.25$ & $2.40 - 3.98$ & $46$ \\
  & $R_{gal} > 10$     & $3.01$ & $0.32$ & $2.15 - 3.95$ & $33$ \\
  & All                & $3.15$ & $0.29$ & $2.11 - 4.16$ & $677$ \\
  \multicolumn{7}{c}{} \\
\multirow{5}{*}{$\mathrm{\log_{10}(F_{160\microns}/F_{\rm radio})}$} & $4 < R_{gal} < 6$ & $3.54$ & $0.29$ & $2.61 - 4.61$ & $326$ \\
  & $7 < R_{gal} < 8$  & $3.54$ & $0.40$ & $2.23 - 4.96$ & $261$ \\
  & $8 < R_{gal} < 10$ & $3.57$ & $0.37$ & $2.45 - 4.69$ & $46$ \\
  & $R_{gal} > 10$     & $3.37$ & $0.29$ & $2.27 - 4.19$ & $33$ \\
  & All                & $3.51$ & $0.34$ & $2.23 - 4.96$ & $665$ \\
\hline
\multirow{5}{*}{$\mathrm{\log_{10}(F_{12\microns}/F_{8\microns})}$} & $4 < R_{gal} < 6$ & $0.03$ & $0.09$ & $-0.27 - 0.33$ & $327$ \\
  & $6 < R_{gal} < 8$  & $0.05$ & $0.08$ & $-0.20 - 0.32$ & $306$ \\
  & $8 < R_{gal} < 10$ & $0.07$ & $0.05$ & $-0.11 - 0.25$ & $56$ \\
  & $R_{gal} > 10$     & $0.15$ & $0.09$ & $-0.03 - 0.35$ & $30$ \\
  & All                & $0.08$ & $0.08$ & $-0.27 - 0.35$ & $719$ \\
  \multicolumn{7}{c}{} \\
\multirow{5}{*}{$\mathrm{\log_{10}(F_{24\microns}/F_{12\microns})}$} & $4 < R_{gal}< 6$ & $0.43$ & $0.17$ & $-0.14 - 0.98$ & $328$ \\
  & $6 < R_{gal} < 8$  & $0.36$ & $0.18$ & $-0.25 - 0.96$ & $304$ \\
  & $8 < R_{gal} < 10$ & $0.25$ & $0.16$ & $-0.15 - 0.75$ & $55$ \\
  & $R_{gal}> 10$      & $0.39$ & $0.20$ & $0.07 - 0.85$  & $27$ \\
  & All                & $0.36$ & $0.18$ & $-0.25 - 0.98$ & $714$ \\
  \multicolumn{7}{c}{} \\
\multirow{5}{*}{$\mathrm{\log_{10}(F_{70\microns}/F_{24\microns})}$} & $4 < R_{gal} < 6$ & $1.25$ & $0.13$ & $0.73 - 1.76$ & $331$ \\
  & $6 < R_{gal} < 8$  & $1.26$ & $0.14$ & $0.78 - 1.73$ & $306$ \\
  & $8 < R_{gal} < 10$ & $1.20$ & $0.09$ & $0.88 - 1.54$ & $54$ \\
  & $R_{gal} > 10$     & $1.16$ & $0.14$ & $0.95 - 1.48$ & $27$ \\
  & All                & $1.22$ & $0.13$ & $0.73 - 1.76$ & $718$ \\
  \multicolumn{7}{c}{} \\
\multirow{5}{*}{$\mathrm{\log_{10}(F_{160\microns}/F_{70\microns})}$} & $4 < R_{gal} < 6$ & $0.32$ & $0.11$ & $-0.07 - 0.74$ & $332$ \\
  & $6 < R_{gal} < 8$  & $0.32$ & $0.14$ & $-0.17 - 0.79$ & $313$ \\
  & $8 < R_{gal} < 10$ & $0.33$ & $0.11$ & $-0.11 - 0.78$ & $56$ \\
  & $R_{gal} > 10$     & $0.31$ & $0.10$ & $0.05 - 0.63$  & $35$ \\
  & All                & $0.32$ & $0.11$ & $-0.17 - 0.79$ & $736$ \\
\enddata
\tablenotetext{*}{$Q1$ and $Q3$ are the $25^th$ and $75^th$ percentile of the data, respectively, and $IQR=(Q3-Q1)$ is the interquartile.}
\end{deluxetable*}

\begin{figure*}[h!]
   \centering
   \includegraphics[width=0.9\textwidth]{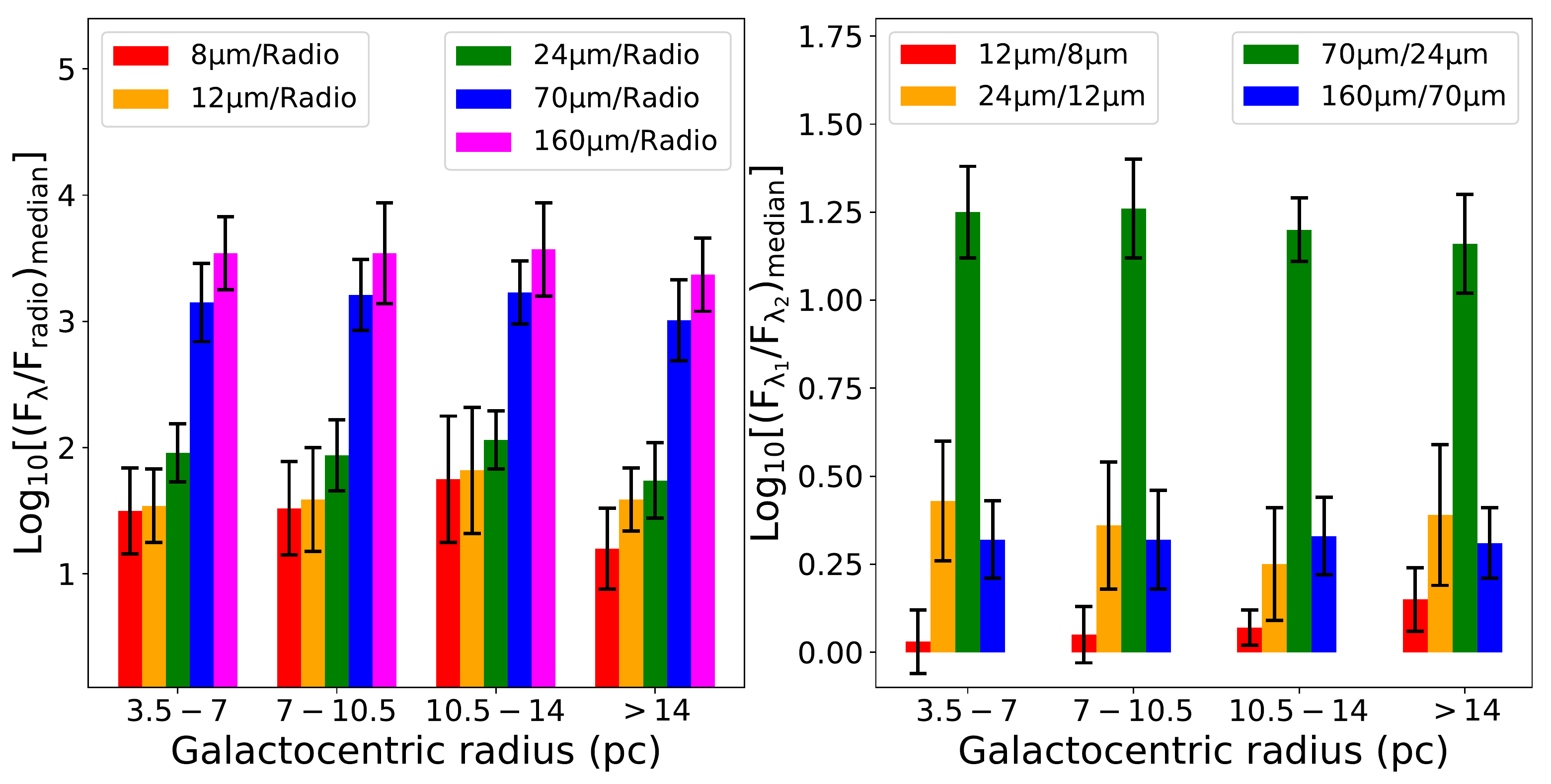}
   \caption{Graphical representation of the median values within the
     $R_{gal}$ bins from Table~\ref{tab:bxplt_res4}. The standard
     deviation of \textit{all} data points are marked with the error
     bars. The term ``radio'' means the combination of the
       $20\,$cm and the $21\,$cm radio continuum data, see Section
       \ref{subsec:combradio}. On the left plot, we show the ratios
     of the IR and radio flux densities.  There is essentially no
     change in the IR to radio flux density ratios with $R_{gal}$, 
     altough the $8$\microns to radio ratio is low for the regions with 
     the largest $R_{gal}$ values. The right plot show the distribution of
     median values of color indices. This plot suggests that all the
     investigated IR flux density ratios are unchanged within the
     errors, independent of Galactocentric
     radius. \label{fig:barplot2}}
\end{figure*}

\noindent{a fuction of Galactocentric radius,
$R_{gal}$.  We summarize the results of these two Figures in
Table~\ref{tab:bxplt_res4}, and graphically in
Figure~\ref{fig:barplot2}.}

Interestingly, the IR to radio ratios do not show a consistent trend 
with $R_{gal}$ (Figure~\ref{fig:boxplot3b}); \hii regions 
throughout the Galaxy have similar IR to radio ratios. Again, the final bin
$R_{gal} > 10\,\kpc$ shows some deviation from this trend compared
with the rest of the Galaxy. The radio continuum is a proxy for the
power of the ionizing source(s), where as the IR arises from various
dust species associated with the regions. A lower IR to radio ratio in
the outer Galaxy may be caused by a lack of dust, a lack of photons to
excite the dust, or both.

\citet{crocker2013} investigated the ratio of $8$\microns to H$\alpha$
for \hii regions in NGC628 as a function of galactocentric
radius. They found that relative to H$\alpha$, the $8$\microns
emission decreases with increasing galactocentric radius. Over the
range of galactocentric radii probed, the H$\alpha$ to $8$\microns
flux density ratio decreases by a factor of $4.0$. \citet{crocker2013}
suggested that at high $R_{gal}$ there are fewer photons absorbed by
PAHs, leading to lower $8$\microns emission. They also suggested that
the negative trend they found may be caused by the metallicity
gradient in the disk, because PAH emission is weaker at lower
metallicities \citep[e.g.][]{engelbracht2005, galliano2005}.

\begin{figure*}
   \centering
   \includegraphics[width=0.6\textwidth]{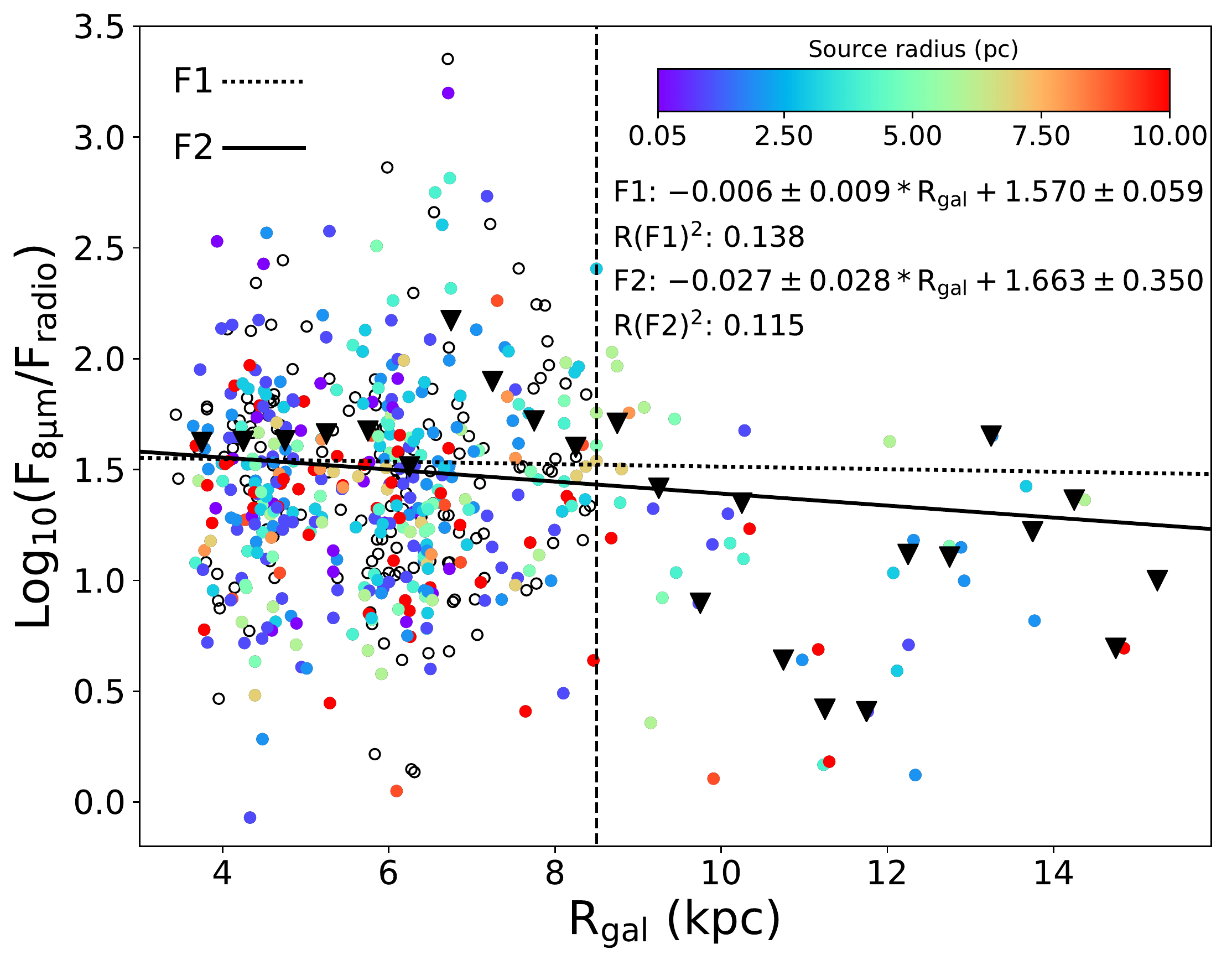}
   \caption{The $\log_{10}(F_{8\microns}/F_{radio})$ flux density
     ratio of $\sim 600$ \hii regions as a function of Galactocentric
     radius. The $F_{radio}$ denotes the combination of the
       $20\,$cm and the $21\,$cm radio continuum data, see Section
       \ref{subsec:combradio}. The empty circles mark the sources
     that lack Heliocentric distances. The vertical dashed line
     denotes the boundary between the inner and outer Galaxy (at
     $8.5\,$kpc), the dotted line shows the best fit to all
       data points and the solid line is the best fit to the $250\,$pc
       binned data (filled triangles). The fits are similar
       within the fit uncertainties. \label{fig:rgal1}}
\end{figure*}

Although we do not have H$\alpha$ data for all regions, radio
continuum traces the same ionized gas. We fit a linear regression line
to the $\log_{10}(F_{8\microns}/F_{radio})$ ratio. We find a slight 
negative trend, in agreement with \citet{crocker2013}. The slope of the 
fit, however, $-0.006 \pm 0.009$, is flat within the uncertainties over 
the range of $R_{gal}$ probed here ($\sim 4-15$\,kpc). Only $\sim 7\%$ of 
the the data points are located in the outer Galaxy ($\rgal > 8.5\,$kpc), 
and nearly all ($80\%$) of the outer Galaxy data points fall below the fit 
line. Since the fit can be biased by the low number of data points in the 
outer Galaxy, we also fit the $500\,$pc binned data (filled triangles on 
Figure~\ref{fig:rgal1}). This fit also shows a weak negaive trend. Additional 
outer Galaxy data points would help to determine the strength of the relationship.

\section{Summary}
\label{sec:sum}
We derived the infrared ($8-160$\microns) and radio ($20\,$cm and
$21\,$cm) flux densities for $\sim 1000$ \hii regions from the WISE
Catalog of Galactic \hii Regions to investigate relationships between
the flux densities, angular sizes, and Galactocentric radii. The
investigated \hii regions are located in the first Galactic quadrant
($17\fd5 < \ell < 65\degree$) where kinematic distances are relatively
accurate. 

Comparing VGPS and MAGPIS flux densities, we were unable to reproduce
the result of \citet{helfand2006} that MAGPIS is overestimating the
flux densities of large sources. This indicates that either MAGPIS is
well-calibrated, or that both the VGPS and MAGPIS underestimate the
flux densities of large sources. By comparing $21\,$cm flux densities
with those derived at $3\,$cm, we showed that radio optical depth
effects are statistically unimportant for our sample.

All measured IR and radio flux densities are highly correlated, 
with a higher scatter at the lower end of the flux densities. With the 
exception of $70$\microns data, the IR emissions have similar strong 
correlation with radio data. Removing the smallest regions from the fit 
increases the correlation between the IR and radio flux densities.

All \hii regions have similar infrared flux density ratios, regardless
of \hii region physical size. WC89 suggested that ultra-compact (UC)
\hii regions alone are well separated from other sources found in the
IRAS Point Source Catalog. Our results show that \hii regions of all
physical sizes can satisfy their criteria, not just UC \hii regions,
in agreement with previous studies
\citep[e.g.][]{anderson2012a,leto2009,chini1987}.  This result implies
that the total number of UC \hii\ regions in the Galaxy as derived
from IR color indices is significantly uncertain.  The 160 to
70\,\micron\ flux density ratio is above unity for nearly all regions,
which implies that emission from large, cold dust grains dominates the
spectral energy distributions.
Using our data, we suggest that the 
$\log_{10}(F_{24\microns}/F_{12\microns}) \ge 0$ and 
$\log_{10}(F_{70\microns}/F_{12\microns}) \ge 1.2$, and 
$\log_{10}(F_{24\microns}/F_{12\microns}) \ge 0$ and 
$\log_{10}(F_{160\microns}/F_{70\microns}) \le 0.67$ IR color indice pairs 
can be used as an approximate IR limits for \hii regions, independent of 
size.

We find a trend of weakly decreasing IR to radio flux density ratios
with increasing $R_{gal}$. This has been noted for external galaxies
\citep{crocker2013}, and we confirm their results using $8.0$\microns
and radio continuum data. Since the IR emission traces dust, and the
trend is seen in all IR wavelengths, this result indicates that there
may be decreasing dust abundance in the outer Galaxy.

\acknowledgements

We thank the anonymous referee for his/her comments, which 
improved the paper.
This work was supported by NASA ADAP grant NNX12AI59G to LDA and NSF
grant AST1516021 to LDA.  PACS has been developed by a
con\-sor\-ti\-um of institutes led by MPE (Germany) and including UVIE
(Austria); KU Leuven, CSL, IMEC (Belgium); CEA, LAM (France); MPIA
(Germany); INAF-IFSI/OAA/OAP/OAT, LENS, SISSA (Italy); IAC
(Spain). This development has been supported by the funding agencies
BMVIT (Austria), ESA-PRODEX (Belgium), CEA/CNES (France), DLR
(Germany), ASI/INAF (Italy), and CICYT/MCYT (Spain). This research has
made use of NASAs Astrophysics Data System Bibliographic Services and
the SIMBAD database operated at CDS, Strasbourg, France.  This
publication makes use of data products from WISE, which is a joint
project of the University of California, Los Angeles, and the Jet
Propulsion Laboratory/California Institute of Technology, funded by
the National Aeronautics and Space Administration. This research has
made use of NASA's Astrophysics Data System Bibliographic Services.

\bibliographystyle{aasjournal}
\bibliography{Hiipaper}

\hspace{0pt}

\clearpage

\appendix
\section{WISE Catalog Web Site}
We have updated the WISE Catalog Web Site\footnote{\url{http://astro.phys.wvu.edu/wise/}} 
with the aperture photometry results given here.

\section{Flux density uncertainties}
\label{appdx:fr_err}

\begin{figure*}
   \centering
   \includegraphics[width=0.9\textwidth]{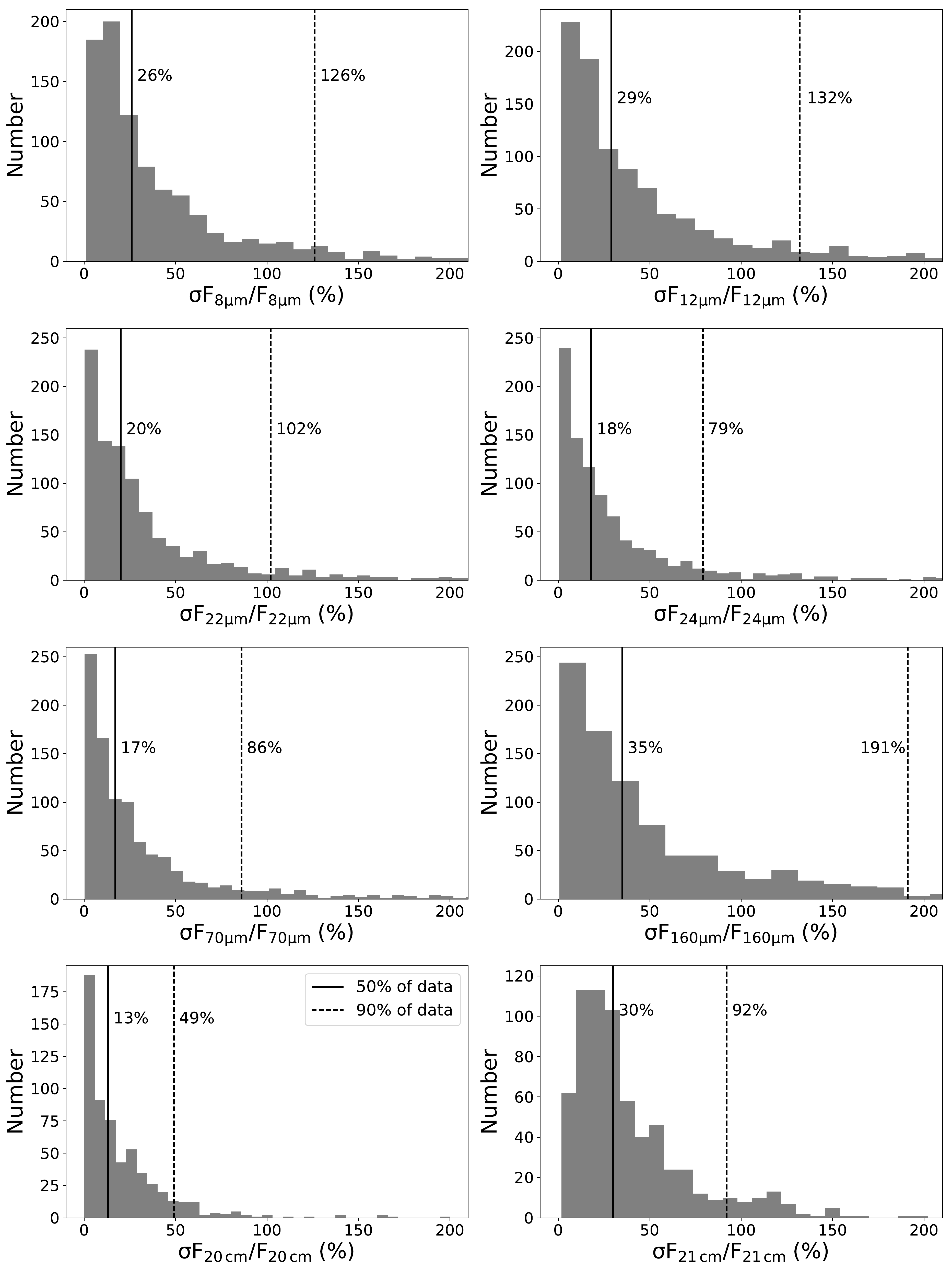}
   \caption{Distribution of fractional uncertainties ($\sigma F_{\lambda}/F_{\lambda}$). 
     The solid and dashed lines represent 
     the fractional error limits at a given wavelength, for the $50\%$ 
     and $90\%$ of the data, respectively (marked in the lower left 
     subplot). \label{fig:appdx1}}
\end{figure*}

\begin{figure*}
   \centering
   \includegraphics[width=0.9\textwidth]{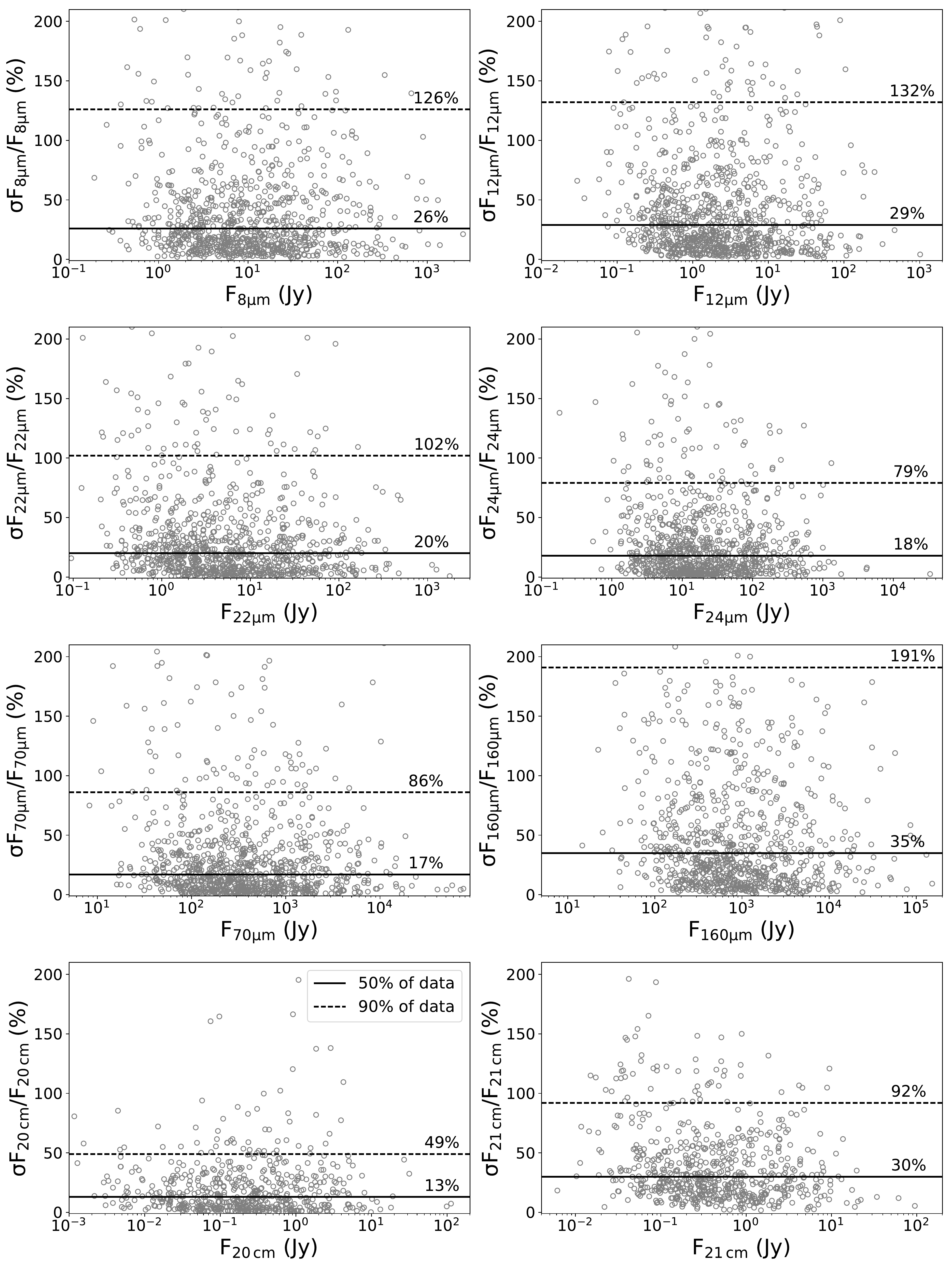}
   \caption{Distribution of fractional uncertainties as a function of flux density. 
     Horizontal lines represent the same as in Figure~\ref{fig:appdx1}. 
     The fractional uncertainties are not strongly correlated with the flux densities. 
     \label{fig:appdx2}}
\end{figure*}

To check the uncertainties of the derived IR ($8$\microns,
$12$\microns, $22$\microns, $24$\microns, $70$\microns, and
$160$\microns,) and radio ($20\,$cm and $21\,$cm) flux densities, we
investigate their relative (or fractional) errors
(Figures~\ref{fig:appdx1} and \ref{fig:appdx2}). The number
distributions of the fractional errors (~\ref{fig:appdx1}) show that,
considering \textit{all} wavelengths, $50\%$ of the data lies under
$\sim 35\%$, while $90\%$ of the data is under $\sim 191\%$ of
fractional uncertainty. We also investigated the relative errors as a
function of flux density values, Figure~\ref{fig:appdx2}. These plots
suggest that the fractional errors are not correlated with their flux
density values, independent of wavelengths.

\begin{deluxetable}{lcc}
\tablecaption{Fractional error limits at $50\%$ and $90\%$ of the data points. \label{tab:frac_errors}}
\tablehead{
\colhead{$\lambda$} & \multicolumn2c{$\sigma F_{\lambda}/F_{\lambda}$ ($\%$)}}
\startdata
 & at $50\%$ of data & at $90\%$ of data \\
\cline{2-3}
$\mathrm{8\microns}$   & $\sim 26\%$ & $\sim 126\%$\\
$\mathrm{12\microns}$  & $\sim 29\%$ & $\sim 132\%$\\
$\mathrm{22\microns}$  & $\sim 20\%$ & $\sim 102\%$\\
$\mathrm{24\microns}$  & $\sim 18\%$ & $\sim 79\%$\\
$\mathrm{70\microns}$  & $\sim 17\%$ & $\sim 86\%$\\
$\mathrm{160\microns}$ & $\sim 35\%$ & $\sim 191\%$\\
$\mathrm{20\,cm}$      & $\sim 13\%$ & $\sim 49\%$\\
$\mathrm{21\,cm}$      & $\sim 30\%$ & $\sim 92\%$
\enddata
\end{deluxetable}

\end{document}